\documentclass[prd,aps,nofootinbib,showpacs,preprintnumbers,amsmath,amssymb,floatfix]{revtex4}

\usepackage{times}
\usepackage{hyperref}

\usepackage{graphicx}% Include figure files
\usepackage{dcolumn}% Align table columns on decimal point
\usepackage{bm}% bold math
\usepackage{slashed}

\newcommand{\mathsym}[1]{{}}
\newcommand{\unicode}[1]{{}}

\begin{document}

\title{$s$-wave baryon nonleptonic decay amplitude in large-$\bm{N_c}$ chiral perturbation theory}

\author{
Rub\'en Flores-Mendieta
}
\affiliation{
Instituto de F{\'\i}sica, Universidad Aut\'onoma de San Luis Potos{\'\i}, \'Alvaro Obreg\'on 64, Zona Centro, San Luis Potos{\'\i}, S.L.P.\ 78000, Mexico
}

\date{\today}

\begin{abstract}
The $s$-wave decay amplitude in the nonleptonic decay of baryons is analyzed within heavy baryon chiral perturbation theory in the large-$N_c$ limit at one-loop order, where $N_c$ is the number of color charges. Loop graphs with octet and decuplet intermediate states are systematically incorporated into the analysis and the effects of the decuplet-octet mass difference are accounted for. There are large-$N_c$ cancellations between different one-loop graphs as a consequence of the large-$N_c$ spin-flavor symmetry of QCD baryons. The predictions of large-$N_c$ baryon chiral perturbation theory are in very good agreement both with the expectations from the $1/N_c$ expansion and with the experimental data.

\end{abstract}


\maketitle

\section{Introduction}

The remarkable success of the $1/N_c$ expansion of QCD---where $N_c$ is the number of color charges \cite{thooft,witten}---and its subsequent combination with heavy-baryon chiral perturbation theory \cite{jen96} to describe several static properties of baryons has been evident over the past two decades.

Initially, a $1/N_c$ expansion of the chiral Lagrangian was formulated in Ref.~\cite{jen96}; since then, the Lagrangian has been useful to evaluate nonanalytic meson-loop corrections to baryon amplitudes in the $1/N_c$ expansion for finite $N_c$. Specifically, the method was originally applied to compute flavor-$\mathbf{27}$ baryon mass splittings at leading order in chiral perturbation theory \cite{jen96}. Later, a number of additional baryon properties were also successfully evaluated, namely, baryon axial-vector couplings \cite{rfm06,rfm12}, baryon magnetic moments \cite{rfm09,rfm14}, baryon vector couplings \cite{rfm14a}, and Dirac form factors \cite{rfm15} to name but a few. 

The approach to compute nonanalytic meson-loop corrections in $1/N_c$ baryon chiral perturbation theory at finite $N_c$ consists in identifying all the pertinent one-loop Feynman diagrams for the process under consideration. These diagrams are given by the product of a baryon operator with well-defined transformation properties under the spin-flavor symmetry times a loop integral, which depends nonanalytically on the light quark masses $m_q$. In this way, the $1/N_c$ and group theoretic structure of the loop corrections are manifest. Although theoretically the procedure is straightforward, in practice the reduction of the baryon operator becomes a rather involved task. With the advent of more powerful technical computing systems, the reduction is possible to an unprecedented level.

All the computations of baryon properties mentioned above generalize the formulas obtained previously in conventional baryon chiral perturbation theory (i.e., without a $1/N_c$ expansion). An extra feature of the approach is that the $1/N_c$ formulas exhibit the $1/N_c$ and flavor-breaking structure of the one-loop corrections so various relations obtained in the limit of exact $SU(3)$ flavor symmetry (for instance, the Coleman-Glashow relations for baryon magnetic moments \cite{rfm09,rfm14}) can be better understood.

In the framework of baryon chiral perturbation theory, the analyses of $s$- and $p$-wave amplitudes have been addressed in Refs.~\cite{bij,jen92,b99,abd}, each of which with some particular focus. References \cite{jen92} and \cite{abd} evaluated the leading nonanalytical corrections including both octet and decuplet baryons as intermediate states, focusing on the $|\Delta I|=1/2$ component of the decay amplitude (i.e., the so-called $|\Delta I|=1/2$ rule was assumed to be valid). There are a few differences in some decay diagrams between these two analyses. Reference \cite{b99} also assumed the validity of the $|\Delta I|=1/2$ rule, but included only octet baryons as intermediate states in the loops so the effects of the decuplet baryons were incorporated into the low-energy constants of the effective Lagrangian; to this purpose, all counterterms to chiral order $\mathcal{O}(p^2)$ and some terms of order $\mathcal{O}(p^3)$ were included. While Refs.~\cite{jen92} and \cite{abd} conclude that good agreement with experiment cannot be simultaneously obtained using $s$- and $p$-wave amplitudes at one-loop level, Ref.~\cite{b99} claims the opposite.

In this paper, the applicability of the combined expansion in $1/N_c$ and chiral corrections is extended to the analysis of decay amplitudes in the nonleptonic decays of baryons. Due to the enormous amount of algebraic calculations involved, it is more appropriate to present first the $s$-wave amplitude here and to leave the $p$-wave one for a further paper. To this end, one-loop graphs with intermediate spin-1/2 octet and spin-3/2 decuplet baryon states are analyzed including the full dependence on the decuplet-octet baryon mass difference, while at the same time including the cancellations that follow from the large-$N_c$ spin-flavor symmetry of baryons.

The organization of the paper is as follows. In Sec.~\ref{sec:combined} the central ideas on the combined formalism are provided in order to introduce the notation and conventions. In Sec.~\ref{sec:nonlep} a theoretical description of baryon nonleptonic decays is presented, with emphasis on the calculation of tree-level $s$-wave amplitudes. In Sec.~\ref{sec:oneloop} the one-loop contributions to the $s$-wave amplitude are evaluated; partial operator reductions already performed in Ref.~\cite{rfm14a} are recognized to be present in the current analysis so they are borrowed and adapted to make up the new results. At this point, a direct comparison with conventional baryon chiral perturbation theory is performed. The comparison is done by identifying the existing relations between the chiral coefficients and the operator coefficients that appear in the present analysis. Both analyses agree in full. In Sec.~\ref{sec:sb} explicit symmetry breaking corrections to linear order in the quark mass $m_s$ are evaluated. As discussed in the text, these contributions are necessary to get a consistent numerical analysis, which is performed in Sec.~\ref{sec:num} through a least-squares fit to data \cite{part}. The analysis is satisfactory. In Sec.~\ref{sec:fin} some concluding remarks are addressed. The paper is complemented by two appendices. In Appendix \ref{sec:reduc} all the new operator reductions required are listed whereas in Appendix \ref{sec:opcoeff} all the coefficients that come along with the baryon operators in the several one-loop contributions are provided.

\section{\label{sec:combined}Baryon chiral perturbation theory in the $1/N_c$ expansion}

The aspects related to the $1/N_c$ expansion for baryons have been discussed in detail in Refs.~\cite{dm315,djm95,jen96}, so in this section a survey to introduce the notation and conventions used is provided. To start with, it should be recalled that in the large-$N_c$ limit the lowest-lying baryons\footnote{The $J^P = 1/2^+$ octet containing the nucleon and the $J^P = 3/2^+$ decuplet containing the $\Delta(1232)$ together make up the ground-state 56-plet of $SU(6)$.} are given by the completely symmetric spin-flavor representation of $N_c$ quarks $SU(2N_f)$ \cite{dm315,gs}. Under $SU(2)\times SU(N_f)$, this representation decomposes into a tower of baryon flavor representations with spins $J=1/2,3/2,\ldots,N_c/2$. Corrections to the large-$N_c$ limit are expressed in terms of $1/N_c$ suppressed operators \cite{dm315}, which yields the $1/N_c$ expansion of QCD.

The $1/N_c$ expansion of a QCD $m$-body quark operator acting on a single baryon state can be written in the most general way as \cite{djm95}
\begin{equation}
\mathcal{O}_{\textrm{QCD}}^{\textrm{$m$-body}} = N_c^m \sum_n c_n \frac{1}{N_c^n} \mathcal{O}_n,
\end{equation}
where the $\mathcal{O}_n$ ($0\leq n \leq N_c$) constitute a complete set of linearly independent operator products which are of $n$th order in the baryon spin-flavor generators $J^k$, $T^c$ and $G^{kc}$, and the $c_n(1/N_c)$ are arbitrary
unknown coefficients with an expansion in $1/N_c$ beginning at order unity. Examples of $1/N_c$ expansions for baryon operators include the $1/N_c$ expansion of the baryon mass operator $\mathcal{M}_{\mathrm{baryon}}$ and the baryon axial vector current $A^{kc}$. The former is given by \cite{djm95}
\begin{equation}
\mathcal{M}_{\mathrm{baryon}} = m_{0}^{0,\mathbf{1}}N_c\openone + \sum_{n=2,4}^{N_c-1} m_{n}^{0,\mathbf{1}} \frac{1}{N_c^{n-1}}J^n, \label{eq:massop}
\end{equation}
where the coefficients $m_{n}^{0,\mathbf{1}}$ are \textit{a priori} unknown parameters of order $\mathcal{O}(\Lambda_\chi)$, and the superscripts attached to them indicate the spin-flavor representation they belong to. The first summand in Eq.~(\ref{eq:massop}) denotes the overall spin-independent mass of the baryon multiplet and the remaining terms, which are spin-dependent, constitute $\mathcal{M}_{\textrm{hyperfine}}$.

The $1/N_c$ expansion of the baryon axial vector current, in turn, can be written for $N_c=3$ as \cite{djm95}
\begin{equation}
A^{kc} = a_1 G^{kc} + \frac{b_2}{N_c} \mathcal{D}_2^{kc} + \frac{b_3}{N_c^2} \mathcal{D}_3^{kc} + \frac{c_3}{N_c^2} \mathcal{O}_3^{kc}, \label{eq:akc}
\end{equation}
where the unknown coefficients $a_1$, $b_n$ and $c_n$ also have expansions in $1/N_c$ beginning at order unity and the leading operators that accompany them are given explicitly by
\begin{eqnarray}
\mathcal{D}_2^{kc} & = & J^kT^c, \label{eq:d2kc} \\
\mathcal{D}_3^{kc} & = & \{J^k,\{J^r,G^{rc}\}\}, \label{eq:d3kc} \\
\mathcal{O}_3^{kc} & = & \{J^2,G^{kc}\} - \frac12 \{J^k,\{J^r,G^{rc}\}\}. \label{eq:o3kc}
\end{eqnarray}
Higher order operators are constructed from the previous ones by anticommuting them with $J^2$.

The chiral Lagrangian for baryons $\mathcal{L}_{\text{baryon}}$, formulated to understand the low-energy dynamics of baryons interacting with the pion nonet $\pi$, $K$, $\eta$, and $\eta^\prime$ in a combined expansion in $1/N_c$ and chiral symmetry breaking, was given explicitly in Ref.~\cite{jen96}. In the baryon rest frame, $\mathcal{L}_{\text{baryon}}$ reads
\begin{equation}
\mathcal{L}_{\text{baryon}} = i \mathcal{D}^0 - \mathcal{M}_{\text{hyperfine}} + \text{Tr} \left(\mathcal{A}^k \lambda^c \right) A^{kc} + \frac{1}{N_c} \text{Tr} \left(\mathcal{A}^k \frac{2I}{\sqrt 6}\right) A^k + \ldots, \label{eq:ncch}
\end{equation}
where the covariant derivative reads
\begin{equation}
\mathcal{D}^0 = \partial^0 \openone + \text{Tr} \left(\mathcal{V}^0 \lambda^c\right) T^c. \label{eq:kin}
\end{equation}
The ellipses in Eq.~(\ref{eq:ncch}) refer to higher partial wave meson couplings which occur at subleading orders in the $1/N_c$ expansion for $N_c>3$ \cite{jen96}. The Lagrangian depends on the meson field $\xi(x)=\exp[i\Pi(x)/f_\pi]$ through the vector and axial-vector currents
\begin{equation}
\mathcal{V}^0 = \frac12 \left(\xi \partial^0 \xi^\dagger + \xi^\dagger \partial^0 \xi\right), \qquad
\mathcal{A}^k = \frac{i}{2} \left(\xi \nabla^k \xi^\dagger - \xi^\dagger \nabla^k \xi\right), \qquad \qquad
\end{equation}
where $\Pi(x)$ represents the nonet of Goldstone boson fields and $f_\pi \approx 93$ $\mathrm{MeV}/c^2$ is the pion decay constant.

Next, explicit flavor symmetry breaking is accounted for in the baryon chiral Lagrangian through terms containing powers of the quark mass matrix. The leading Lagrangian with a single insertion of the quark mass matrix can be written as \cite{jen96}
\begin{equation}
\mathcal{L}_{\text{baryon}}^{\mathcal{M}} = \mathrm{Tr} \left[ [\xi \mathcal{M}(\theta) \xi + \xi^\dag \mathcal{M}^\dag(\theta) \xi^\dag ] \frac{\lambda^a}{2} \right] \mathcal{H}^a + \frac{1}{N_c} \mathrm{Tr} \left[ [\mathcal{M}(\theta) \Sigma + \mathcal{M}^\dag(\theta) \Sigma^\dag ] \frac{I}{\sqrt{6}} \right] \mathcal{H}^0, \label{eq:lbarm}
\end{equation}
where $a=3,8,9$, and the explicit symmetry breaking perturbations to the baryon Hamiltonian read \cite{jen96}
\begin{equation}
\mathcal{H}^0 = b_{(0)}^{0,\mathbf{1}} N_c\openone + b_{(2)}^{0,\mathbf{1}} \frac{1}{N_c} J^2,
\end{equation}
and
\begin{equation}
\mathcal{H}^a = b_{(1)}^{0,\mathbf{8}} T^a + b_{(2)}^{0,\mathbf{8}} \frac{1}{N_c} \{J^r,G^{ra}\} + b_{(3)}^{0,\mathbf{8}} \frac{1}{N_c^2} \{J^2,T^a\}, \label{eq:Ha}
\end{equation}
for $N_c=3$. Here $b_{(n)}^{0,\mathbf{rep}}$ are unknown parameters which come along with $n$-body operators within flavor representation $\mathbf{rep}$. Additional details about $\mathcal{M}(\theta)$ can be found in the original paper \cite{jen96}.

\section{\label{sec:nonlep}Baryon nonleptonic decays}

The dominant decays of baryons are the nonleptonic modes
\begin{equation}
B_i (p_i) \to B_f (p_f) + \pi(k),
\end{equation}
where $B_i$ and $B_f$ are the decaying and emitted baryons and $\pi$ is the emitted pion, respectively, with momenta $p_i$, $p_f$ and $k$. These $\Delta S=1$ processes are quite useful in understanding the electroweak interaction in hadrons.

The decay amplitude for the nonleptonic decays of spin-$1/2$ baryons can be written as \cite{part}
\begin{equation}
{\mathcal M} = G_F m_{\pi^+}^2 \bar{u}_{B_f} (A - B \gamma_5)u_{B_i},
\end{equation}
where $G_F$ is the Fermi constant, $m_{\pi^+}$ is the pion mass, and $A$ and $B$ are parity-violating $s$-wave and parity-conserving $p$-wave decay amplitudes. $A$ and $B$ are related to the amplitudes $s$ and $p$ by
\begin{equation}
s = A, \qquad p =B \frac{|\mathbf{p}_f|}{E_f + M_f},
\end{equation}
where $M_f$, $E_f$, and $\mathbf{p}_f$ are the mass, energy, and three-momentum of the final baryon; the usual observables, e.g., the partial decay rate $\Gamma$ and the decay asymmetry $\alpha$, can therefore be expressed as $\Gamma \propto |s|^2 + |p|^2$ and $\alpha \propto (|s|^2 + |p|^2)^{-1}$.

As far as isospin is concerned, both the $s$- and $p$-wave components consist of contributions describing $\Delta I=1/2$ and $\Delta I=3/2$ transitions. An unexpected experimental result is that the former transitions are more favored than the latter ones by nearly a factor of 20 to 1. This enhancement is also seen in kaon nonleptonic decays. Thus the so-called $\Delta I=1/2$ rule seems to be a rather universal feature of nonleptonic decays and will be considered to be valid hereafter.

\begin{figure}
\scalebox{0.5}{\includegraphics{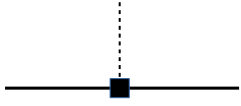}}
\caption{\label{fig:tree}Tree graph for $s$-wave decay amplitudes in baryon nonleptonic decays. A solid square represents a $\Delta S=1$ weak vertex, and a solid (dashed) line denotes a baryon (pion).
}
\end{figure}

The present paper focuses only on the analysis of the $s$-wave decay amplitude within large-$N_c$ chiral perturbation theory. At tree level, the only graph that contributes is depicted in Fig.~\ref{fig:tree}. For definiteness, this amplitude can be obtained using a soft pion theorem as \cite{djm95}
\begin{equation}
\mathcal{A}(B_i \to B_f + \pi^c) = \frac{i}{f_\pi} \langle B_f |[Q_5^c, \mathcal{H}_W]|B_i \rangle, \label{eq:swave}
\end{equation}
where $c$ is an explicit flavor index, $Q_5^c$ is the axial charge, and $\mathcal{H}_W$ is the weak Hamiltonian. The latter contains pieces which transform as $(\mathbf{8},\mathbf{1})$ and $(\mathbf{27},\mathbf{1})$ under $SU(3)_L \times SU(3)_R$. The $\mathbf{8}$ component dominates the $\mathbf{27}$ component so this fact is usually referred to as octet dominance. Under this assumption, $[Q_5^c, \mathcal{H}_W]=[Q^c, \mathcal{H}_W]$, where $Q^c$ is the vector charge \cite{djm95}.

Octet dominance assumption also implies that $\mathcal{H}_W$ transforms as the $(0,6+i7)$ component of a $(0,\mathbf{8})$ representation of the spin-flavor symmetry $SU(2)\times SU(3)$ \cite{djm95}. The $1/N_c$ expansion of a $(0,\mathbf{8})$ operator has been discussed in detail in Ref.~\cite{jl}. A simple operator analysis reveals that only $n$-body operators with a single factor of either $T^c$ or $G^{ic}$ appear. The allowed 1- and 2-body operators are
\begin{equation}
O_1^c = T^c,
\end{equation}
and
\begin{equation}
O_2^c = \{J^r, G^{rc}\},
\end{equation}
whereas higher-order operators are obtained as $O_n^c = \{J^2,O_{n-2}^c\}$ for $n\geq 3$. Thus, the $1/N_c$ expansion for $\mathcal{H}_W$ reads
\begin{equation}
\mathcal{H}_W = h_1 T^u + \frac{h_2}{N_c} \{J^r, G^{ru}\}, \label{eq:hw}
\end{equation}
up to corrections of relative order $\mathcal{O}(1/N_c^2)$. Here, $h_i$ are undetermined parameters with dimensions of mass. Hereafter, the flavor index $u$ will stand for $u=6+i7$ so any operator of the form $W^u$ should be understood as $W^6+iW^7$. As in previous works (see for instance Ref.~\cite{rfm14a}), the naive estimate that matrix elements of $T^c$ and $G^{kc}$ are both of order $N_c$, which is the largest they can be, is also implemented. The estimate is legitimate provided the analysis is restricted to the lowest-lying baryon states. Within this naive power counting, $A^{kc}$ and $T^c$ are both order $N_c$ and so is $\mathcal{H}_W $ of Eq.~(\ref{eq:hw}).

The vector charge is given by $Q^c=T^c$ to all orders in the $1/N_c$ expansion \cite{rfm98}. Thus, the commutator $[Q^c,\mathcal{H}_W]$ reduces to
\begin{equation}
[T^c,\mathcal{H}_W] = h_1 if^{cue} T^e + \frac{h_2}{N_c} if^{cue} \{J^r, G^{re}\}. \label{eq:comm}
\end{equation}

Substituting Eq.~(\ref{eq:comm}) into (\ref{eq:swave}) yields the decay amplitude at tree level $ \mathcal{A}_{\mathrm{tree}}^{(s)}$; it is given by
\begin{equation}
-i f_\pi \mathcal{A}_{\mathrm{tree}}^{(s)}(B_i \to B_f + \pi^c) = h_1 \langle B_f|if^{cue} T^e|B_i \rangle + \frac{h_2}{N_c} \langle B_f|if^{cue} \{J^r,G^{re}\}|B_i \rangle, \label{eq:swaveexp}
\end{equation}
where the flavor index $c$ will stand for $c=1\mp i2$ and $c=3$ for $\pi^\pm$ and $\pi^0$, respectively. For completeness, any operator of the form $W^c$ should be understood as $(W^1\mp iW^2)/\sqrt{2}$ and $W^3$ for $c=1\mp i2$ and $c=3$, respectively. For the observed processes the expressions read
\begin{equation}
-i\sqrt{2}f_\pi \mathcal{A}_{\mathrm{tree}}^{(s)}(\Sigma^+ \to n + \pi^+) = 0, \label{eq:asn}
\end{equation}
\begin{equation}
-i\sqrt{2}f_\pi \mathcal{A}_{\mathrm{tree}}^{(s)}(\Sigma^+ \to p + \pi^0) = \frac{1}{\sqrt{2}} \left[ h_1 - \frac{1}{2N_c} h_2 \right],
\end{equation}
\begin{equation}
-i\sqrt{2}f_\pi \mathcal{A}_{\mathrm{tree}}^{(s)}(\Sigma^- \to n + \pi^-) = -h_1 + \frac{1}{2N_c} h_2,
\end{equation}
\begin{equation}
-i\sqrt{2}f_\pi \mathcal{A}_{\mathrm{tree}}^{(s)}(\Lambda \to p + \pi^-) = -\sqrt{\frac32} \left[ h_1 + \frac{3}{2N_c} h_2 \right],
\end{equation}
\begin{equation}
-i\sqrt{2}f_\pi \mathcal{A}_{\mathrm{tree}}^{(s)}(\Lambda \to n + \pi^0) = \frac{\sqrt{3}}{2} \left[ h_1 + \frac{3}{2N_c} h_2 \right],
\end{equation}
\begin{equation}
-i\sqrt{2}f_\pi \mathcal{A}_{\mathrm{tree}}^{(s)}(\Xi^- \to \Lambda + \pi^-) = \sqrt{\frac32} \left[ h_1 + \frac{1}{2N_c} h_2 \right],
\end{equation}
\begin{equation}
-i\sqrt{2}f_\pi \mathcal{A}_{\mathrm{tree}}^{(s)}(\Xi^0 \to \Lambda + \pi^0) = -\frac{\sqrt{3}}{2} \left[ h_1 + \frac{1}{2N_c} h_2 \right]. \label{eq:axl}
\end{equation}

The $s$-wave amplitudes at tree level for the nonleptonic decays of octet baryons into octet baryons can be fully described only by two parameters $h_1$ and $h_2$. Adding higher-order operators in the $1/N_c$ expansion (\ref{eq:hw}) results into redefinitions of the already existing parameters, e.g., $h_1\to h_1+h_3/6$ and so on.

The right-hand sides of Eqs.~(\ref{eq:asn})-(\ref{eq:axl}) can be straightforwardly compared to their counterparts, $\alpha_{B_iB_f}^{(s)}$, obtained within heavy baryon chiral perturbation theory, which can be found in Eqs.~(3.7) of Ref.~\cite{jen92}. The operator coefficients $h_1$ and $h_2$ are related to the chiral coefficients $h_D$ and $h_F$, for $N_c=3$, by
\begin{equation}
h_1 = \frac23 h_D - h_F, \qquad \qquad h_2 = -2 h_D. \label{eq:chiralcoeff}
\end{equation}

Isospin symmetry of the strong interactions implies three relations among the seven amplitudes, namely,
\begin{subequations}
\label{eq:isosrel}
\begin{equation}
-\mathcal{A}_{\mathrm{tree}}^{(s)}(\Sigma^+ \to n + \pi^+) + \sqrt{2} \mathcal{A}_{\mathrm{tree}}^{(s)}(\Sigma^+ \to p + \pi^0) + \mathcal{A}_{\mathrm{tree}}^{(s)}(\Sigma^- \to n + \pi^-) = 0,
\end{equation}
\begin{equation}
\mathcal{A}_{\mathrm{tree}}^{(s)}(\Lambda \to p + \pi^-) + \sqrt{2} \mathcal{A}_{\mathrm{tree}}^{(s)}(\Lambda \to n + \pi^0) = 0,
\end{equation}
\begin{equation}
\mathcal{A}_{\mathrm{tree}}^{(s)}(\Xi^- \to \Lambda + \pi^-) + \sqrt{2} \mathcal{A}_{\mathrm{tree}}^{(s)}(\Xi^0 \to \Lambda + \pi^0) = 0,
\end{equation}
\end{subequations}
so there are effectively four independent amplitudes; the preferred study cases are those with a charged pion in the final state \cite{jen92}, namely, $\mathcal{A}_{\mathrm{tree}}^{(s)}(\Sigma^+ \to n + \pi^+)$, $\mathcal{A}_{\mathrm{tree}}^{(s)}(\Sigma^- \to n + \pi^-)$, $\mathcal{A}_{\mathrm{tree}}^{(s)}(\Lambda \to p + \pi^-)$, and $\mathcal{A}_{\mathrm{tree}}^{(s)}(\Xi^- \to \Lambda + \pi^-)$.\footnote{Isospin relations (\ref{eq:isosrel}) hold also for the $p$-wave decay amplitudes, of course.} These amplitudes can be combined to eliminate $h_1$ and $h_2$, leading to the celebrated Lee-Suwagara relation \cite{lee,sugawara}
\begin{equation}
\frac{3}{\sqrt{6}} \mathcal{A}_{\mathrm{tree}}^{(s)}(\Sigma^- \to n + \pi^-) + \mathcal{A}_{\mathrm{tree}}^{(s)}(\Lambda \to p + \pi^-) + 2 \mathcal{A}_{\mathrm{tree}}^{(s)}(\Xi^- \to \Lambda + \pi^-) = 0, \label{eq:leesuw}
\end{equation}
which holds in the limit of exact $SU(3)$ flavor symmetry.

Equation (\ref{eq:swaveexp}) can also be used to compute the tree-level $s$-wave amplitude for the nonleptonic decays of decuplet baryons to decuplet baryons. Specifically, the $\Omega^-$ baryon is the only member of the baryon decuplet that decays predominantly through the weak interaction. For the known processes the amplitudes read
\begin{equation}
-i \sqrt{2} f_\pi \mathcal{A}_{\mathrm{tree}}^{(s)}(\Omega^- \to {\Xi^*}^0 + \pi^-) = \frac{3}{\sqrt{3}} \left[ h_1 + \frac56(h_2+h_3) \right],
\end{equation}
and
\begin{equation}
-i \sqrt{2} f_\pi \mathcal{A}_{\mathrm{tree}}^{(s)}(\Omega^- \to {\Xi^*}^- + \pi^0) = -\frac{3}{\sqrt{6}} \left[ h_1 + \frac56(h_2+h_3) \right] .
\end{equation}

The above expressions are related by isospin as
\begin{equation}
\mathcal{A}_{\mathrm{tree}}^{(s)}(\Omega^- \to {\Xi^*}^0 + \pi^-) + \sqrt{2} \mathcal{A}_{\mathrm{tree}}^{(s)}(\Omega^- \to {\Xi^*}^- + \pi^0) = 0.
\end{equation}

The inclusion of the third operator coefficient $h_3$ is necessary in order to account for the third chiral coefficient $h_C$ introduced in heavy baryon chiral perturbation theory \cite{jen92}. For $N_c=3$ they are related by
\begin{equation}
h_3 = \frac25 [h_C+3(h_D+h_F)]. \label{eq:h3}
\end{equation}

\section{\label{sec:oneloop}One-loop corrections to the s-wave amplitude in baryon nonleptonic decays}

\begin{figure}
\scalebox{0.6}{\includegraphics{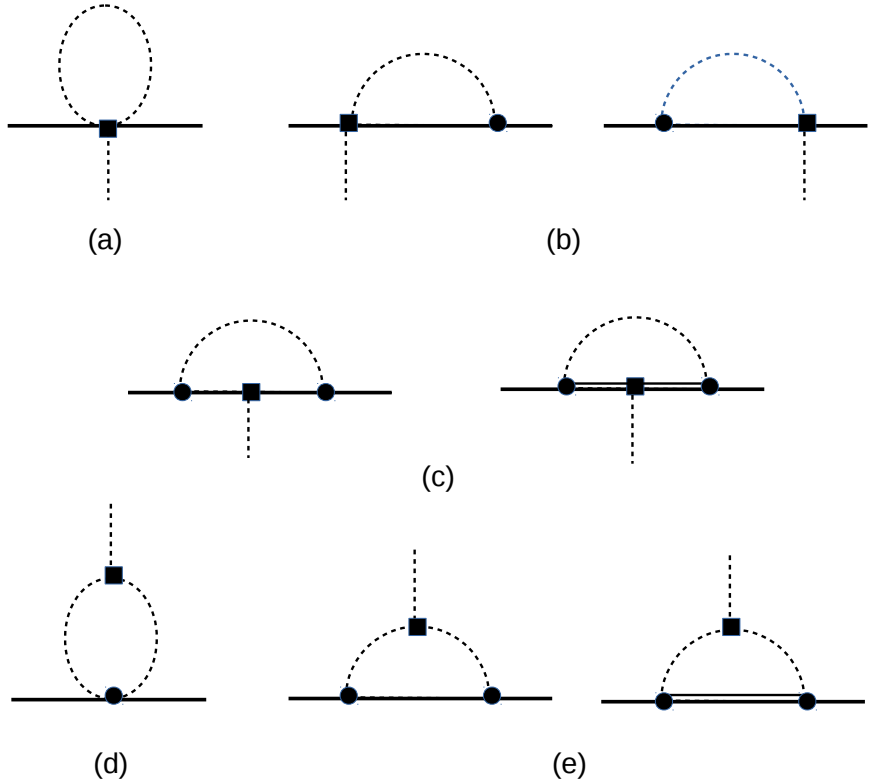}}
\caption{\label{fig:1loop}One-loop graphs for $s$-wave decay amplitudes in baryon nonleptonic decays. A solid square represents a $\Delta S=1$ weak vertex, a solid circle represents a strong vertex, and a solid (dashed) line denotes a baryon (pion). Wave function renormalization graphs are not shown, but are taken into account in the analysis.}
\end{figure}

The most general one-loop graphs that contribute to the $s$-wave amplitudes in the nonleptonic decays of baryons are displayed in Fig.~\ref{fig:1loop}. The approach to evaluate one-loop corrections to a baryon operator from Feynman diagrams like the ones in Figs.~\ref{fig:1loop}(a)-\ref{fig:1loop}(c) have been dealt with in Ref.~\cite{fmhjm}. The analysis, general enough to apply to any baryon operator transforming as a flavor octet, was first specialized to the baryon axial-vector current \cite{fmhjm,rfm06,rfm12}, later on to the baryon magnetic moment \cite{rfm09,rfm14} and more recently to the baryon vector current \cite{rfm14a}. With only minor adaptations, the very same approach can be implemented here to evaluate corrections to the $s$-wave amplitude of baryon nonleptonic decays. The computation of these diagrams is discussed below.

\subsection{One-loop corrections from Fig.~\ref{fig:1loop}(a)}

The contribution to the $s$-wave amplitude from Fig.~\ref{fig:1loop}(a) is given by
\begin{equation}
-i f_\pi \delta \mathcal{A}_{\mathrm{2a}}^{(s)}(B_i \to B_f + \pi^c) = -\frac12 \langle B_f|[T^a,[T^b,[T^c,\mathcal{H}_W]]] S^{ab} |B_i\rangle, \label{eq:2a}
\end{equation}
where $S^{ab}$ is the symmetric tensor with two octet indices given in Eq.~(58) of Ref.~\cite{rfm14a}, which in turn keeps a similar structure with the one introduced in Eq.~(4.18) of Ref.~\cite{jen96}; i.e., it contains flavor singlet, flavor $\mathbf{8}$, and flavor $\mathbf{27}$ representations as
\begin{equation}
S^{ab} = I_{\mathrm{a},\mathbf{1}} \delta^{ab} + I_{\mathrm{a},\mathbf{8}} d^{ab8} + I_{\mathrm{a},\mathbf{27}} \left[ \delta^{a8} \delta^{b8} - \frac18 \delta^{ab} - \frac35 d^{ab8} d^{888}\right], \label{eq:gamsym}
\end{equation}
where
\begin{subequations}
\begin{eqnarray}
I_{\mathrm{a},\mathbf{1}} & = & \frac18 \left[3I_{\mathrm{a}}(m_\pi,\mu) + 4I_{\mathrm{a}}(m_K,\mu) + I_{\mathrm{a}}(m_\eta,\mu) \right], \label{eq:G1} \\
I_{\mathrm{a},\mathbf{8}} & = & \frac{2\sqrt 3}{5} \left[\frac32 I_{\mathrm{a}}(m_\pi,\mu) - I_{\mathrm{a}}(m_K,\mu) - \frac12 I_{\mathrm{a}}(m_\eta,\mu) \right], \label{eq:G8} \\
I_{\mathrm{a},\mathbf{27}} & = & \frac13 I_{\mathrm{a}}(m_\pi,\mu) - \frac43 I_{\mathrm{a}}(m_K,\mu) + I_{\mathrm{a}}(m_\eta,\mu). \label{eq:G27}
\end{eqnarray}
\label{eq:loopggs}
\end{subequations}
and $I_{\mathrm{a}}(m,\mu)$ is the integral over the loop (cf.\ Eq.~(A22) of Ref.~\cite{rfm14a}),
\begin{equation}
I_{\mathrm{a}}(m,\mu) = \frac{m^2}{16\pi^2 f_\pi^2} \left[-\lambda_\epsilon - 1 + \ln \frac{m^2}{\mu^2} \right], \label{eq:ia}
\end{equation}
where $\mu$ is the scale of dimensional regularization and the ultraviolet (UV) divergence is given by the term proportional to
\begin{equation}
\lambda_\epsilon = \frac{2}{\epsilon} - \gamma + \ln(4\pi),
\end{equation}
with $\gamma \simeq 0.577216$ the Euler constant and $2\epsilon = 4-d$.

The use of the explicit form of $[T^c,\mathcal{H}_W]$ given in Eq.~(\ref{eq:comm}) directly into Eq.~(\ref{eq:2a}) exhibits the existence of two double commutators in the operator structure, namely, $[T^a,[T^b,T^e]]$ and $[T^a,[T^b,\{J^r,G^{re}\}]]$. The former has been previously evaluated in Ref.~\cite{rfm14a} and displayed in Eqs.~(60)-(62) of that reference for flavor singlet, flavor $\mathbf{8}$, and flavor $\mathbf{27}$ representations, respectively. The latter is the new contribution to be evaluated. A straightforward calculation yields the following:

(1) Flavor singlet contribution
\begin{equation}
[T^a,[T^a,\{J^r,G^{rc}\}]] = N_f \{J^r,G^{rc}\}. \label{eq:sind}
\end{equation}

(2) Flavor $\mathbf{8}$ contribution
\begin{equation}
d^{ab8} [T^a,[T^b,\{J^r,G^{rc}\}]] = \frac{N_f}{2} d^{c8e} \{J^r,G^{re}\}. \label{eq:octd}
\end{equation}

(3) Flavor $\mathbf{27}$ contribution
\begin{equation}
[T^8,[T^8,\{J^r,G^{rc}\}]] = f^{c8e} f^{8eg} \{J^r,G^{rg}\}. \label{eq:27d}
\end{equation}

It should be stressed that flavor singlet and flavor octet contributions must be subtracted off Eq.~(\ref{eq:27d}) so a truly flavor $\mathbf{27}$ contribution remains.

The matrix elements of the baryon operators can be straightforwardly obtained. The evaluation simplifies considerably by using the relations
\begin{subequations}
\label{eq:mtx}
\begin{equation}
if^{(1+i2)(6+i7)e} W^e = \frac{1}{\sqrt{2}} W^{4+i5}, 
\end{equation}
and
\begin{equation}
if^{(1-i2)(6+i7)e} W^e = 0,
\end{equation}
\end{subequations}
for any operator $W^c$. Thus, the matrix elements of baryon operators describing the $s$-wave amplitudes in nonleptonic processes can be related to the ones describing the vector current in strangeness changing semileptonic processes. The latter can be found in Ref.~\cite{rfm14a} and will not be repeated here.

After collecting partial results, the corrections from Fig.~\ref{fig:1loop}(a) to the $s$-wave baryon nonleptonic decay amplitude read
\begin{equation}
-i \sqrt{2} f_\pi \delta \mathcal{A}_{\mathrm{2a}}^{(s)}(\Sigma^+ \to n + \pi^+) = 0,
\end{equation}
\begin{eqnarray}
-i \sqrt{2} f_\pi \delta \mathcal{A}_{\mathrm{2a}}^{(s)}(\Sigma^- \to n + \pi^-) & = & \frac{1}{16}(6h_1-h_2) [I_{\mathrm{a}}(m_\pi,\mu) + 2 I_{\mathrm{a}}(m_K,\mu) + I_{\mathrm{a}}(m_\eta,\mu)] \nonumber \\
& = & \mbox{} -\frac{3}{8}(-h_D+h_F) [I_{\mathrm{a}}(m_\pi,\mu) + 2 I_{\mathrm{a}}(m_K,\mu) + I_{\mathrm{a}}(m_\eta,\mu)], \label{eq:1asall}
\end{eqnarray}
\begin{eqnarray}
-i \sqrt{2} f_\pi \delta \mathcal{A}_{\mathrm{2a}}^{(s)}(\Lambda \to p + \pi^-) & = & \frac{3}{16} \sqrt{\frac32} (2h_1+h_2) [I_{\mathrm{a}}(m_\pi,\mu) + 2 I_{\mathrm{a}}(m_K,\mu) + I_{\mathrm{a}}(m_\eta,\mu)] \nonumber \\
& = & \mbox{} -\frac{1}{8} \sqrt{\frac32} (h_D + 3 h_F) [I_{\mathrm{a}}(m_\pi,\mu) + 2 I_{\mathrm{a}}(m_K,\mu) + I_{\mathrm{a}}(m_\eta,\mu)], \label{eq:1alall}
\end{eqnarray}
\begin{eqnarray}
-i \sqrt{2} f_\pi \delta \mathcal{A}_{\mathrm{2a}}^{(s)}(\Xi^- \to \Lambda + \pi^-) & = & -\frac{1}{16} \sqrt{\frac32} (6h_1+h_2) [I_{\mathrm{a}}(m_\pi,\mu) + 2 I_{\mathrm{a}}(m_K,\mu) + I_{\mathrm{a}}(m_\eta,\mu)] \nonumber \\
& = & \mbox{} -\frac{1}{8} \sqrt{\frac32} (h_D - 3 h_F) [I_{\mathrm{a}}(m_\pi,\mu) + 2 I_{\mathrm{a}}(m_K,\mu) + I_{\mathrm{a}}(m_\eta,\mu)], \label{eq:1axall}
\end{eqnarray}
and
\begin{eqnarray}
-i \sqrt{2} f_\pi \delta \mathcal{A}_{\mathrm{2a}}^{(s)}(\Omega^- \to {\Xi^*}^0 + \pi^-) & = & -\frac{1}{16} \sqrt{\frac32} [6h_1+5(h_2+h_3)] [I_{\mathrm{a}}(m_\pi,\mu) + 2 I_{\mathrm{a}}(m_K,\mu) + I_{\mathrm{a}}(m_\eta,\mu)] \nonumber \\
& = & \mbox{} -\frac{\sqrt{3}}{8} h_C [I_{\mathrm{a}}(m_\pi,\mu) + 2 I_{\mathrm{a}}(m_K,\mu) + I_{\mathrm{a}}(m_\eta,\mu)].
\end{eqnarray}
Notice that relation (\ref{eq:mtx}(b) explains why $\delta \mathcal{A}_{\mathrm{2a}}^{(s)}(\Sigma^+ \to n + \pi^+)$ vanishes.

It is also worth noticing that
\begin{equation}
\frac{3}{\sqrt{6}} \delta \mathcal{A}_{\mathrm{2a}}^{(s)}(\Sigma^- \to n + \pi^-) + \delta \mathcal{A}_{\mathrm{2a}}^{(s)}(\Lambda \to p + \pi^-) + 2 \delta \mathcal{A}_{\mathrm{2a}}^{(s)}(\Xi^- \to \Lambda + \pi^-) = 0, \label{eq:leesuw2a}
\end{equation}
so loop graph \ref{fig:1loop}(a) does not modify the Lee-Suwagara relation (\ref{eq:leesuw}).

On the other hand, by retaining only the chiral logs in the loop integral (\ref{eq:ia}), neglecting the pion mass and using the Gell-Mann--Okubo relation to express $m_\eta^2$ as $4m_K^2/3$ yields
\begin{equation}
-i \sqrt{2} f_\pi \delta \mathcal{A}_{\mathrm{2a}}^{(s)}(\Sigma^- \to n + \pi^-) = -\frac54 (-h_D+h_F) \frac{m_K^2}{16\pi^2 f_\pi^2} \ln \frac{m_K^2}{\mu^2}, \label{eq:1as}
\end{equation}
\begin{equation}
-i \sqrt{2} f_\pi \delta \mathcal{A}_{\mathrm{2a}}^{(s)}(\Lambda \to p + \pi^-) = -\frac{5}{4\sqrt{6}}(h_D+3h_F) \frac{m_K^2}{16\pi^2 f_\pi^2} \ln \frac{m_K^2}{\mu^2}, \label{eq:1al}
\end{equation}
\begin{equation}
-i \sqrt{2} f_\pi \delta \mathcal{A}_{\mathrm{2a}}^{(s)}(\Xi^- \to \Lambda + \pi^-) = -\frac{5}{4\sqrt{6}}(h_D-3h_F) \frac{m_K^2}{16\pi^2 f_\pi^2} \ln \frac{m_K^2}{\mu^2}, \label{eq:1ax}
\end{equation}
and
\begin{equation}
-i \sqrt{2} f_\pi \delta \mathcal{A}_{\mathrm{2a}}^{(s)}(\Omega^- \to {\Xi^*}^0 + \pi^-) = -\frac{5}{4\sqrt{6}} h_C \frac{m_K^2}{16\pi^2 f_\pi^2} \ln \frac{m_K^2}{\mu^2}.
\end{equation}

Equations (\ref{eq:1asall})--(\ref{eq:1axall}) and their reduced forms (\ref{eq:1as})--(\ref{eq:1ax}) can be compared with the heavy-baryon chiral perturbation results of Refs.~\cite{b99} and \cite{jen92,abd}, respectively.
A full agreement is obtained when including all the appropriate $Z$-factors \textit{and} replacing the pseudoscalar decay constant in the chiral limit $\mathring{F}$ by $f_\pi$ (cf.\ Eq.~(34) of Ref.~\cite{b99}).

\subsection{One-loop corrections from Figs.~\ref{fig:1loop}(b) and \ref{fig:1loop}(c)}

The correction to the $s$-wave amplitude arising from Figs.~\ref{fig:1loop}(b) and \ref{fig:1loop}(c) can be written as (cf.\ Eq.~(14) of Ref.~\cite{rfm12})
\begin{eqnarray}
-i f_\pi \delta \mathcal{A}_{\mathrm{2b}}^{(s)}(B_i \to B_f + \pi^c) & = & \frac12 \langle B_f|[A^{ja},[A^{jb},[T^c,\mathcal{H}_W]]] Q_{(1)}^{ab}|B_i\rangle - \frac12 \langle B_f|\{A^{ja},[[T^c,\mathcal{H}_W],[\mathcal{M},A^{jb}]]\} Q_{(2)}^{ab}|B_i\rangle \nonumber \\
&  & \mbox{} + \frac16 \Big(\langle B_f|[A^{ja},[[\mathcal{M},[\mathcal{M},A^{jb}]],[T^c,\mathcal{H}_W]]]Q_{(3)}^{ab}|B_i\rangle \nonumber \\
&  & \mbox{\hglue1.0truecm} - \frac12 \langle B_f|[[\mathcal{M},A^{ja}],[[\mathcal{M},A^{jb}],[T^c,\mathcal{H}_W]]]Q_{(3)}^{ab}|B_i\rangle \Big) + \ldots , \label{eq:vc2b}
\end{eqnarray}
where $A^{ja}$ and $A^{jb}$ stand for the meson-baryon vertices, $\mathcal{M}$ is the baryon mass operator and $Q_{(n)}^{ab}$ is a symmetric tensor which is written in terms of the corresponding loop integral $I_{\mathrm{b}}^{(n)}(m,\Delta,\mu)$, where $\Delta$ is the decuplet-octet mass difference. $Q_{(n)}^{ab}$ also decomposes into flavor singlet, flavor $\mathbf{8}$, and flavor $\mathbf{27}$ representations as \cite{jen96}
\begin{equation}
Q_{(n)}^{ab} = I_{\mathrm{b},\mathbf{1}}^{(n)} \delta^{ab} + I_{\mathrm{b},\mathbf{8}}^{(n)} d^{ab8} + I_{\mathrm{b},\mathbf{27}}^{(n)} \left[ \delta^{a8} \delta^{b8} - \frac18 \delta^{ab} - \frac35 d^{ab8} d^{888}\right], \label{eq:pisym}
\end{equation}
where
\begin{subequations}
\begin{eqnarray}
I_{\mathrm{b},\mathbf{1}}^{(n)} & = & \frac18 \left[3I_{\mathrm{b}}^{(n)}(m_\pi,0,\mu) + 4I_{\mathrm{b}}^{(n)}(m_K,0,\mu) + I_{\mathrm{b}}^{(n)}(m_\eta,0,\mu) \right], \label{eq:F1} \\
I_{\mathrm{b},\mathbf{8}}^{(n)} & = & \frac{2\sqrt 3}{5} \left[\frac32 I_{\mathrm{b}}^{(n)}(m_\pi,0,\mu) - I_{\mathrm{b}}^{(n)}(m_K,0,\mu) - \frac12 I_{\mathrm{b}}^{(n)}(m_\eta,0,\mu) \right], \label{eq:F8} \\
I_{\mathrm{b},\mathbf{27}}^{(n)} & = & \frac13 I_{\mathrm{b}}^{(n)}(m_\pi,0,\mu) - \frac43 I_{\mathrm{b}}^{(n)}(m_K,0,\mu) + I_{\mathrm{b}}^{(n)}(m_\eta,0,\mu). \label{eq:F27}
\end{eqnarray}
\end{subequations}

$I_{\mathrm{b}}^{(n)}(m,0,\mu)$ stands for the degeneracy limit $\Delta \to 0$ of the function $I_{\mathrm{b}}^{(n)}(m,\Delta,\mu)$, which is defined as \cite{fmhjm}
\begin{equation}
I_{\mathrm{b}}^{(n)}(m,\Delta,\mu) \equiv \frac{\partial^n}{\partial \Delta^n} I_{\mathrm{b}}(m,\Delta,\mu). \label{eq:fn}
\end{equation}

The function $I_{\mathrm{b}}(m,\Delta,\mu)$, given in Eq.~(A6) of Ref.~\cite{rfm14a}. Its first derivative, for the sake of completeness, reads
\begin{equation}
16 \pi^2 f_\pi^2 I_{\mathrm{b}}^{(1)}(m,\Delta,\mu) = (m^2-2\Delta^2)\left[ \lambda_\epsilon + 1 - \ln \frac{m^2}{\mu^2} \right] - 2\Delta^2 - \left\{ \begin{array}{ll} \displaystyle 4\Delta\sqrt{m^2-\Delta^2} \left[ \frac{\pi}{2}-\tan^{-1} \left[ \frac{\Delta}{\sqrt{m^2-\Delta^2}}\right]\right], & |\Delta|< m \\[6mm]
\displaystyle 2\Delta \sqrt{\Delta^2-m^2} \ln \left[ \frac{\Delta-\sqrt{\Delta^2-m2}}{\Delta+\sqrt{\Delta^2-m^2}} \right], & |\Delta| > m. \end{array} \right. \label{eq:ibp}
\end{equation}
Therefore, in the $\Delta\to 0$ limit, it reduces to
\begin{equation}
I_{\mathrm{b}}^{(1)}(m,0,\mu) = \frac{m^2}{16\pi^2 f_\pi^2} \left[ \lambda_\epsilon + 1 -\ln \frac{m^2}{\mu^2} \right].
\end{equation}

The final expression for the correction to the decay amplitude from Figs.~\ref{fig:1loop}(b) and \ref{fig:1loop}(c) can be organized as
\begin{equation}
\delta \mathcal{A}_{\mathrm{2b}}^{(s)}(B_i \to B_f + \pi^c) = \delta \mathcal{A}_{\mathrm{2b},\mathbf{1}}^{(s)}(B_i \to B_f + \pi^c) + \delta \mathcal{A}_{\mathrm{2b},\mathbf{8}}^{(s)}(B_i \to B_f + \pi^c) + \delta \mathcal{A}_{\mathrm{2b},\mathbf{27}}^{(s)}(B_i \to B_f + \pi^c), \label{eq:delta2a}
\end{equation}
where \textit{all} the contributions from flavor singlet, flavor $\mathbf{8}$ and flavor $\mathbf{27}$ representations, for $N_c=3$, can be cast into
\begin{equation}
-if_\pi \delta \mathcal{A}_{\mathrm{2b},\mathbf{1}}^{(s)}(B_i \to B_f + \pi^c) = \sum_{m=1}^{8} \left( a_{m}^{\mathbf{1}} I_{\mathrm{b},\mathbf{1}}^{(1)} + \Delta b_{m}^{\mathbf{1}} I_{\mathrm{b},\mathbf{1}}^{(2)} + \Delta^2 c_{m}^{\mathbf{1}} I_{\mathrm{b},\mathbf{1}}^{(3)} + \ldots \right) \langle B_f| f^{cue} X_m^e|B_i\rangle, \label{eq:delta1}
\end{equation}
\begin{equation}
-if_\pi \delta \mathcal{A}_{\mathrm{2b},\mathbf{8}}^{(s)}(B_i \to B_f + \pi^c) = \sum_{m=1}^{31} \left( a_{m}^{\mathbf{8}} I_{\mathrm{b},\mathbf{8}}^{(1)} + \Delta b_{m}^{\mathbf{8}} I_{\mathrm{b},\mathbf{8}}^{(2)} + \Delta^2 c_{m}^{\mathbf{8}} I_{\mathrm{b},\mathbf{8}}^{(3)} + \ldots \right) \langle B_f| f^{cue} Y_m^e|B_i\rangle, \label{eq:delta8}
\end{equation}
and
\begin{equation}
-if_\pi \delta \mathcal{A}_{\mathrm{2b},\mathbf{27}}^{(s)}(B_i \to B_f + \pi^c) = \sum_{m=1}^{46} \left( a_{m}^{\mathbf{27}} I_{\mathrm{b},\mathbf{27}}^{(1)} + \Delta b_{m}^{\mathbf{27}} I_{\mathrm{b},\mathbf{27}}^{(2)} + \Delta^2 c_{m}^{\mathbf{27}} I_{\mathrm{b},\mathbf{27}}^{(3)} + \ldots \right) \langle B_f| f^{cue} Z_m^e |B_i\rangle, \label{eq:delta27}
\end{equation}
where the ellipses refer to operators that appear for $N_c>3$. It is understood that flavor singlet and flavor $\mathbf{8}$ contributions must be subtracted off Eq.~(\ref{eq:delta27}) in order to have a truly flavor $\mathbf{27}$ contribution. The operator coefficients $a_{m}^{\mathbf{rep}}$, $b_{m}^{\mathbf{rep}}$, and $c_{m}^{\mathbf{rep}}$ for flavor representation $\mathbf{rep}$ are listed in Appendix \ref{sec:opcoeff} whereas the corresponding operator bases are
\begin{eqnarray}
&  & X_{1}^c = T^c, \nonumber \\
&  & X_{2}^c = \{J^r,G^{rc}\}, \nonumber \\
&  & X_p^c = \{J^2,X_{p-2}^c \}, \qquad p=3,\ldots, \label{eq:bx}
\end{eqnarray}
\begin{eqnarray}
&  & Y_1^c = \delta^{c8}, \nonumber \\
&  & Y_2^c = d^{c8e} T^e, \nonumber \\
&  & Y_3^c = \{T^c,T^8\}, \nonumber \\
&  & Y_4^c = \{G^{rc},G^{r8}\}, \nonumber \\
&  & Y_5^c = d^{c8e} \{J^r,G^{re}\}, \nonumber \\
&  & Y_6^c = \delta^{c8} J^2, \nonumber \\
&  & Y_7^c = \{J^2,Y_2^c\}, \nonumber \\
&  & Y_8^c = \{T^c,\{J^r,G^{r8}\}\}, \nonumber \\
&  & Y_9^c = \{T^8,\{J^r,G^{rc}\}\}, \nonumber \\
&  & Y_{o}^c = \{J^2,Y_{o-7}^c\}, \qquad o=10,\ldots,12, \nonumber \\
&  & Y_{13}^c = f^{cab} f^{8be}\{J^2,\{G^{ra},G^{re}\}\}, \nonumber \\
&  & Y_{14}^c = d^{cab} d^{8be}\{J^2,\{G^{ra},G^{re}\}\}, \nonumber \\
&  & Y_{15}^c = \{\{J^r,G^{rc}\},\{J^m,G^{m8}\}\}, \nonumber \\
&  & Y_{p}^c = \{J^2,Y_{p-9}^c\}, \qquad p=16,\ldots,23, \nonumber \\
&  & Y_{q}^c = \{J^2,Y_{q-8}^c\}, \qquad q=24,\ldots,31, \label{eq:by}
\end{eqnarray}
and
\begin{eqnarray}
&  & Z_{1}^c = f^{c8e} f^{8eg} T^g, \nonumber \\
&  & Z_{2}^c = f^{c8e} f^{8eg} \{J^r,G^{rg}\}, \nonumber \\
&  & Z_{3}^c = d^{c8e} d^{8eg} \{J^r,G^{rg}\}, \nonumber \\
&  & Z_{4}^c = \delta^{c8} \{J^r,G^{r8}\}, \nonumber \\
&  & Z_{5}^c = \delta^{88} \{J^r,G^{rc}\}, \nonumber \\
&  & Z_{6}^c = d^{c8e} \{G^{re},G^{r8}\}, \nonumber \\
&  & Z_{7}^c = d^{88e} \{G^{rc},G^{re}\}, \nonumber \\
&  & Z_{8}^c = d^{c88} J^2, \nonumber \\
&  & Z_{9}^c = f^{c8e}f^{egh} \{T^g,\{G^{r8},G^{rh}\}\},\nonumber \\
&  & Z_{o}^c = \{J^2,Z_{o-9}^c\}, \qquad o=10\ldots,17, \nonumber \\
&  & Z_{18}^c = \{\{J^r,G^{rc}\},\{G^{m8},G^{m8}\}\}, \nonumber \\
&  & Z_{19}^c = \{\{G^{rc},G^{r8}\},\{J^m,G^{m8}\}\}, \nonumber \\
&  & Z_{20}^c = d^{c8e} \{\{J^r,G^{re}\},\{J^m,G^{m8}\}\}, \nonumber \\
&  & Z_{21}^c = d^{88e} \{\{J^r,G^{rc}\},\{J^m,G^{me}\}\}, \nonumber \\
&  & Z_{p}^c = \{J^2,Z_{p-13}^c\}, \qquad p=22\ldots,34, \nonumber \\
&  & Z_{35}^c = \{J^2,Z_{22}^c\}, \nonumber \\
&  & Z_{36}^c = \{J^2,Z_{23}^c\}, \nonumber \\
&  & Z_{q}^c = \{J^2,Z_{q-12}^c\}, \qquad q=37\ldots,46. \label{eq:bz}
\end{eqnarray}

As in the previous case, the matrix elements of the operators in the operator bases can be easily obtained. In each case, only the leading ones are required because the rest are obtained in most cases by anticommuting with $J^2$. Also, relations (\ref{eq:mtx}) can be used so the matrix elements can be found in Ref.~\cite{rfm14a}.

\subsubsection{Total correction from Figs.~\ref{fig:1loop}(b) and \ref{fig:1loop}(c)}

Gathering together partial results, the final expressions for the correction to the $s$-wave amplitude in baryon nonleptonic decays from Figs.~\ref{fig:1loop}(b) and \ref{fig:1loop}(c), for $N_c=3$, can be organized as
\begin{eqnarray}
-i\sqrt{2}f_\pi \delta \mathcal{A}_{\mathrm{2b}}^{(s)}(\Sigma^- \to n + \pi^-) & = & \left[ h_1 \left( \frac{7}{32} a_1^2 + \frac{1}{16} a_1b_2 + \frac{7}{48} a_1b_3 - \frac{1}{32} b_2^2 + \frac{1}{48} b_2b_3 + \frac{7}{288} b_3^2 \right) \right. \nonumber \\
&  & \mbox{} + \left. h_2 \left( \frac{73}{192} a_1^2 + \frac{7}{96} a_1b_2 + \frac{73}{288} a_1b_3 + \frac{1}{192} b_2^2 + \frac{7}{288} b_2b_3 + \frac{73}{1728} b_3^2 \right) \right] I_{\mathrm{b}}^{(1)}(m_\pi,0,\mu) \nonumber \\
&  & \mbox{} + \left[ h_1 \left( \frac{3}{16} a_1^2 + \frac18 a_1b_2 + \frac18 a_1b_3 - \frac{1}{16} b_2^2 + \frac{1}{24} b_2b_3 + \frac{1}{48} b_3^2 \right) \right. \nonumber \\
&  & \mbox{} + \left. h_2 \left( \frac{37}{96} a_1^2 + \frac{1}{16} a_1b_2 + \frac{37}{144} a_1b_3 + \frac{1}{96} b_2^2 + \frac{1}{48} b_2b_3 + \frac{37}{864} b_3^2 \right) \right] I_{\mathrm{b}}^{(1)}(m_K,0,\mu) \nonumber \\
&  & \mbox{} + \left[ h_1 \left( -\frac{1}{32} a_1^2 + \frac{1}{16} a_1b_2 - \frac{1}{48} a_1b_3 - \frac{1}{32} b_2^2 + \frac{1}{48} b_2b_3 - \frac{1}{288} b_3^2 \right) \right. \nonumber \\
&  & \mbox{} + \left. h_2 \left( \frac{1}{192} a_1^2 - \frac{1}{96} a_1b_2 + \frac{1}{288} a_1b_3 + \frac{1}{192} b_2^2 - \frac{1}{288} b_2b_3 + \frac{1}{1728} b_3^2\right) \right] I_{\mathrm{b}}^{(1)}(m_\eta,0,\mu) \nonumber \\
&  & \mbox{} + \left[ h_1 \left( -\frac12 a_1^2 - \frac12 a_1c_3 - \frac18 c_3^2 \right) + h_2 \left( \frac34 a_1^2 + \frac34 a_1c_3 + \frac{3}{16} c_3^2 \right) \right] I_{\mathrm{b}}^{(1)}(m_\pi,\Delta,\mu) \nonumber \\
&  & \mbox{} + \left[ h_1 \left( -\frac34 a_1^2 - \frac34 a_1c_3 - \frac{3}{16} c_3^2 \right) + h_2 \left( \frac{11}{24} a_1^2 + \frac{11}{24} a_1c_3 + \frac{11}{96} c_3^2 \right) \right] I_{\mathrm{b}}^{(1)}(m_K,\Delta,\mu) \nonumber \\
&  & \mbox{} + \left[ h_1 \left( -\frac14 a_1^2 - \frac14 a_1c_3 - \frac{1}{16} c_3^2 \right) + h_2 \left( \frac{1}{24} a_1^2 + \frac{1}{24} a_1c_3 + \frac{1}{96} c_3^2 \right) \right] I_{\mathrm{b}}^{(1)}(m_\eta,\Delta,\mu), \nonumber \\
\end{eqnarray}
\begin{eqnarray}
-i \frac{2}{\sqrt{3}} f_\pi \delta \mathcal{A}_{\mathrm{2b}}^{(s)}(\Lambda \to p + \pi^-) & = & \left[ h_1 \left( -\frac{17}{32} a_1^2 - \frac{3}{16} a_1b_2 - \frac{17}{48} a_1b_3 - \frac{1}{32} b_2^2 - \frac{1}{16} b_2b_3 - \frac{17}{288} b_3^2 \right) \right. \nonumber \\
&  & \mbox{} + \left. h_2 \left( -\frac{131}{192} a_1^2 - \frac{17}{96} a_1b_2 - \frac{131}{288} a_1b_3 - \frac{1}{64} b_2^2 - \frac{17}{288} b_2b_3 - \frac{131}{1728} b_3^2 \right) \right] I_{\mathrm{b}}^{(1)}(m_\pi,0,\mu) \nonumber \\
&  & \mbox{} + \left[ h_1 \left( -\frac{13}{16} a_1^2 - \frac38 a_1b_2 - \frac{13}{24} a_1b_3 - \frac{1}{16} b_2^2 - \frac18 b_2b_3 - \frac{13}{144} b_3^2 \right) \right. \nonumber \\
&  & \mbox{} + \left. h_2 \left( -\frac{47}{96} a_1^2 - \frac{13}{48} a_1b_2 - \frac{47}{144} a_1b_3 - \frac{1}{32} b_2^2 - \frac{13}{144} b_2b_3 - \frac{47}{864} b_3^2 \right) \right] I_{\mathrm{b}}^{(1)}(m_K,0,\mu) \nonumber \\
&  & \mbox{} + \left[ h_1 \left( -\frac{9}{32} a_1^2 - \frac{3}{16} a_1b_2 - \frac{3}{16} a_1b_3 - \frac{1}{32} b_2^2 - \frac{1}{16} b_2b_3 - \frac{1}{32} b_3^2 \right) \right. \nonumber \\
&  & \mbox{} + \left. h_2 \left( -\frac{9}{64} a_1^2 - \frac{3}{32} a_1b_2 - \frac{3}{32} a_1b_3 - \frac{1}{64} b_2^2 - \frac{1}{32} b_2b_3 - \frac{1}{64} b_3^2 \right) \right] I_{\mathrm{b}}^{(1)}(m_\eta,0,\mu) \nonumber \\
&  & \mbox{} + \left[ h_1 \left( \frac14 a_1^2 + \frac14 a_1c_3 + \frac{1}{16} c_3^2 \right) + h_2 \left( \frac{19}{24} a_1^2 + \frac{19}{24} a_1c_3 + \frac{19}{96} c_3^2 \right) \right] I_{\mathrm{b}}^{(1)}(m_\pi,\Delta,\mu) \nonumber \\
&  & \mbox{} + \left[ h_1 \left( \frac14 a_1^2 + \frac14 a_1c_3 + \frac{1}{16} c_3^2 \right) + h_2 \left( \frac{11}{24} a_1^2 + \frac{11}{24} a_1c_3 + \frac{11}{96} c_3^2 \right) \right] I_{\mathrm{b}}^{(1)}(m_K,\Delta,\mu), \nonumber \\
\end{eqnarray}
\begin{eqnarray}
-i \frac{2}{\sqrt{3}} f_\pi \delta \mathcal{A}_{\mathrm{2b}}^{(s)}(\Xi^- \to \Lambda + \pi^-) & = & \left[ h_1 \left( \frac{9}{32} a_1^2 + \frac{1}{16} a_1b_2 + \frac{3}{16} a_1b_3 + \frac{1}{32} b_2^2 + \frac{1}{48} b_2b_3 + \frac{1}{32} b_3^2 \right) \right. \nonumber \\
&  & \mbox{} + \left. h_2 \left( -\frac{7}{192} a_1^2 + \frac{3}{32} a_1b_2 - \frac{7}{288} a_1b_3 + \frac{1}{192} b_2^2 + \frac{1}{32} b_2b_3 - \frac{7}{1728} b_3^2 \right) \right] I_{\mathrm{b}}^{(1)}(m_\pi,0,\mu) \nonumber \\
&  & \mbox{} + \left[ h_1 \left( \frac{5}{16} a_1^2 + \frac18 a_1b_2 + \frac{5}{24} a_1b_3 + \frac{1}{16} b_2^2 + \frac{1}{24} b_2b_3 + \frac{5}{144} b_3^2 \right) \right] \nonumber \\
&  & \mbox{} + \left. h_2 \left( \frac{29}{96} a_1^2 + \frac{5}{48} a_1b_2 + \frac{29}{144} a_1b_3 + \frac{1}{96} b_2^2 + \frac{5}{144} b_2b_3 + \frac{29}{864} b_3^2 \right) \right] I_{\mathrm{b}}^{(1)}(m_K,0,\mu) \nonumber \\
&  & \mbox{} + \left[ h_1 \left( \frac{1}{32} a_1^2 + \frac{1}{16} a_1b_2 + \frac{1}{48} a_1b_3 + \frac{1}{32} b_2^2 + \frac{1}{48} b_2b_3 + \frac{1}{288} b_3^2 \right) \right. \nonumber \\
&  & \mbox{} + \left. h_2 \left( \frac{1}{192} a_1^2 + \frac{1}{96} a_1b_2 + \frac{1}{288} a_1b_3 + \frac{1}{192} b_2^2 + \frac{1}{288} b_2b_3 + \frac{1}{1728} b_3^2 \right) \right] I_{\mathrm{b}}^{(1)}(m_\eta,0,\mu) \nonumber \\
&  & \mbox{} + h_2 \left( -\frac23 a_1^2 - \frac23 a_1c_3 - \frac16 c_3^2 \right) I_{\mathrm{b}}^{(1)}(m_\pi,\Delta,\mu) \nonumber \\
&  & \mbox{} + \left[ h_1 \left( \frac14 a_1^2 + \frac14 a_1c_3 + \frac{1}{16} c_3^2 \right) + h_2 \left( -\frac58 a_1^2 - \frac58 a_1c_3 - \frac{5}{32} c_3^2 \right) \right] I_{\mathrm{b}}^{(1)}(m_K,\Delta,\mu) \nonumber \\
&  & \mbox{} + \left[ h_1 \left( \frac14 a_1^2 + \frac14 a_1c_3 + \frac{1}{16} c_3^2 \right) + h_2 \left( \frac{1}{24} a_1^2 + \frac{1}{24} a_1c_3 + \frac{1}{96} c_3^2 \right) \right] I_{\mathrm{b}}^{(1)}(m_\eta,\Delta,\mu), \nonumber \\
\end{eqnarray}
and
\begin{eqnarray}
-i \sqrt{\frac23} f_\pi \delta \mathcal{A}_{\mathrm{2b}}^{(s)}(\Omega^- \to {\Xi^*}^0 + \pi^-) & = & \left[ h_1 \left( \frac{5}{32} a_1^2 + \frac{5}{16} a_1b_2 + \frac{25}{48} a_1b_3 + \frac{5}{32} b_2^2 + \frac{25}{48} b_2b_3 + \frac{125}{288} b_3^2 \right) \right. \nonumber \\
&  & \mbox{} + \left. h_2 \left( \frac{25}{192} a_1^2 + \frac{25}{96} a_1b_2 + \frac{125}{288} a_1b_3 + \frac{25}{192} b_2^2 + \frac{125}{288} b_2b_3 + \frac{625}{1728} b_3^2 \right) \right] I_{\mathrm{b}}^{(1)}(m_\pi,0,\mu) \nonumber \\
&  & \mbox{} + \left[ h_1 \left( \frac{5}{16} a_1^2 + \frac58 a_1b_2 + \frac{25}{24} a_1b_3 + \frac{5}{16} b_2^2 + \frac{25}{24} b_2b_3 + \frac{125}{144} b_3^2 \right) \right. \nonumber \\
&  & \mbox{} + \left. h_2 \left( \frac{25}{96} a_1^2 + \frac{25}{48} a_1b_2 + \frac{125}{144} a_1b_3 + \frac{25}{96} b_2^2 + \frac{125}{144} b_2b_3 + \frac{625}{864} b_3^2 \right) \right] I_{\mathrm{b}}^{(1)}(m_K,0,\mu) \nonumber \\
&  & \mbox{} + \left[ h_1 \left( \frac{5}{32} a_1^2 + \frac{5}{16} a_1b_2 + \frac{25}{48} a_1b_3 + \frac{5}{32} b_2^2 + \frac{25}{48} b_2b_3 + \frac{125}{288} b_3^2 \right) \right. \nonumber \\
&  & \mbox{} + \left. h_2 \left( \frac{25}{192} a_1^2 + \frac{25}{96} a_1b_2 + \frac{125}{288} a_1b_3 + \frac{25}{192} b_2^2 + \frac{125}{288} b_2b_3 + \frac{625}{1728} b_3^2 \right) \right] I_{\mathrm{b}}^{(1)}(m_\eta,0,\mu) \nonumber \\
&  & \mbox{} + \left[ h_1 \left( \frac18 a_1^2 + \frac18 a_1c_3 + \frac{1}{32} c_3^2 \right) + h_2 \left( \frac{5}{48} a_1^2 + \frac{5}{48} a_1c_3 + \frac{5}{192} c_3^2 \right) \right] I_{\mathrm{b}}^{(1)}(m_\pi,-\Delta,\mu) \nonumber \\
&  & \mbox{} + \left[ h_1 \left( \frac14 a_1^2 + \frac14 a_1c_3 + \frac{1}{16} c_3^2 \right) + h_2 \left( \frac38 a_1^2 + \frac38 a_1c_3 + \frac{3}{32} c_3^2 \right) \right] I_{\mathrm{b}}^{(1)}(m_K,-\Delta,\mu) \nonumber \\
&  & \mbox{} + \left[ h_1 \left( \frac18 a_1^2 + \frac18 a_1c_3 + \frac{1}{32} c_3^2 \right) + h_2 \left( \frac{5}{48} a_1^2 + \frac{5}{48} a_1c_3 + \frac{5}{192} c_3^2 \right) \right] I_{\mathrm{b}}^{(1)}(m_\eta,-\Delta,\mu), \nonumber \\
\end{eqnarray}
where $I_{\mathrm{b}}^{(1)}(m,\Delta,\mu)$ has been obtained from the Maclaurin series expansion
\begin{equation}
I_{\mathrm{b}}^{(1)}(m,\Delta,\mu) = I_{\mathrm{b}}^{(1)}(m,0,\mu) + I_{\mathrm{b}}^{(2)}(m,0,\mu) \Delta + \frac12 I_{\mathrm{b}}^{(3)}(m,0,\mu) \Delta^2 + \ldots
\end{equation}

The above expressions can be rewritten in terms of the chiral coefficients of Ref.~\cite{jen92}, namely,
\begin{eqnarray}
-i\sqrt{2}f_\pi \delta \mathcal{A}_{\mathrm{2b}}^{(s)}(\Sigma^- \to n + \pi^-) & = & \left[ h_D \left( - \frac{17}{8} D^2 + \frac14 DF - \frac98 F^2 \right) + h_F \left( \frac18 D^2 - \frac94 DF + \frac98 F^2 \right) \right] I_{\mathrm{b}}^{(1)}(m_\pi,0,\mu) \nonumber \\
&  & \mbox{} + \left[ h_D \left( - \frac{13}{4} D^2 + \frac52 DF - \frac94 F^2 \right) + h_F \left( \frac54 D^2 - \frac92 DF + \frac94 F^2 \right) \right] I_{\mathrm{b}}^{(1)}(m_K,0,\mu) \nonumber \\
&  & \mbox{} + \left[ h_D \left( - \frac98 D^2 + \frac94 DF - \frac98 F^2 \right) + h_F \left(\frac98 D^2 - \frac94 DF + \frac98 F^2 \right) \right] I_{\mathrm{b}}^{(1)}(m_\eta,0,\mu) \nonumber \\
&  & \mbox{} + \left( -\frac{11}{6} h_D + \frac12 h_F \right) \mathcal{C}^2 I_{\mathrm{b}}^{(1)}(m_\pi,\Delta,\mu) + \left( -\frac{17}{12} h_D + \frac34 h_F \right) \mathcal{C}^2 I_{\mathrm{b}}^{(1)}(m_K,\Delta,\mu) \nonumber \\
&  & \mbox{} + \left( -\frac14 h_D + \frac14 h_F \right) \mathcal{C}^2 I_{\mathrm{b}}^{(1)}(m_\eta,\Delta,\mu),
\end{eqnarray}
\begin{eqnarray}
-i \frac{2}{\sqrt{3}} f_\pi \delta \mathcal{A}_{\mathrm{2b}}^{(s)}(\Lambda \to p + \pi^-) & = & \left[ h_D \left( \frac{19}{8} D^2 + \frac94 DF + \frac38 F^2 \right) + h_F \left( \frac98 D^2 + \frac34 DF + \frac98 F^2 \right) \right] I_{\mathrm{b}}^{(1)}(m_\pi,0,\mu) \nonumber \\
&  & \mbox{} + \left[ h_D \left( - \frac14 D^2 + \frac52 DF + \frac34 F^2 \right) + h_F \left( \frac54 D^2 + \frac32 DF + \frac94 F^2 \right) \right] I_{\mathrm{b}}^{(1)}(m_K,0,\mu) \nonumber \\
&  & \mbox{} + \left[ h_D \left( \frac{1}{24} D^2 + \frac14 DF + \frac38 F^2 \right) + h_F \left( \frac18 D^2 + \frac34 DF + \frac98 F^2 \right) \right] I_{\mathrm{b}}^{(1)}(m_\eta,0,\mu) \nonumber \\
&  & \mbox{} + \left( -\frac{17}{12} h_D - \frac14 h_F \right) \mathcal{C}^2 I_{\mathrm{b}}^{(1)}(m_\pi,\Delta,\mu) + \left( - \frac34 h_D - \frac14 h_F \right) \mathcal{C}^2 I_{\mathrm{b}}^{(1)}(m_K,\Delta,\mu),
\end{eqnarray}
\begin{eqnarray}
-i \frac{2}{\sqrt{3}} f_\pi \delta \mathcal{A}_{\mathrm{2b}}^{(s)}(\Xi^ - \to \Lambda + \pi^-) & = & \left[ h_D \left( \frac{19}{8} D^2 - \frac94 DF + \frac38 F^2 \right) + h_F \left( -\frac98 D^2 + \frac34DF - \frac98 F^2 \right) \right] I_{\mathrm{b}}^{(1)}(m_\pi,0,\mu) \nonumber \\
&  & \mbox{} + \left[ h_D \left( -\frac14D^2 - \frac52 DF + \frac34 F^2 \right) + h_F \left( -\frac54 D^2 + \frac32 DF - \frac94 F^2 \right) \right] I_{\mathrm{b}}^{(1)}(m_K,0,\mu) \nonumber \\
&  & \mbox{} + \left[ h_D \left( \frac{1}{24} D^2 - \frac14 DF + \frac38 F^2 \right) + h_F \left( -\frac18 D^2 + \frac34 DF - \frac98 F^2 \right) \right] I_{\mathrm{b}}^{(1)}(m_\eta,0,\mu) \nonumber \\
&  & \mbox{} + \frac43 h_D \mathcal{C}^2 I_{\mathrm{b}}^{(1)}(m_\pi,\Delta,\mu) + \left( \frac{17}{12} h_D - \frac14 h_F \right) \mathcal{C}^2 I_{\mathrm{b}}^{(1)}(m_K,\Delta,\mu) \nonumber \\
&  & \mbox{} + \left( \frac{1}{12} h_D - \frac14 h_F \right) \mathcal{C}^2 I_{\mathrm{b}}^{(1)}(m_\eta,\Delta,\mu),
\end{eqnarray}
and
\begin{eqnarray}
- i \sqrt{\frac23} f_\pi \delta \mathcal{A}_{\mathrm{2b}}^{(s)}(\Omega^- \to {\Xi^*}^0 + \pi^-) & = & -\frac{5}{72} (h_D + h_F) \mathcal{H}^2 \left[ I_{\mathrm{b}}^{(1)}(m_\pi,0,\mu) + 2 I_{\mathrm{b}}^{(1)}(m_K,0,\mu) + I_{\mathrm{b}}^{(1)}(m_\eta,0,\mu) \right] \nonumber \\
&  & \mbox{} - \frac18 (h_D + h_F) \mathcal{C}^2 I_{\mathrm{b}}^{(1)}(m_\pi,-\Delta,\mu) - \frac{1}{12} (7h_D + 3 h_F) \mathcal{C}^2 I_{\mathrm{b}}^{(1)}(m_K,-\Delta,\mu) \nonumber \\
&  & \mbox{} - \frac18 (h_D + h_F) \mathcal{C}^2 I_{\mathrm{b}}^{(1)}(m_\eta,-\Delta,\mu).
\end{eqnarray}

Again, by retaining only the chiral logs in the loop integral (\ref{eq:ibp}), neglecting both the pion mass and the decuplet-octet baryon mass difference $\Delta$, and using the Gell-Mann--Okubo relation leads to
\begin{eqnarray}
-i\sqrt{2} f_\pi \delta \mathcal{A}_{\mathrm{2b}}^{(s)}(\Sigma^- \to n + \pi^-) & = & \left[ h_D \left(\frac74 \mathcal{C}^2 + \frac{19}{4} D^2 - \frac{11}{2} D F + \frac{15}{4} F^2 \right) \right. \nonumber \\
&  & \mbox{} + \left. h_F \left( -\frac{13}{12} \mathcal{C}^2 - \frac{11}{4} D^2 + \frac{15}{2} D F - \frac{15}{4} F^2 \right) \right] \frac{m_K^2}{16 \pi^2 f_\pi^2} \ln \frac{m_K^2}{\mu^2}, \label{eq:ch1}
\end{eqnarray}
\begin{eqnarray}
-i \frac{2}{\sqrt{3}} f_\pi \delta \mathcal{A}_{\mathrm{2b}}^{(s)}(\Lambda \to p + \pi^-) & = & \left[ h_D \left(\frac34 \mathcal{C}^2 + \frac{7}{36} D^2 - \frac{17}{6} DF - \frac54 F^2 \right) \right. \nonumber \\
&  & \mbox{} + \left. h_F \left( \frac14 \mathcal{C}^2 - \frac{17}{12} D^2 - \frac52 DF - \frac{15}{4} F^2 \right) \right] \frac{m_K^2}{16 \pi^2 f_\pi^2} \ln \frac{m_K^2}{\mu^2}, \label{eq:ch2}
\end{eqnarray}
\begin{eqnarray}
-i \frac{2}{\sqrt{3}} f_\pi \delta \mathcal{A}_{\mathrm{2b}}^{(s)}(\Xi^ - \to \Lambda + \pi^-) & = & \left[ h_D \left(-\frac{55}{36} \mathcal{C}^2 + \frac{7}{36} D^2 + \frac{17}{6} DF - \frac54 F^2 \right) \right. \nonumber \\
&  & \mbox{} + \left. h_F \left( \frac{7}{12} \mathcal{C}^2 + \frac{17}{12} D^2 - \frac52 DF + \frac{15}{4} F^2 \right) \right] \frac{m_K^2}{16 \pi^2 f_\pi^2} \ln \frac{m_K^2}{\mu^2}, \label{eq:ch3}
\end{eqnarray}
\begin{equation}
- i \sqrt{\frac23} f_\pi \delta \mathcal{A}_{\mathrm{2b}}^{(s)}(\Omega^- \to {\Xi^*}^0 + \pi^-) = \left[ h_D \left( \frac34 \mathcal{C}^2 + \frac{25}{108} \mathcal{H}^2 \right) + h_F \left( \frac{5}{12} \mathcal{C}^2 + \frac{25}{108} \mathcal{H}^2 \right) \right] \frac{m_K^2}{16 \pi^2 f_\pi^2} \ln \frac{m_K^2}{\mu^2}, \label{eq:ch4}
\end{equation}

Equations (\ref{eq:ch1})--(\ref{eq:ch3}) can be compared to their counterparts displayed in Eq.~(3.6) of Ref.~\cite{jen92}. The expressions match identically for
\begin{equation}
h_C = -3(h_D+h_F). \label{eq:rappx}
\end{equation}
The correction to this relation is order $1/N_c^2$. Although the above result is unexpected because in the chiral Lagrangian the coefficients are presumably independent, it could have been anticipated in the $h_3\to 0$ limit in Eq.~(\ref{eq:h3}). A relation that keeps a close similarity to (\ref{eq:rappx}) has been derived between the parameters $b_D$, $b_F$, and $c$ introduced in the chiral Lagrangian for the octet and decuplet baryons to first order in the quark mass matrix, namely, $b_D+b_F = -c/3$, which is valid up to corrections of order $1/N_c^2$ \cite{jen96}.

To close this section, corrections to the Lee-Sugawara relation (\ref{eq:leesuw}) can be readily be computed; they are made up from flavor $\mathbf{8}$ and flavor $\mathbf{27}$ contributions. The resultant expression is
\begin{eqnarray}
&  & -i\frac{2}{\sqrt{3}} f_\pi \left[ \frac{3}{\sqrt{6}} \delta \mathcal{A}_{\mathrm{2b}}^{(s)}(\Sigma^- \to n + \pi^-) + \delta \mathcal{A}_{\mathrm{2b}}^{(s)}(\Lambda \to p + \pi^-) + 2 \delta \mathcal{A}_{\mathrm{2b}}^{(s)}(\Xi^- \to \Lambda + \pi^-) \right] = \nonumber \\
&  & \mbox{} \left[ h_1 \left( \frac14 a_1^2 + \frac16 a_1b_3 + \frac{1}{36} b_3^2 \right) + h_2 \left( -\frac38 a_1^2 + \frac{1}{12} a_1b_2 - \frac14 a_1b_3 + \frac{1}{36} b_2b_3 - \frac{1}{24} b_3^2 \right) \right]
I_{\mathrm{b}}^{(1)}(m_\pi,0,\mu) \nonumber \\
&  & \mbox{} + h_2 \left( \frac{1}{2} a_1^2 + \frac13 a_1b_3 + \frac{1}{18} b_3^2 \right) I_{\mathrm{b}}^{(1)}(m_K,0,\mu) \nonumber \\
&  & \mbox{} + \left[ h_1 \left( -\frac14 a_1^2 - \frac16 a_1b_3 - \frac{1}{36} b_3^2 \right) + h_2 \left( -\frac18 a_1^2 - \frac{1}{12} a_1b_2 - \frac{1}{12} a_1b_3 - \frac{1}{36} b_2b_3 - \frac{1}{72} b_3^2 \right) \right] I_{\mathrm{b}}^{(1)}(m_\eta,0,\mu) \nonumber \\
&  & \mbox{} + \left[ h_1 \left( -\frac14 a_1^2 - \frac14 a_1c_3 - \frac{1}{16} c_3^2 \right) + h_2 \left( \frac{5}{24} a_1^2 + \frac{5}{24} a_1c_3 + \frac{5}{96} c_3^2 \right) \right] I_{\mathrm{b}}^{(1)}(m_\pi,\Delta,\mu) \nonumber \\
&  & \mbox{} + h_2 \left( -\frac13 a_1^2 - \frac13 a_1c_3 - \frac{1}{12} c_3^2 \right) I_{\mathrm{b}}^{(1)}(m_K,\Delta,\mu) \nonumber \\
&  & \mbox{} + \left[ h_1 \left( \frac14 a_1^2 + \frac14 a_1c_3 + \frac{1}{16} c_3^2 \right) + h_2 \left( \frac18 a_1^2 +\frac18 a_1c_3 + \frac{1}{32} c_3^2 \right) \right] I_{\mathrm{b}}^{(1)}(m_\eta,\Delta,\mu), \label{eq:ls2}
\end{eqnarray}
or in terms of the chiral coefficients,
\begin{eqnarray}
&  & -i\frac{2}{\sqrt{3}} f_\pi \left[ \frac{3}{\sqrt{6}} \delta \mathcal{A}_{\mathrm{2b}}^{(s)}(\Sigma^- \to n + \pi^-) + \delta \mathcal{A}_{\mathrm{2b}}^{(s)}(\Lambda \to p + \pi^-) + 2 \delta \mathcal{A}_{\mathrm{2b}}^{(s)}(\Xi^- \to \Lambda + \pi^-) \right] = \nonumber \\
&  & \left[ h_D(5D^2 - 2DF) - h_FD^2 \right] I_{\mathrm{b}}^{(1)}(m_\pi,0,\mu) - 4h_DD^2 I_{\mathrm{b}}^{(1)}(m_K,0,\mu) + \left[h_D(-D^2+2 DF) + h_FD^2 \right] I_{\mathrm{b}}^{(1)}(m_\eta,0,\mu) \nonumber \\
&  & \mbox{} - \frac{1}{12} (7h_D-3h_F)\mathcal{C}^2 I_{\mathrm{b}}^{(1)}(m_\pi,\Delta,\mu) + \frac23 h_D \mathcal{C}^2 I_{\mathrm{b}}^{(1)}(m_K,\Delta,\mu) - \frac{1}{12} (h_D+3h_F) \mathcal{C}^2 I_{\mathrm{b}}^{(1)}(m_\eta,\Delta,\mu).
\end{eqnarray}

Notice that the leading term $h_1a_1^2$ in Eq.~(\ref{eq:ls2}) cancels exactly in the limit $\Delta\to 0$, so the dependence on the regularization scale $\mu$ appears at subleading order.

\subsection{One-loop corrections from Figs.~\ref{fig:1loop}(d) and \ref{fig:1loop}(e)}

Loop diagrams Figs.~\ref{fig:1loop}(d) and \ref{fig:1loop}(e) involve a vertex from the term
\begin{equation}
h_\pi \frac{f_\pi^2}{4} \mathrm{Tr} \, \left[ h \partial_\mu \Sigma \partial^\mu \Sigma^\dag \right],
\end{equation}
where the dimensionless parameter $h_\pi=3.2\times 10^{-7}$ is determined from $\Delta S=1$ kaon decays, $\Sigma = \xi^2$, and $h$ is the matrix that selects out $s\to d$ transitions. The contributions of these diagrams to the decay amplitudes have been evaluated in Refs.~\cite{jen92,b99}. For $s$-wave amplitudes, they are found be to proportional to the mass difference between the initial and final baryons so their contributions are marginal.

In the combined formalism, diagram \ref{fig:1loop}(d) can be written as
\begin{equation}
-i f_\pi \delta \mathcal{A}_{\mathrm{2d}}^{(s)}(B_i \to B_f + \pi^c) = ih_\pi f^{cue} \sum_{\textsf{j}} A^{ia} \mathcal{P}_{\textsf{j}} A^{ib} P^{abe}(\Delta_{\textsf{j}}), \label{eq:loop1d}
\end{equation}
where $A^{ia}$ and $A^{jb}$ are used at the meson-baryon vertices and $\mathcal{P}_{\mathsf{j}}$ is the baryon projector for spin $J=\mathsf{j}$ \cite{jen96}
\begin{equation}
\frac{i\mathcal{P}_{\mathsf{j}}}{k^0-\Delta_{\mathsf{j}}}, \label{eq:barprop}
\end{equation}
and $\Delta_{\mathsf{j}}$ stands for the difference of the hyperfine mass splitting between the intermediate baryon with spin $J=\mathsf{j}$ and the external baryon, namely,
\begin{equation}
\Delta_{\mathsf{j}} = \mathcal{M}_{\textrm{hyperfine}}|_{J^2=\mathsf{j}(\mathsf{j}+1)}-\mathcal{M}_{\textrm{hyperfine}}|_{J^2=\mathsf{j}_{\textrm{ext}}(\mathsf{j}_{\textrm{ext}}+1)}.
\end{equation}
$P^{abe}(\Delta_\mathsf{j})$ is an antisymetric tensor (cf.\ Eq.~(28) of Ref.~\cite{rfm14a}) which depends on the loop integral listed in Eq.~(A3) of that reference.

Similarly, the one-loop contribution to the $s$-wave amplitude arising from the Feynman diagram of Fig.~\ref{fig:1loop}(e) reads
\begin{equation}
-i f_\pi \delta \mathcal{A}_{\mathrm{2e}}^{(s)}(B_i \to B_f + \pi^c) = -ih_\pi f^{cud}f^{dae} f^{beg} T^g R^{ab}, \label{eq:vc1c}
\end{equation}
where the tensor $R^{ab}$ can be written in terms of the loop integral (A19) of Ref.~\cite{rfm14a}.

 For graph \ref{fig:1loop}(d), for instance, away from the chiral limit quark mass splittings must be considered in the inversion of the baryon quadratic terms in the Lagrangian \cite{jen96}. This means that baryon mass splittings which are comparable to the meson octet masses should be retained in the baryon propagator. The leading quark mass splitting proportional to $T^8$ in Eq.~(\ref{eq:Ha}) and $\Delta$ are two of such quantities. In the loop integral (A3) of Ref.~\cite{rfm14a}, retaining the $T^8$ quark mass splitting $\Delta_s$ can be achieved through the replacement \cite{jen96}
\begin{equation}
I(m_1,m_2,\Delta,\mu) \to \frac12 [I(m_1,m_2,\Delta-\Delta_s,\mu) + I(m_1,m_2,\Delta+\Delta_s,\mu)],
\end{equation}
which would have some effects on the kaon loop graphs for $\Delta S = \pm 1$ transitions. Performing a detailed analysis on this subject is rather involved for such tiny contributions. Thus the findings of Refs.~\cite{jen92,b99} will be considered here and those diagrams will not be taken into account.

\section{\label{sec:sb}Explicit flavor symmetry breaking corrections to the $s$-wave amplitudes}

The basic idea of renormalization comes from the observation that in one-loop graphs the divergences amount to shifts in the parameters of the action. Loop integrals Eqs.~(\ref{eq:ia}) and (\ref{eq:ibp}) possess an ultraviolet (UV) divergent term proportional to $\lambda_\epsilon$, rather involved polynomial terms in the meson masses and decuplet-octet mass difference squared and nonanalytical contributions.

The analysis of all the counterterms at chiral order $\mathcal{O}(p^3)$ that renormalize the low-energy constants that describe $s$- and $p$-wave amplitudes in hyperon nonleptonic decays has been presented in detail in Ref.~\cite{b99}. 
These contributions include explicit symmetry breaking terms, double-derivative terms and relativistic corrections. After a few considerations the authors conclude that the actual number of significant counterterms is ten, four of which come from both explicit symmetry breaking and double-derivative terms, and another four from relativistic corrections, apart from the two lowest-order ones $h_D$ and $h_F$. For $s$-wave amplitudes relativistic corrections do not participate. The authors perform a fit to data to determine all ten parameters from both $s$- and $p$-wave data and find a satisfactory fit.

In the combined formalism under consideration here, a more pragmatic approach will be followed in order to evaluate the counterterms of order $\mathcal{O}(m_q)$: Only explicit symmetry breaking terms coming from $\mathcal{L}_{\text{baryon}}^{\mathcal{M}}$ of Eq.~(\ref{eq:lbarm}) will be retained. Flavor SB in QCD is due to the strange quark mass $m_s$ and transforms as a flavor octet \cite{djm95}. To linear order in SB, the correction is obtained from the tensor product $(0,\mathbf{8})\times (0,\mathbf{8})$ so that the $SU͑(2͒)\times SU(3)$ representations involved are $(0,\mathbf{1})$, $(0,\mathbf{8})$, $(0,\mathbf{10+\overline{10}})$ and $(0,\mathbf{27})$ \cite{jl,rfm98}. To second-order SB the representation $(0,\mathbf{64})$ also appears. The most general expressions for the $1/N_c$ expansions for a spin-0 flavor octet operator including first- and second-order SB have been presented in Ref.~\cite{fm17}. Thus, SB will be incorporated into the present analysis following the lines of that reference.

Explicit SB to the $s$-wave amplitude can thus be expressed as
\begin{equation}
-i f_\pi \delta \mathcal{A}_{\mathrm{SB}}^{(s)}(B_i \to B_f + \pi^c) = \langle B_f|i f^{cue} O_\mathrm{SB}^e|B_i \rangle 
\end{equation}
where $O_\mathrm{SB}^a$ is the most general spin-0 flavor octet operator containing flavor SB. The $1/N_c$ expansion of this operator with first-order SB is \cite{fm17}
\begin{eqnarray}
\lambda O_\mathrm{SB}^a & = & \lambda N_c a_{(0)}^\mathbf{1} \delta^{a8} + \lambda a_{(1)}^\mathbf{8} d^{a8b} T^b + \lambda a_{(2)}^\mathbf{8} \frac{1}{N_c} d^{a8b} \{J^r,G^{rb}\} + \lambda a_{(3)}^\mathbf{10+\overline{10}} \frac{1}{N_c^2} \left( \{T^a,\{J^r,G^{r8}\}\} - \{T^8,\{J^r,G^{ra}\}\} \right) \nonumber \\
&  & \mbox{} + \lambda a_{(2)}^\mathbf{27} \frac{1}{N_c} \left[ \{T^a,T^8\} - \frac{N_f-2}{2N_f(N_f^2-1)} N_c(N_c+2N_f) \delta^{a8} - \frac{2}{N_f^2-1} \delta^{a8} J^2 - \frac{N_f-4}{N_f^2-4}(N_c + N_f) d^{a8b} T^b \right. \nonumber \\
&  & \mbox{\hglue1.9truecm} \left. - \frac{2N_f}{N_f^2-4} d^{a8b} \{J^r,G^{rb}\} \right] \nonumber \\
&  & \mbox{} + \lambda a_{(3)}^\mathbf{27} \frac{1}{N_c^2} \left[ \{T^a,\{J^r,G^{r8}\}\} + \{T^8,\{J^r,G^{ra}\}\} - \frac{4}{N_f(N_f+1)}(N_c+N_f) \delta^{a8} J^2 \right. \nonumber \\
&  & \mbox{\hglue1.9truecm} \left. - \frac{2}{N_f+2}(N_c + N_f) d^{a8b} \{J^r,G^{rb}\} - \frac{2}{N_f+2} d^{a8b} \{J^2,T^b \} \right], \label{eq:qfirst}
\end{eqnarray}
whereas the corresponding $1/N_c$ expansion with second-order SB reads
\begin{eqnarray}
\lambda^2 O_\mathrm{SB}^a & = & \lambda^2 b_{(0)}^{\mathbf{1}}N_c d^{a88} \openone + \lambda^2 b_{(1)}^{\mathbf{8}} \delta^{a8} T^8 + \lambda^2 e_{(1)}^{\mathbf{8}} f^{a8b} f^{8bg} T^g + \lambda^2 g_{(1)}^{\mathbf{8}} d^{a8b} d^{8bg} T^g \nonumber \\
&  & \mbox{} + \lambda^2 h_{(1)}^{\mathbf{8}} (i f^{abg} d^{8b8} T^g-i d^{ab8} f^{8bg} T^g-i f^{a8b} d^{bg8} T^g) + \lambda^2 b_{(2)}^{\mathbf{8}} \frac{1}{N_c} \delta^{a8} \{J^r,G^{r8}\} + \lambda^2 e_{(2)}^{\mathbf{8}} \frac{1}{N_c} f^{a8b} f^{8bg} \{J^r,G^{rg}\} \nonumber \\
&  & \mbox{} + \lambda^2 g_{(2)}^{\mathbf{8}} \frac{1}{N_c} d^{a8b} d^{8bg} \{J^r,G^{rg}\} + \lambda^2 h_{(2)}^{\mathbf{8}} \frac{1}{N_c} (i f^{abg} d^{8b8} - i d^{ab8} f^{8bg} - i f^{a8b} d^{bg8}) \{J^r,G^{rg}\} \nonumber \\
&  & \mbox{} + \lambda^2 b_{(2)}^{\mathbf{10+\overline{10}}}\frac{1}{N_c^2} d^{a8b} \left(\{T^b,\{J^r,G^{r8}\}\} - \{T^8,\{J^r ,G^{rb}\}\} \right) \nonumber \\
&  & \mbox{} + \lambda^2 b_{(2)}^{\mathbf{27}} \frac{1}{N_c} \left[ d^{a8b}\{T^b,T^8\} - \frac{N_f-4}{N_f^2-4}(N_c+N_f) d^{a8b} d^{8bg} T^g - \frac{2N_f}{N_f^2-4} d^{a8b} d^{8bg} \{J^r ,G^{rg}\} \right] \nonumber \\
&  & \mbox{} + \lambda^2 b_{(3)}^{\mathbf{27}} \frac{1}{N_c^2} \left[ d^{a8b} \left( \{T^b,\{J^r,G^{r8}\}\} + \{T^8,\{J^r ,G^{rb}\}\} \right) - \frac{2}{N_f+2}(N_c+N_f) d^{a8b} d^{8bg} \{J^r ,G^{rg}\} \right. \nonumber \\
&  & \mbox{\hglue2.0truecm} \left. - \frac{2}{N_f+2} d^{a8b} d^{8bg} \{J^2,T^g\} \right] \nonumber \\
&  & \mbox{} + \lambda^2 b_{(3)}^{\mathbf{64}} \frac{1}{N_c^2} \left[ \{T^a,\{T^8,T^8\}\} -\frac{N_f-2}{N_f(N_f^2-1)}N_c(N_c+2N_f) \delta^{88} T^a - \frac{1}{2} f^{a8b}f^{8bg} T^g \right. \nonumber \\
&  & \mbox{\hglue2.0truecm} - \frac{N_f-4}{2(N_f^2-4)}(N_c+N_f) d^{a8b}\{T^b,T^8\} - \frac{N_f-4}{2(N_f^2-4)}(N_c+N_f) d^{88b} \{T^a,T^b\} - \frac{2}{N_f^2-1} \delta^{88} \{J^2,T^a\} \nonumber \\
&  & \mbox{\hglue2.0truecm} \left. - \frac{N_f}{N_f^2-4} d^{a8b}\{T^8,\{J^r,G^{rb}\}\} - \frac{N_f}{N_f^2-4} d^{88b}\{T^a,\{J^r,G^{rb}\}\} \right], \label{eq:qsecond}
\end{eqnarray}
where $a_{(n)}^\mathbf{rep},\ldots, g_{(n)}^{\mathbf{rep}}$ are unknown coefficients that accompany $n$-body operators that explicitly break flavor symmetry and $\lambda\sim m_s$ is a (dimensionless) measure of $SU(3)$ SB introduced to keep track of the number of times the perturbation enters. There are five nontrivial coefficients for first-order SB and ten more for second-order SB. These coefficients should formally encode both the contributions from the counterterms and the analytical parts of the loop integral.

The total corrections from explicit SB to the $s$-wave amplitudes of the processes analyzed here read
\begin{eqnarray}
-i \sqrt{2} f_\pi \delta \mathcal{A}_\mathrm{SB}^{(s)}(\Sigma^- \to n + \pi^-) & = & \frac{1}{2 \sqrt{3}} \lambda a_{(1)}^\mathbf{8} - \frac{1}{12 \sqrt{3}} \lambda a_{(2)}^\mathbf{8} - \frac{1}{3 \sqrt{3}} \lambda a_{(3)}^\mathbf{10+\overline{10}} - \frac{1}{5\sqrt{3}} \lambda a_{(2)}^\mathbf{27} - \frac{2}{15 \sqrt{3}} \lambda a_{(3)}^\mathbf{27} - \frac34 \lambda^2 e_{(1)}^{\mathbf{8}} \nonumber \\
&  & \mbox{} - \frac{1}{12} \lambda^2 g_{(1)}^{\mathbf{8}} + \frac12 \lambda^2 h_{(1)}^{\mathbf{8}} + \frac18 \lambda^2 e_{(2)}^{\mathbf{8}} + \frac{1}{72} \lambda^2 g_{(2)}^{\mathbf{8}} -\frac{1}{12} \lambda^2 h_{(2)}^{\mathbf{8}} + \frac{1}{18} \lambda^2 b_{(2)}^{\mathbf{10+\overline{10}}} + \frac{1}{30} \lambda^2 b_{(2)}^{\mathbf{27}} \nonumber \\
&  & \mbox{} + \frac{1}{45} \lambda^2 b_{(3)}^{\mathbf{27}} - \frac{1}{30} \lambda^2 b_{(3)}^{\mathbf{64}},
\end{eqnarray}
\begin{eqnarray}
-i \sqrt{2} f_\pi \delta \mathcal{A}_\mathrm{SB}^{(s)}(\Lambda \to p + \pi^-) & = & \frac{1}{2 \sqrt{2}} \lambda a_{(1)}^\mathbf{8} + \frac{1}{4 \sqrt{2}} \lambda a_{(2)}^\mathbf{8} + \frac{1}{3 \sqrt{2}} \lambda a_{(3)}^\mathbf{10+\overline{10}} - \frac{3}{5\sqrt{2}} \lambda a_{(2)}^\mathbf{27} - \frac{\sqrt{2}}{5} \lambda a_{(3)}^\mathbf{27} - \frac34 \sqrt{\frac{3}{2}} \lambda^2 e_{(1)}^{\mathbf{8}} \nonumber \\
&  & \mbox{} - \frac{1}{4 \sqrt{6}} \lambda^2 g_{(1)}^{\mathbf{8}} + \frac{1}{2} \sqrt{\frac{3}{2}} \lambda^2 h_{(1)}^{\mathbf{8}} - \frac38 \sqrt{\frac{3}{2}} \lambda^2 e_{(2)}^{\mathbf{8}} - \frac{1}{8 \sqrt{6}} \lambda^2 g_{(2)}^{\mathbf{8}} + \frac{1}{4} \sqrt{\frac{3}{2}} \lambda^2 h_{(2)}^{\mathbf{8}} \nonumber \\
&  & \mbox{} - \frac{1}{6 \sqrt{6}} \lambda^2 b_{(2)}^{\mathbf{10+\overline{10}}} + \frac{1}{10} \sqrt{\frac32} \lambda^2 b_{(2)}^{\mathbf{27}} + \frac{1}{5 \sqrt{6}} \lambda^2 b_{(3)}^{\mathbf{27}},
\end{eqnarray}
\begin{eqnarray}
-i \sqrt{2} f_\pi \delta \mathcal{A}_\mathrm{SB}^{(s)}(\Xi^ - \to \Lambda + \pi^-) & = & - \frac{1}{2\sqrt{2}} \lambda a_{(1)}^\mathbf{8} - \frac{1}{12 \sqrt{2}} \lambda a_{(2)}^\mathbf{8} - \frac{1}{3\sqrt{2}} \lambda a_{(3)}^\mathbf{10+\overline{10}} - \frac{3}{5\sqrt{2}} \lambda a_{(2)}^\mathbf{27} - \frac{\sqrt{2}}{5} \lambda a_{(3)}^\mathbf{27} + \frac34 \sqrt{\frac{3}{2}} \lambda^2 e_{(1)}^{\mathbf{8}} \nonumber \\
&  & \mbox{} + \frac{1}{4\sqrt{6}} \lambda^2 g_{(1)}^{\mathbf{8}} - \frac12 \sqrt{\frac32} \lambda^2 h_{(1)}^{\mathbf{8}} + \frac18 \sqrt{\frac32} \lambda^2 e_{(2)}^{\mathbf{8}} + \frac{1}{24\sqrt{6}} \lambda^2 g_{(2)}^{\mathbf{8}} - \frac{1}{4\sqrt{6}} \lambda^2 h_{(2)}^{\mathbf{8}} \nonumber \\
&  & \mbox{} + \frac{1}{6\sqrt{6}} \lambda^2 b_{(2)}^{\mathbf{10+\overline{10}}} + \frac{1}{10} \sqrt{\frac32} \lambda^2 b_{(2)}^{\mathbf{27}} + \frac{1}{5\sqrt{6}} \lambda^2 b_{(3)}^{\mathbf{27}},
\end{eqnarray}
and
\begin{eqnarray}
-i \sqrt{2} f_\pi \delta \mathcal{A}_\mathrm{SB}^{(s)}(\Omega^- \to {\Xi^*}^0 + \pi^-) & = & - \frac12 \lambda a_{(1)}^\mathbf{8} - \frac{5}{12} \lambda a_{(2)}^\mathbf{8} - \frac65 \lambda a_{(2)}^\mathbf{27} - 2 \lambda a_{(3)}^\mathbf{27} + \frac{3 \sqrt{3}}{4} \lambda^2 e_{(1)}^{\mathbf{8}} + \frac{1}{4\sqrt{3}} \lambda^2 g_{(1)}^{\mathbf{8}} - \frac{\sqrt{3}}{2} \lambda^2 h_{(1)}^{\mathbf{8}} \nonumber \\
&  & \mbox{} + \frac{5\sqrt{3}}{8} \lambda^2 e_{(2)}^{\mathbf{8}} + \frac{5}{24\sqrt{3}}  \lambda^2 g_{(2)}^{\mathbf{8}} - \frac{5}{4\sqrt{3}} \lambda^2 h_{(2)}^{\mathbf{8}} + \frac{\sqrt{3}}{5} \lambda^2 b_{(2)}^{\mathbf{27}}
+ \frac{1}{\sqrt{3}} \lambda^2 b_{(3)}^{\mathbf{27}} \nonumber \\
&  & \mbox{} + \frac{2 \sqrt{3}}{5} \lambda^2 b_{(3)}^{\mathbf{64}}.
\end{eqnarray}

In passing, with explicit SB the Lee-Sugawara relation (\ref{eq:leesuw}) becomes
\begin{eqnarray}
&  & -i \frac{3f_\pi}{\sqrt{6}} \mathcal{A}_{\mathrm{tree}}^{(s)}(\Sigma^- \to n + \pi^-) -i f_\pi \mathcal{A}_{\mathrm{tree}}^{(s)}(\Lambda \to p + \pi^-) - 2i f_\pi \mathcal{A}_{\mathrm{tree}}^{(s)}(\Xi^- \to \Lambda + \pi^-) =
 \nonumber \\
&  & \mbox{}  - \frac13 \lambda a_{(3)}^\mathbf{10+\overline{10}} - \lambda a_{(2)}^\mathbf{27} - \frac23 \lambda a_{(3)}^\mathbf{27} + \frac{1}{6\sqrt{3}} \lambda^2 b_{(2)}^{\mathbf{10+\overline{10}}} + \frac{1}{2\sqrt{3}} \lambda^2 b_{(2)}^{\mathbf{27}} + \frac{1}{3\sqrt{3}} \lambda^2 b_{(3)}^{\mathbf{27}} - \frac{1}{20\sqrt{3}}  \lambda^2 b_{(3)}^{\mathbf{64}},
\end{eqnarray}
so neither singlet not octet flavor representations produces corrections to this relation.

\section{\label{sec:num}Numerical analysis}

At this stage, a least-squares fit can be readily performed to compare theory and experiment. The available experimental data about baryon nonleptonic decays are given in the form of lifetimes, branching ratios, and decay asymmetries \cite{part}, which can be used to determine the $s$- and $p$-wave amplitudes. This information is listed in the second column of Table \ref{t:amplitudes1.0}.

As for the free parameters in the analysis, there are four operator coefficients from the $1/N_c$ expansion of the baryon axial current Eq.~(\ref{eq:akc}), namely, $a_1$, $b_2$, $b_3$, and $c_3$, three more from the $1/N_c$ expansion of the weak Hamiltonian Eq.~\ref{eq:hw}), namely, $h_1$, $h_2$, and $h_3$, and 15 additional ones from SB, for a total of 22. So, in order to get a meaningful fit, most of these parameters should be estimated and/or determined from other sources, otherwise predictive power is lost. In particular, the first set of coefficients, $a_1$, $b_2$, $b_3$, and $c_3$, has been already determined in Ref.~\cite{rfm14a} from baryon semileptonic decays, practically under the same footing as in the present case. In other words, these parameters have been determined from a comparison between theory and experiment, where the theoretical expressions have been obtained in the combined formalism at one-loop order, including explicit SB and the effects of the baryon decuplet-octet mass difference in the loop integrals. The values that will be borrowed, labeled as Fit 1b in Table IV of that reference, are $a_1=0.95\pm 0.14$, $b_2=-1.10\pm 0.19$, $b_3=1.10\pm 0.09$, and $c_3=1.07\pm 0.15$. For the second set, the limit $h_3\to 0$ will be assumed, which is equivalent to using the approximation (\ref{eq:rappx}). The expected error introduced with this assumption is order $\mathcal{O}(1/N_c^2)$, so $h_1$ and $h_2$ are left as free parameters. Finally, regarding SB, one can still resort to a naive large-$N_c$ counting. By assuming that $\lambda \sim 0.3$, then first- and second-order SB should be comparable to $1/N_c$ and $1/N_c^2$ corrections for the physical value $N_c=3$, respectively, so the latter can be safely omitted. Thus, all in all, there are seven free parameters, still a large number compared to the available amplitudes listed in Table \ref{t:amplitudes1.0}.

A further assumption still can be made: Loop integrals do not contain contributions from the (antisymetric) $\mathbf{10}+\overline{\mathbf{10}}$ representation. Thus, a kind of a restricted fit can be performed by retaining only the coefficient $a_{(3)}^\mathbf{10+\overline{10}}$ from first-order SB introduced in Eq.~(\ref{eq:qfirst}). A more detailed analysis requires the inclusion of all SB terms, at least to first order, which could be done in a simultaneous analysis with $s$- and $p$-waves alike.

In a similar fashion, for definiteness, the meson masses used are the experimental ones \cite{part}, the pion decay constant is $f_\pi=93\, \mathrm{MeV}$, the scale of dimensional regularization is set to $\mu = 1.0 \, \mathrm{GeV}$, and the decuplet-octet baryon mass difference is $\Delta = 0.232\, \mathrm{GeV}$. As for the loop integrals (\ref{eq:ia}) and (\ref{eq:fn}), only the nonanalytical terms will be retained. Furthermore, a theoretical error of 0.1 (equivalent to corrections of order $1/N_c^2$) will be added in quadrature to the experimental errors to avoid potential biases.

\begingroup
\begin{table}
\caption{\label{t:amplitudes1.0} Values of $s$-wave amplitudes in baryon nonleptonic decays, $\mathcal{A}^{(s)}(B_i \to B_j + \pi)$. The values are given in dimensionless units of $G_F m_{\pi^+}^2$. The scale of dimensional regularization $\mu$ is set to 1.0 GeV.}
\begin{ruledtabular}
\begin{tabular}{lccccccc}
Process & $-i\mathcal{A}_{\mathrm{exp}}^{(s)}$ & $-i\mathcal{A}_{\mathrm{th}}^{(s)}$ & $-i\mathcal{A}_{\mathrm{tree}}^{(s)}$ & $-i\delta\mathcal{A}_{\mathrm{2a}}^{(s)}$ & $-i\delta\mathcal{A}_{\mathrm{2b}}^{(s)}$ & $-i\delta\mathcal{A}_{\mathrm{loop}}^{(s)}$ & $-i\delta\mathcal{A}_\mathrm{SB}^{(s)}$ \\ \hline
$\Sigma^+ \to n + \pi^+$         & $ 0.06 \pm 0.01$ & $ 0.00$ & $ 0.00$ & $ 0.00$ & $ 0.00$ & $ 0.00$ & $ 0.00$ \\
$\Sigma^+ \to p + \pi^0$         & $-1.43 \pm 0.05$ & $-1.35$ & $-2.84$ & $-0.88$ & $ 1.86$ & $ 0.98$ & $ 0.51$ \\ 
$\Sigma^- \to n + \pi^-$         & $ 1.88 \pm 0.01$ & $ 1.92$ & $ 4.02$ & $ 1.25$ & $-2.63$ & $-1.38$ & $-0.72$ \\
$\Lambda \to n + \pi^0$          & $-1.04 \pm 0.02$ & $-1.02$ & $-0.94$ & $-0.29$ & $ 0.83$ & $ 0.54$ & $-0.62$ \\
$\Lambda \to p + \pi^-$          & $ 1.42 \pm 0.01$ & $ 1.44$ & $ 1.33$ & $ 0.41$ & $-1.18$ & $-0.77$ & $ 0.88$ \\
$\Xi^0 \to \Lambda + \pi^0$      & $ 1.51 \pm 0.02$ & $ 1.44$ & $ 2.21$ & $ 0.69$ & $-2.08$ & $-1.39$ & $ 0.62$ \\
$\Xi^- \to \Lambda + \pi^-$      & $-1.98 \pm 0.01$ & $-2.04$ & $-3.13$ & $-0.97$ & $ 2.94$ & $ 1.97$ & $-0.88$ \\
$\Omega^- \to {\Xi^*}^0 + \pi^-$ &                  & $ 3.65$ & $ 0.68$ & $ 0.21$ & $ 2.76$ & $ 2.97$ & $ 0.00$ \\
\end{tabular}
\end{ruledtabular}
\end{table}
\endgroup

With all the above considerations, the best-fit parameters are
\begin{equation}
h_1 = -3.29 \pm 0.15, \qquad h_2 = 4.42 \pm 0.33, \qquad a_{(3)}^\mathbf{10+\overline{10}} = 3.75 \pm 0.60,
\end{equation}
which are given in units of $\sqrt{2} f_\pi G_F m_{\pi^+}^2$ and the quoted errors are a consequence of the theoretical error added. For this constrained fit, $\chi^2=1.59$ for 3 degrees of freedom.

The information concerning the output of the fit is collected in Table \ref{t:amplitudes1.0}. The predicted amplitude $\mathcal{A}_{\mathrm{th}}^{(s)}$ is constituted by adding up tree-level $\mathcal{A}_{\mathrm{tree}}^{(s)}$, one-loop $\delta \mathcal{A}_{\mathrm{loop}}^{(s)}$, and SB $\delta \mathcal{A}_{\mathrm{SB}}^{(s)}$ contributions. Loop contributions are evaluated from the graphs displayed in Figs.~\ref{fig:1loop}(a) and \ref{fig:1loop}(b) and \ref{fig:1loop}(c), which correspond to $\delta\mathcal{A}_{\mathrm{2a}}^{(s)}$ and $\delta\mathcal{A}_{\mathrm{2b}}^{(s)}$, respectively.

The total amplitudes are in good agreement with the observed ones. Notice that, individually, loop corrections roughly represent (in absolute value) $30\%$, $50\%$, and $60\%$ of the lowest-order result for $\Sigma^-n$, $\Lambda p$, and $\Xi^-\Lambda$ processes, respectively. However, SB effects in the latter two cases partially cancel loop effects, so the combined corrections amount to respectively $-52\%$, $8\%$, and $-34\%$. The prediction for $\Sigma^-n$ is somewhat higher than expected. As for 
$\Omega^-{\Xi^*}^0$ process, SB corrections make up most of the predicted value and exceeds by far the tree-level value.

On the other hand, the numerical evaluation of the Lee-Sugawara relation yields
\begin{equation}
-\frac{3i}{\sqrt{6}} \delta \mathcal{A}_{\mathrm{2b}}^{(s)}(\Sigma^- \to n + \pi^-) - i \delta \mathcal{A}_{\mathrm{2b}}^{(s)}(\Lambda \to p + \pi^-) - 2 i\delta \mathcal{A}_{\mathrm{2b}}^{(s)}(\Xi^- \to \Lambda + \pi^-) = -0.28,
\end{equation}
which is in good agreement with the experimental one of approximately $-0.24$.

Further variations of the fit can be done by fixing the $\mu$ scale to different values to test the sensitivity of the output. For instance, for $\mu=0.8$ GeV, the fit yields
\begin{equation}
h_1 = -6.07 \pm 0.31, \qquad h_2 = 8.68 \pm 0.57, \qquad a_{(3)}^\mathbf{10+\overline{10}} = 6.23 \pm 0.72,
\end{equation}
whereas for $\mu=1.2$ GeV the output is
\begin{equation}
h_1 = -2.42 \pm 0.11, \qquad h_2 = 2.46 \pm 0.25, \qquad a_{(3)}^\mathbf{10+\overline{10}} = 2.97 \pm 0.57, \label{eq:fit1.2}
\end{equation}

\begingroup
\begin{table}
\caption{\label{t:amplitudes1.2} Values of $s$-wave amplitudes in baryon nonleptonic decays, $\mathcal{A}^{(s)}(B_i \to B_j + \pi)$. The values are given in dimensionless units of $G_F m_{\pi^+}^2$. The scale of dimensional regularization $\mu$ is set to 1.2 GeV.}
\begin{ruledtabular}
\begin{tabular}{lccccccc}
Process & $-i\mathcal{A}_{\mathrm{exp}}^{(s)}$ & $-i\mathcal{A}_{\mathrm{th}}^{(s)}$ & $-i\mathcal{A}_{\mathrm{tree}}^{(s)}$ & $-i\delta\mathcal{A}_{\mathrm{2a}}^{(s)}$ & $-i\delta\mathcal{A}_{\mathrm{2b}}^{(s)}$ & $-i\delta\mathcal{A}_{\mathrm{loop}}^{(s)}$ & $-i\delta\mathcal{A}_\mathrm{SB}^{(s)}$ \\ \hline
$\Sigma^+ \to n + \pi^+$         & $ 0.06 \pm 0.01$ & $ 0.00$ & $ 0.00$ & $ 0.00$ & $ 0.00$ & $ 0.00$ & $ 0.00$ \\
$\Sigma^+ \to p + \pi^0$         & $-1.43 \pm 0.05$ & $-1.36$ & $-2.07$ & $-0.81$ & $ 1.12$ & $ 0.31$ & $ 0.40$ \\ 
$\Sigma^- \to n + \pi^-$         & $ 1.88 \pm 0.01$ & $ 1.92$ & $ 2.94$ & $ 1.14$ & $-1.59$ & $-0.45$ & $-0.57$ \\
$\Lambda \to n + \pi^0$          & $-1.04 \pm 0.02$ & $-1.02$ & $-0.75$ & $-0.29$ & $ 0.52$ & $ 0.23$ & $-0.50$ \\
$\Lambda \to p + \pi^-$          & $ 1.42 \pm 0.01$ & $ 1.44$ & $ 1.06$ & $ 0.41$ & $-0.73$ & $-0.32$ & $ 0.70$ \\
$\Xi^0 \to \Lambda + \pi^0$      & $ 1.51 \pm 0.02$ & $ 1.44$ & $ 1.65$ & $ 0.64$ & $-1.34$ & $-0.70$ & $ 0.49$ \\
$\Xi^- \to \Lambda + \pi^-$      & $-1.98 \pm 0.01$ & $-2.04$ & $-2.33$ & $-0.91$ & $ 1.90$ & $ 0.99$ & $-0.70$ \\
$\Omega^- \to {\Xi^*}^0 + \pi^-$ &                  & $ 2.29$ & $ 0.30$ & $ 0.11$ & $ 1.88$ & $ 1.99$ & $ 0.00$ \\
\end{tabular}
\end{ruledtabular}
\end{table}
\endgroup

For $\mu=0.8$ GeV the fit leads to a breakdown of the expansion, with nonphysical corrections over $100\%$ of the tree value. On the contrary, for $\mu=1.2$ GeV, the $\chi^2$ value remains unchanged but the best-fit parameters improve the expected contributions to the amplitudes compared to the case for $\mu=1.0$ GeV. Now, loop corrections represent (in absolute value) $15\%$, $30\%$, and $42\%$ of the lowest-order value for $\Sigma^-n$, $\Lambda p$, and $\Xi^-\Lambda$ processes, respectively. The net effects of both loop and SB corrections represent $-35\%$, $36\%$, and $-12\%$, respectively, which are in more accord with expectations. For $\Omega^-{\Xi^*}^0$ process, still loop corrections make up most of the predicted value; this is a direct consequence of the smallness of the value $h_C=-3(h_D+h_F)$, with $h_D=-1.55$ and $h_F=1.38$ obtained from (\ref{eq:fit1.2}). To be conclusive, a measurement of this amplitude would be welcome  in the future; this could avoid biasing the fit to the octet-octet measured amplitudes in the fit. As for the Lee-Sugawara relations, it remains unchanged so its value is still $-0.28$.

Previous analyses within chiral perturbation theory \cite{bij,jen92,b99,abd} have found mixed results. In Ref.~\cite{bij} pion loops were omitted and the Gell-Mann--Okubo relation was used; no baryon wave-function renormalization graphs were included. The analysis suggested a breakdown of the chiral expansion. In Ref.~\cite{jen92} also pion loops were omitted and the Gell-Mann--Okubo relation was used, but the effects of decuplet baryons in the loops were taken into account; only nonanalytical pieces in the loop integrals were retained and no counterterms were included. Reference \cite{abd} followed the lines of \cite{jen92}, but retained the pion loops. The analyses of these two references were carried out by fitting 3 parameters, namely, $b_D$, $b_F$, and $b_C$; the latter was not well determined in any analysis. In Ref.~\cite{b99} the decuplet degrees of freedom were integrated out and the counterterms were included up to chiral order $\mathcal{O}(p^3)$; a good convergence of the chiral expansion was obtained in a simultaneous fit of both $s$- and $p$-wave amplitudes.

Here, the numerical analysis provides an overall good description of baryon nonleptonic decays. One-loop corrections (including the mass difference between intermediate and external baryons where applicable) and SB effects are included systematically into the analysis. And most importantly, all baryon operators that appear at the physical value $N_c$ are evaluated.

\section{\label{sec:fin}Concluding remarks}

In this paper, the $s$-wave amplitudes in baryon nonleptonic decays were evaluated in heavy baryon chiral perturbation theory in the large-$N_c$ limit at one-loop order. All baryon operators present for $N_c=3$ were considered and the mass difference between decuplet and octet intermediate baryon states in the loop integrals was accounted for. Explicit flavor symmetry breaking effects were also included.

The calculation was performed following the lines of previous analyses in conventional heavy baryon chiral perturbation theory \cite{jen92,b99,abd}. First, the validity of the $\Delta I=1/2$ rule was taken for granted, i.e., the $\Delta I=3/2$ component of the decay amplitude was neglected. Second, the assumption of octet dominance was also made, i.e., it was assumed that the $\mathbf{8}$ component dominates the $\mathbf{27}$ component in the weak Hamiltonian.

The method was simple. A baryon operator, gathering together tree, one-loop and SB corrections was constructed. The operator had a well-defined $1/N_c$ expansion and correctly picked the octet component of the $\Delta S=1$ transitions under consideration. The matrix elements of that operator between $SU(6)$ symmetric baryon states yielded the $s$-wave amplitudes. At tree level, there were three unknown operator coefficients $h_i$; the first two went to octet-octet and the third one went to decuplet-decuplet transitions. These coefficients could be directly related to the chiral coefficients $h_D$, $h_F$, and $h_C$. At one-loop order, two kinds of Feynman diagrams were evaluated. One of them was linear in the $h_i$ coefficients and the other one depended not only linearly on $h_i$ but also quadratically on the operator coefficients introduced in the $1/N_c$ expansion of the baryon axial vector current. SB effects, on the other hand, at first- and second-order in the perturbative parameter, introduced several unknowns.

From the theoretical point of view, working out all baryon operators for $N_c=3$ had several advantages. The most striking one was that it allowed a direct comparison with heavy baryon chiral perturbation theory results \textit{term by term}. This comparison also revealed that the analysis of both octet-octet and decuplet-decuplet transitions could be described with only \textit{two parameters} from the weak Hamiltonian, rather than the usual three, so that the third one was related by Eq.~(\ref{eq:rappx}). Although it was at first an unexpected result, a similar one has already been obtained for the coefficients introduced in the chiral Lagrangian to first order in the quark mass matrix (Eq.~(3.61) of Ref.~\cite{jen96}).

There are some key differences between the present analysis and those discussed above. Here, the pion loop is retained and the effects of decuplet baryons in the loops are evaluated but, unlike Refs.~\cite{jen92,abd}, the mass difference between intermediate and external baryons in the loops is considered, which changes the numerics considerably. For instance, $I_\mathrm{b}^{(1)}(m_K,0,\mu)=0.25$ whereas $I_\mathrm{b}^{(1)}(m_K,\Delta,\mu)=-0.18$. And last but not least, {\it only} two parameters from the lowest-order Lagrangian, $h_1$ and $h_2$ (or equivalently $h_D$ and $h_F$) and one parameter from explicit SB are used in the fit to data. An overall good description of $s$-wave amplitudes is obtained and the outputs are listed in Tables \ref{t:amplitudes1.0} and \ref{t:amplitudes1.2}.

Some improved inputs in the analysis will be welcome in the near future. Some of them could be found in the operator coefficients from the axial current, which by the way can be used to determine the low-energy constants of the chiral Lagrangian $D$, $F$, $\mathcal{C}$, and $\mathcal{H}$. At present, the only calculation for the renormalization of the axial vector current in the context of large-$N_c$ chiral perturbation theory which accounts for the mass difference between octet and decuplet baryons available is the one of Ref.~\cite{rfm14a}. This calculation, however, does not include all baryon operators for $N_c=3$. This might be inconvenient because it has been shown that loop corrections to the axial vector currents are exceptionally sensitive to deviations of the ratios of baryon-pion axial vector couplings from $SU(6)$ values \cite{fmhjm}. A major improvement in that calculation is desirable. This, however, represents a non-negligible effort and will be attempted elsewhere.

To close this paper, it is known for a fact that theory can lead to a good determination of either $s$- or $p$-waves, but not both simultaneously. An intriguing question is whether the analysis of $p$-wave amplitudes, computed under the same footing as $s$-wave amplitudes, can yield a stable fit by using, among others, the above-mentioned two parameters from the weak Hamiltonian. This task, however, will be attempted in the near future.

\acknowledgments

This work was partially supported by Consejo Nacional de Ciencia y Tecnolog{\'\i}a and Fondo de Apoyo a la Investigaci\'on (Universidad Aut\'onoma de San Luis Potos{\'\i}), Mexico.

\appendix

\section{\label{sec:reduc}Reduction of baryon operators}

Equation (\ref{eq:vc2b}) contains $n$-body operators\footnote{An $n$-body operator is one with $n$ $q$'s and $n$ $q^\dagger$'s. It can be written as a polynomial of order $n$ in $J^i$, $T^a$, and $G^{ia}$ \cite{djm95}.}, with $n>N_c$, which are complicated commutators and/or anticommutators of the one-body operators $J^k$, $T^c$, and $G^{kc}$. All these complicated operator structures should be reduced and rewritten as linear combinations of the operator bases (\ref{eq:bx})--(\ref{eq:bz}), with $n \leq N_c$. The reduction, although lengthy and tedious in view of the considerable amount of group theory involved, is nevertheless doable because the operator bases are complete and independent. All the baryon operator reductions required for $N_c=3$ are listed here.

\subsection{$[A^{ia},[A^{ia},\{J^r,G^{rc}\}]]$}

\begin{equation}
[G^{ia},[G^{ia},\{J^r,G^{rc}\}]] = - \frac32 (N_c+ N_f) T^c + \frac14 (7N_f+4) \{J^r,G^{rc}\},
\end{equation}

\begin{equation}
[G^{ia},[\mathcal{D}_2^{ia},\{J^r,G^{rc}\}]] + [\mathcal{D}_2^{ia},[G^{ia},\{J^r,G^{rc}\}]] = (N_c+N_f) \{J^r,G^{rc}\} + \frac12 (N_f-2) \{J^2,T^c\},
\end{equation}

\begin{eqnarray}
&  & [G^{ia},[\mathcal{D}_3^{ia},\{J^r,G^{rc}\}]] + [\mathcal{D}_3^{ia},[G^{ia},\{J^r,G^{rc}\}]] = [N_c(N_c+2N_f)+2N_f] \{J^r,G^{rc}\} - (N_c+N_f) \{J^2,T^c\} \nonumber \\
&  & \mbox{} + N_f \{J^2,\{J^r,G^{rc}\}\},
\end{eqnarray}

\begin{eqnarray}
&  & [G^{ia},[\mathcal{O}_3^{ia},\{J^r,G^{rc}\}]] + [\mathcal{O}_3^{ia},[G^{ia},\{J^r,G^{rc}\}]] = - 6(N_c+N_f) T^c - \frac32 [N_c(N_c+2N_f)-4N_f] \{J^r,G^{rc}\} \nonumber \\
&  & \mbox{} - \frac72 (N_c+N_f) \{J^2,T^c\} + 3(N_f+2) \{J^2,\{J^r,G^{rc}\}\},
\end{eqnarray}

\begin{equation}
[\mathcal{D}_2^{ia},[\mathcal{D}_2^{ia},\{J^r,G^{rc}\}]] = \frac12 N_f \{J^2,\{J^r,G^{rc}\}\}, 
\end{equation}

\begin{equation}
[\mathcal{D}_2^{ia},[\mathcal{D}_3^{ia},\{J^r,G^{rc}\}]] + [\mathcal{D}_3^{ia},[\mathcal{D}_2^{ia},\{J^r,G^{rc}\}]] = 2 (N_c+N_f) \{J^2,\{J^r,G^{rc}\}\} + (N_f-2) \{J^2,\{J^2,T^c\}\},
\end{equation}

\begin{equation}
[\mathcal{D}_2^{ia},[\mathcal{O}_3^{ia},\{J^r,G^{rc}\}]] + [\mathcal{O}_3^{ia},[\mathcal{D}_2^{ia},\{J^r,G^{rc}\}]] = 0,
\end{equation}

\begin{eqnarray}
&  & [\mathcal{D}_3^{ia},[\mathcal{D}_3^{ia},\{J^r,G^{rc}\}]] = [N_c(N_c+2N_f)+2N_f] \{J^2,\{J^r,G^{rc}\}\} - (N_c+N_f) \{J^2,\{J^2,T^c\}\} \nonumber \\
&  & \mbox{} + N_f \{J^2,\{J^2,\{J^r,G^{rc}\}\}\},
\end{eqnarray}

\begin{equation}
[\mathcal{D}_3^{ia},[\mathcal{O}_3^{ia},\{J^r,G^{rc}\}]] + [\mathcal{O}_3^{ia},[\mathcal{D}_3^{ia},\{J^r,G^{rc}\}]] = 0,
\end{equation}

\begin{eqnarray}
&  & [\mathcal{O}_3^{ia},[\mathcal{O}_3^{ia},\{J^r,G^{rc}\}]] = - 6(N_c+N_f) T^c - \frac32 [N_c(N_c+2N_f)-4N_f] \{J^r,G^{rc}\} - \frac{19}{2} (N_c+N_f) \{J^2,T^c\} \nonumber \\
&  & \mbox{} - \frac14 [5N_c(N_c+2N_f)-38N_f-24] \{J^2,\{J^r,G^{rc}\}\} - \frac94 (N_c+ N_f) \{J^2,\{J^2,T^c\}\} \nonumber \\
&  & \mbox{} + \frac12 (3N_f+10) \{J^2,\{J^2,\{J^r,G^{rc}\}\}\},
\end{eqnarray}

\subsection{$d^{ab8} [A^{ia},[A^{ib},\{J^r,G^{rc}\}]]$}

\begin{eqnarray}
&  & d^{ab8} [G^{ia},[G^{ib},\{J^r,G^{rc}\}]] = - \frac{3 N_c(N_c+2N_f)}{4N_f} \delta^{c8} - \frac34 (N_c+N_f) d^{c8e} T^e - \frac34 \{T^c,T^8\} + (N_f+1) \{G^{rc},G^{r8}\} \nonumber \\
&  & \mbox{} + \frac18 (3N_f+4) d^{c8e} \{J^r,G^{re}\} + \frac{N_f+1}{N_f} \delta^{c8} J^2,
\end{eqnarray}

\begin{eqnarray}
&  & d^{ab8} \left([G^{ia},[\mathcal{D}_2^{ib},\{J^r,G^{rc}\}]] + [\mathcal{D}_2^{ia},[G^{ib},\{J^r,G^{rc}\}]]\right) = \frac12 (N_c+ N_f) d^{c8e} \{J^r,G^{re}\} + \frac14(N_f-2) d^{c8e} \{J^2,T^e\} \nonumber \\
&  & \mbox{} - \frac12 \{T^c,\{J^r,G^{r8}\}\} + \frac12 \{T^8,\{J^r,G^{rc}\}\},
\end{eqnarray}

\begin{eqnarray}
&  & d^{ab8} \left([G^{ia},[\mathcal{D}_3^{ib},\{J^r,G^{rc}\}]] + [\mathcal{D}_3^{ia},[G^{ib},\{J^r,G^{rc}\}]]\right) = N_f d^{c8e} \{J^r,G^{re}\} - \frac{N_c(N_c+2N_f)}{N_f} \delta^{c8} J^2 \nonumber \\
&  & \mbox{} - \frac12 (N_c+N_f) d^{c8e} \{J^2,T^e\} + (N_c+N_f) \{T^8,\{J^r,G^{rc}\}\} + \frac14 (3N_f-2) \{J^2,\{T^c,T^8\}\} + (N_f-4) \{J^2,\{G^{rc},G^{r8}\}\} \nonumber \\
&  & \mbox{} + \frac12 (N_f-2) d^{c8e} \{J^2,\{J^r,G^{re}\}\} + N_f f^{ceg} f^{8eh} \{J^2,\{G^{rg},G^{rh}\}\} - N_f d^{ceg} d^{8eh} \{J^2,\{G^{rg},G^{rh}\}\} \nonumber \\
&  & \mbox{} - N_f \{\{J^r,G^{rc}\},\{J^m,G^{m8}\}\},
\end{eqnarray}

\begin{eqnarray}
&  & d^{ab8} \left([G^{ia},[\mathcal{O}_3^{ib},\{J^r,G^{rc}\}]] + [\mathcal{O}_3^{ia},[G^{ib},\{J^r,G^{rc}\}]]\right) = - \frac{3(N_c^2+2N_cN_f)}{N_f} \delta^{c8} - 3(N_c+N_f) d^{c8e} T^e - 3 \{T^c,T^8\} \nonumber \\
&  & \mbox{} + 4(N_f+1) \{G^{rc},G^{r8}\} + (N_f-2) d^{c8e} \{J^r,G^{re}\} + \frac{N_c(N_c+2N_f)(6N_f^2+5N_f+12)+8N_f(2N_f-1)}{2N_f^2} \delta^{c8} J^2 \nonumber \\
&  & \mbox{} + \frac{(N_c+N_f)(6N_f^2+5N_f+12)}{4N_f} d^{c8e} \{J^2,T^e\} - \frac32 (N_c+N_f) \{T^8,\{J^r,G^{rc}\}\} - \frac{15N_f^2+50N_f+24}{8N_f} \{J^2,\{T^c,T^8\}\} \nonumber \\
&  & \mbox{} + \frac{7N_f^2+24N_f+16}{2N_f} \{J^2,\{G^{rc},G^{r8}\}\} - \frac{N_f^2-N_f-4}{2N_f} d^{c8e} \{J^2,\{J^r,G^{re}\}\} \nonumber \\
&  & \mbox{} - \frac{9N_f^2+20N_f+16}{2N_f} f^{ceg} f^{8eh} \{J^2,\{G^{rg},G^{rh}\}\} + \frac12 (N_f+4) d^{ceg} d^{8eh} \{J^2,\{G^{rg},G^{rh}\}\} \nonumber \\
&  & \mbox{} + \frac12 (N_f+4) \{\{J^r,G^{rc}\},\{J^m,G^{m8}\}\},
\end{eqnarray}

\begin{equation}
d^{ab8} [\mathcal{D}_2^{ia},[\mathcal{D}_2^{ib},\{J^r,G^{rc}\}]] = \frac14 N_f d^{c8e} \{J^2,\{J^r,G^{re}\}\},
\end{equation}

\begin{eqnarray}
&  & d^{ab8} \left( [\mathcal{D}_2^{ia},[\mathcal{D}_3^{ib},\{J^r,G^{rc}\}]] + [\mathcal{D}_3^{ia},[\mathcal{D}_2^{ib},\{J^r,G^{rc}\}]] \right) = (N_c+ N_f) d^{c8e} \{J^2,\{J^r,G^{re}\}\} + \frac12 (N_f-2) d^{c8e} \{J^2,\{J^2,T^e\}\} \nonumber \\
&  & \mbox{} - \{J^2,\{T^c,\{J^r,G^{r8}\}\}\} + \{J^2,\{T^8,\{J^r,G^{rc}\}\}\},
\end{eqnarray}

\begin{equation}
d^{ab8} \left( [\mathcal{D}_2^{ia},[\mathcal{O}_3^{ib},\{J^r,G^{rc}\}]] + [\mathcal{O}_3^{ia},[\mathcal{D}_2^{ib},\{J^r,G^{rc}\}]] \right) = 0,
\end{equation}

\begin{eqnarray}
&  & d^{ab8} [\mathcal{D}_3^{ia},[\mathcal{D}_3^{ib},\{J^r,G^{rc}\}]] = \frac{3N_c^2(N_c+2N_f)^2}{N_f} \delta^{c8} J^2 + \frac32 N_c(N_c+N_f)(N_c+2N_f) d^{c8e} \{J^2,T^e\} \nonumber \\
&  & \mbox{} - \frac32 N_c(N_c+2N_f) \{J^2,\{T^c,T^8\}\} + 2N_c(N_c+2N_f) \{J^2,\{G^{rc},G^{r8}\}\} - [N_c(N_c+2N_f)-N_f] d^{c8e} \{J^2,\{J^r,G^{re}\}\} \nonumber \\
&  & \mbox{} - 4N_c(N_c+2N_f) f^{ceg} f^{8eh} \{J^2,\{G^{rg},G^{rh}\}\} + (N_c+N_f) d^{c8e} \{J^2,\{J^2,T^e\}\} + (N_c+N_f) \{J^2,\{T^8,\{J^r,G^{rc}\}\}\} \nonumber \\
&  & \mbox{} - 2 \{J^2,\{J^2,\{T^c,T^8\}\}\} + 2 N_f \{J^2,\{J^2,\{G^{rc},G^{r8}\}\}\} + \frac12 N_f d^{c8e} \{J^2,\{J^2,\{J^r,G^{re}\}\}\} \nonumber \\
&  & \mbox{} - 4 f^{ceg} f^{8eh} \{J^2,\{J^2,\{G^{rg},G^{rh}\}\}\} - N_f \{J^2,\{\{J^r,G^{rc}\},\{J^m,G^{m8}\}\}\},
\end{eqnarray}

\begin{equation}
d^{ab8} \left([\mathcal{D}_3^{ia},[\mathcal{O}_3^{ib},\{J^r,G^{rc}\}]] + [\mathcal{O}_3^{ia},[\mathcal{D}_3^{ib},\{J^r,G^{rc}\}]]\right) = 0,
\end{equation}

\begin{eqnarray}
&  & d^{ab8} [\mathcal{O}_3^{ia},[\mathcal{O}_3^{ib},\{J^r,G^{rc}\}]] = - \frac{3N_c(N_c+2N_f)}{N_f} \delta^{c8} - 3(N_c+N_f) d^{c8e} T^e - 3 \{T^c,T^8\} + 4(N_f+1) \{G^{rc},G^{r8}\} \nonumber \\
&  & \mbox{} + (N_f-2) d^{c8e} \{J^r,G^{re}\} \nonumber \\
&  & \mbox{} + \frac{9N_c^2(N_c+2N_f)^2N_f(N_f+1)+N_c(N_c+2N_f)(60N_f^2-38N_f+24) + 16N_f(2N_f-1)}{4N_f^2} \delta^{c8} J^2 \nonumber \\
&  & \mbox{} + \frac{(N_c+N_f)[9N_cN_f(N_c+2N_f)(N_f+1)+60N_f^2-38N_f+24]}{8N_f} d^{c8e} \{J^2,T^e\} - \frac32 (N_c+N_f) \{T^8,\{J^r,G^{rc}\}\} \nonumber \\
&  & \mbox{} - \frac{9N_cN_f(N_c+2N_f)(N_f+1)+63N_f^2+50N_f+24}{8N_f} \{J^2,\{T^c,T^8\}\} \nonumber \\
&  & \mbox{} + \frac{3N_cN_f(N_c+2N_f)(N_f+1)+31N_f^2+2N_f+16}{2N_f} \{J^2,\{G^{rc},G^{r8}\}\} \nonumber \\
&  & \mbox{} - \frac{3N_cN_f(N_c+2N_f)(N_f+1)+13N_f^2-2N_f+8}{4N_f} d^{c8e} \{J^2,\{J^r,G^{re}\}\} \nonumber \\
&  & \mbox{} - \frac{6N_cN_f(N_c+2N_f)(N_f+1)+41N_f^2+4N_f+16}{2N_f} f^{ceg} f^{8eh} \{J^2,\{G^{rg},G^{rh}\}\} \nonumber \\
&  & \mbox{} + \frac12 (N_f+4) d^{ceg} d^{8eh} \{J^2,\{G^{rg},G^{rh}\}\} + \frac12 (N_f+4) \{\{J^r,G^{rc}\},\{J^m,G^{m8}\}\} \nonumber \\
&  & \mbox{} + \frac34 (N_c+N_f)(N_f+1) d^{c8e} \{J^2,\{J^2,T^e\} - \frac54 (N_c+N_f) \{J^2,\{T^8,\{J^r,G^{rc}\}\}\} \nonumber \\
&  & \mbox{} - \frac34 (N_f+4) \{J^2,\{J^2,\{T^c,T^8\}\}\} + \frac32 (N_f+4) \{J^2,\{J^2,\{G^{rc},G^{r8}\}\}\} \nonumber \\
&  & \mbox{} - \frac14 N_f d^{c8e} \{J^2,\{J^2,\{J^r,G^{re}\}\}\} - (2N_f+5) f^{ceg} f^{8eh} \{J^2,\{J^2,\{G^{rg},G^{rh}\}\}\} \nonumber \\
&  & \mbox{} + \frac14 (N_f+8) \{J^2,\{\{J^r,G^{rc}\},\{J^m,G^{m8}\}\}\},
\end{eqnarray}

\subsection{$[A^{i8},[A^{i8},\{J^r,G^{rc}\}]]$}

\begin{eqnarray}
&  & [G^{i8},[G^{i8},\{J^r,G^{rc}\}]] = \frac14 f^{c8e} f^{8eg} \{J^r,G^{rg}\} + \frac12 d^{c8e} d^{8eg} \{J^r,G^{rg}\} - \frac{1}{N_f} \delta^{c8} \{J^r,G^{r8}\} + \frac{1}{N_f} \delta ^{88} \{J^r,G^{rc}\} \nonumber \\
&  & \mbox{} - 2 d^{c8e} \{G^{re},G^{r8}\} + d^{88e} \{G^{rc},G^{re}\} + \frac{1}{N_f} d^{c88} J^2,
\end{eqnarray}

\begin{equation}
[G^{i8},[\mathcal{D}_2^{i8},\{J^r,G^{rc}\}]] + [\mathcal{D}_2^{i8},[G^{i8},\{J^r,G^{rc}\}]] = - \frac38 (N_f-4)f^{c8e} f^{8eg} T^g + f^{c8e} f^{egh} \{T^g,\{G^{r8},G^{rh}\}\} + \frac12 f^{c8e} f^{8eg} \{J^2,T^g\}, 
\end{equation}

\begin{eqnarray}
&  & [G^{i8},[\mathcal{D}_3^{i8},\{J^r,G^{rc}\}]] + [\mathcal{D}_3^{i8},[G^{i8},\{J^r,G^{rc}\}]] = \frac32 f^{c8e} f^{8eg} \{J^r,G^{rg}\} + f^{c8e} f^{8eg} \{J^2,\{J^r,G^{rg}\}\} \nonumber \\
&  & \mbox{} - 2 d^{c8e} \{J^2,\{G^{re},G^{r8}\}\} + 2 d^{88e} \{J^2,\{G^{rc},G^{re}\}\} + 2 \{\{J^r,G^{rc}\},\{G^{m8},G^{m8}\}\} - 2 \{\{G^{rc},G^{r8}\},\{J^m,G^{m8}\}\} \nonumber \\
&  & \mbox{} + d^{c8e} \{\{J^r,G^{re}\},\{J^m,G^{m8}\}\} - d^{88e} \{\{J^r,G^{rc}\},\{J^m,G^{me}\}\}, \nonumber \\
\end{eqnarray}

\begin{eqnarray}
&  & [G^{i8},[\mathcal{O}_3^{i8},\{J^r,G^{rc}\}]] + [\mathcal{O}_3^{i8},[G^{i8},\{J^r,G^{rc}\}]] = - \frac54 f^{c8e} f^{8eg} \{J^r,G^{rg}\} + 2 d^{c8e} d^{8eg} \{J^r,G^{rg}\} - \frac{4}{N_f} \delta^{c8} \{J^r,G^{r8}\} \nonumber \\
&  & \mbox{} + \frac{4}{N_f} \delta^{88} \{J^r,G^{rc}\} - 8 d^{c8e} \{G^{re},G^{r8}\} + 4 d^{88e} \{G^{rc},G^{re}\} + \frac{4}{N_f} d^{c88} J^2 + d^{c8e} d^{8eg} \{J^2,\{J^r,G^{rg}\}\} \nonumber \\
&  & \mbox{} - \frac{2}{N_f} \delta^{c8} \{J^2,\{J^r,G^{r8}\}\} + \frac{2}{N_f} \delta^{88} \{J^2,\{J^r,G^{rc}\}\} - 5 d^{c8e} \{J^2,\{G^{re},G^{r8}\}\} + d^{88e} \{J^2,\{G^{rc},G^{re}\}\} \nonumber \\
&  & \mbox{} + \frac{2}{N_f} d^{c88} \{J^2, J^2\} - 3 \{\{J^r,G^{rc}\},\{G^{m8},G^{m8}\}\} + 3 \{\{G^{rc},G^{r8}\},\{J^m,G^{m8}\}\} + \frac12 d^{c8e} \{\{J^r,G^{re}\},\{J^m,G^{m8}\}\} \nonumber \\
&  & \mbox{} + \frac12 d^{88e} \{\{J^r,G^{rc}\},\{J^m,G^{me}\}\}, \nonumber \\
\end{eqnarray}

\begin{equation}
[\mathcal{D}_2^{i8},[\mathcal{D}_2^{i8},\{J^r,G^{rc}\}]] = \frac12 f^{c8e} f^{8eg} \{J^2,\{J^r,G^{rg}\}\},
\end{equation}

\begin{eqnarray}
&  & [\mathcal{D}_2^{i8},[\mathcal{D}_3^{i8},\{J^r,G^{rc}\}]] + [\mathcal{D}_3^{i8},[\mathcal{D}_2^{i8},\{J^r,G^{rc}\}]] = - \frac34 (N_f-4) f^{c8e} f^{8eg} \{J^2,T^g\} + f^{c8e} f^{8eg} \{J^2,\{J^2,T^g\}\} \nonumber \\
&  & \mbox{} + 2 f^{c8e} f^{egh} \{J^2,\{T^g,\{G^{r8},G^{rh}\}\}\},
\end{eqnarray}

\begin{equation}
[\mathcal{D}_2^{i8},[\mathcal{O}_3^{i8},\{J^r,G^{rc}\}]] + [\mathcal{O}_3^{i8},[\mathcal{D}_2^{i8},\{J^r,G^{rc}\}]] = 0,
\end{equation}

\begin{eqnarray}
&  & [\mathcal{D}_3^{i8},[\mathcal{D}_3^{i8},\{J^r,G^{rc}\}]] = \frac32 f^{c8e} f^{8eg} \{J^2,\{J^r,G^{rg}\}\} + f^{c8e} f^{8eg} \{J^2,\{J^2,\{J^r,G^{rg}\}\}\} - 2 d^{c8e} \{J^2,\{J^2,\{G^{re},G^{r8}\}\}\} \nonumber \\
&  & \mbox{} + 2 d^{88e} \{J^2,\{J^2,\{G^{rc},G^{re}\}\}\} + 2 \{J^2,\{\{J^r,G^{rc}\},\{G^{m8},G^{m8}\}\}\} - 2 \{J^2,\{\{G^{rc},G^{r8}\},\{J^m,G^{m8}\}\}\} \nonumber \\
&  & \mbox{} + d^{c8e} \{J^2,\{\{J^r,G^{re}\},\{J^m,G^{m8}\}\}\} - d^{88e} \{J^2,\{\{J^r,G^{rc}\},\{J^m,G^{me}\}\}\},
\end{eqnarray}

\begin{equation}
[\mathcal{D}_3^{i8},[\mathcal{O}_3^{i8},\{J^r,G^{rc}\}]] + [\mathcal{O}_3^{i8},[\mathcal{D}_3^{i8},\{J^r,G^{rc}\}]] = 0, 
\end{equation}

\begin{eqnarray}
&  & [\mathcal{O}_3^{i8},[\mathcal{O}_3^{i8},\{J^r,G^{rc}\}]] = - \frac54 f^{c8e} f^{8eg} \{J^r,G^{rg}\} + 2 d^{c8e} d^{8eg} \{J^r,G^{rg}\} - \frac{4}{N_f} \delta^{c8} \{J^r,G^{r8}\} + \frac{4}{N_f} \delta^{88} \{J^r,G^{rc}\} \nonumber \\
&  & \mbox{} - 8 d^{c8e} \{G^{re},G^{r8}\} + 4 d^{88e} \{G^{rc},G^{re}\} + \frac{4}{N_f} d^{c88} J^2 - \frac78 f^{c8e} f^{8eg} \{J^2,\{J^r,G^{rg}\}\} + 3 d^{c8e} d^{8eg} \{J^2,\{J^r,G^{rg}\}\} \nonumber \\
&  & \mbox{} - \frac{6}{N_f} \delta^{c8} \{J^2,\{J^r,G^{r8}\}\} + \frac{6}{N_f} \delta^{88} \{J^2,\{J^r,G^{rc}\}\} - 13 d^{c8e} \{J^2,\{G^{re},G^{r8}\}\} + 5 d^{88e} \{J^2,\{G^{rc},G^{re}\}\} \nonumber \\
&  & \mbox{} + \frac{6}{N_f} d^{c88} \{J^2,J^2\} - 3 \{\{J^r,G^{rc}\},\{G^{m8},G^{m8}\}\} + 3 \{\{G^{rc},G^{r8}\},\{J^m,G^{m8}\}\} + \frac12 d^{c8e} \{\{J^r,G^{re}\},\{J^m,G^{m8}\}\} \nonumber \\
&  & \mbox{} + \frac12 d^{88e} \{\{J^r,G^{rc}\},\{J^m,G^{me}\}\} + \frac12 d^{c8e} d^{8eg} \{J^2,\{J^2,\{J^r,G^{rg}\}\}\} - \frac{1}{N_f} \delta^{c8} \{J^2,\{J^2,\{J^r,G^{r8}\}\}\} \nonumber \\
&  & \mbox{} + \frac{1}{N_f} \delta^{88} \{J^2,\{J^2,\{J^r,G^{rc}\}\}\} - \frac72 d^{c8e} \{J^2,\{J^2,\{G^{re},G^{r8}\}\}\} + \frac12 d^{88e} \{J^2,\{J^2,\{G^{rc},G^{re}\}\}\} \nonumber \\
&  & \mbox{} + \frac{1}{N_f} d^{c88} \{J^2,\{J^2,J^2\}\} - \frac52 \{J^2,\{\{J^r,G^{rc}\},\{G^{m8},G^{m8}\}\}\} + \frac52 \{J^2,\{\{G^{rc},G^{r8}\},\{J^m,G^{m8}\}\}\} \nonumber \\
&  & \mbox{} + \frac14 d^{88e} \{J^2,\{\{J^r,G^{rc}\},\{J^m,G^{me}\}\}\},
\end{eqnarray}

\subsection{$\{A^{ja},[\{J^r,G^{rc}\},[J^2,A^{ja}]]\}$}

\begin{eqnarray}
&  & \{G^{ja},[\{J^r,G^{rc}\},[J^2,G^{ja}]]\} = 3(N_c+N_f) T^c + \frac12 [N_c(N_c+2N_f)-7N_f] \{J^r,G^{rc}\} + \frac12 (N_c+N_f) \{J^2,T^c\} \nonumber \\
&  & \mbox{} - 2 \{J^2,\{J^r,G^{rc}\}\},
\end{eqnarray}

\begin{eqnarray}
&  & \{G^{ja},[\{J^r,G^{rc}\},[J^2,\mathcal{O}_3^{ja}]]\} + \{\mathcal{O}_3^{ja},[\{J^r,G^{rc}\},[J^2,G^{ja}]]\} = 12 (N_c+N_f) T^c + 3 [N_c(N_c+2N_f)-4N_f] \{J^r,G^{rc}\} \nonumber \\
&  & \mbox{} + 13 (N_c+N_f) \{J^2,T^c\} + [N_c(N_c+2N_f)-13N_f-12] \{J^2,\{J^r,G^{rc}\}\} + (N_c+N_f) \{J^2,\{J^2,T^c\}\} \nonumber \\
&  & \mbox{} - 4 \{J^2,\{J^2,\{J^r,G^{rc}\}\}\},
\end{eqnarray}

\begin{eqnarray}
&  & \{\mathcal{O}_3^{ja},[\{J^r,G^{rc}\},[J^2,\mathcal{O}_3^{ja}]]\} = 12 (N_c+N_f) T^c + 3 [N_c(N_c+2N_f)-4N_f] \{J^r,G^{rc}\} + 25 (N_c+N_f) \{J^2,T^c\} \nonumber \\
&  & \mbox{} + [4N_c(N_c+2N_f)-25N_f-12] \{J^2,\{J^r,G^{rc}\}\} + 11 (N_c+N_f) \{J^2,\{J^2,T^c\}\} \nonumber \\
&  & \mbox{} + \frac12 [N_c(N_c+2N_f)-19N_f-32] \{J^2,\{J^2,\{J^r,G^{rc}\}\}\} + \frac12 (N_c+N_f) \{J^2,\{J^2,\{J^2,T^c\}\}\} \nonumber \\
&  & \mbox{} - 2 \{J^2,\{J^2,\{J^2,\{J^r,G^{rc}\}\}\}\}
\end{eqnarray}

\subsection{$d^{ab8} \{A^{ja},[\{J^r,G^{rc}\},[J^2,A^{jb}]]\}$}

\begin{eqnarray}
&  & d^{ab8} \{G^{ja},[\{J^r,G^{rc}\},[J^2,G^{jb}]]\} = \frac{3N_c(N_c+2N_f)}{2N_f} \delta^{c8} + \frac32 (N_c+N_f) d^{c8e} T^e + \frac32 \{T^c,T^8\} - 2 (N_f+1) \{G^{rc},G^{r8}\} \nonumber \\
&  & \mbox{} - \frac14 (3N_f-4) d^{c8e} \{J^r,G^{re}\} - \frac{N_c(N_c+2N_f)(N_f+3)+4N_f^2-2N_f}{N_f^2} \delta^{c8} J^2 - \frac{(N_c+N_f)(N_f+3)}{2N_f} d^{c8e} \{J^2,T^e\} \nonumber \\
&  & \mbox{} + \frac12 (N_c+N_f) \{T^8,\{J^r,G^{rc}\}\} + \frac{7N_f+6}{4N_f} \{J^2,\{T^c,T^8\}\} - \frac{3N_f+4}{N_f} \{J^2,\{G^{rc},G^{r8}\}\} - \frac{1}{N_f} d^{c8e} \{J^2,\{J^r,G^{re}\}\} \nonumber \\
&  & \mbox{} - \frac{3N_f+4}{N_f} f^{cab} f^{8be} \{J^2,\{G^{ra},G^{re}\}\} - d^{cab} d^{8be} \{J^2,\{G^{ra},G^{re}\}\} - \{\{J^r,G^{rc}\},\{J^m,G^{m8}\}\},
\end{eqnarray}

\begin{eqnarray}
&  & d^{ab8} (\{G^{ja},[\{J^r,G^{rc}\},[J^2,\mathcal{O}_3^{jb}]]\} + \{\mathcal{O}_3^{ja},[\{J^r,G^{rc}\},[J^2,G^{jb}]]\}) = \frac{6N_c(N_c+2N_f)}{N_f} \delta^{c8} + 6 (N_c+N_f) d^{c8e} T^e \nonumber \\
&  & \mbox{} + 6 \{T^c,T^8\} - 8 (N_f+1) \{G^{rc},G^{r8}\} - 2 (N_f-2) d^{c8e} \{J^r,G^{re}\} \nonumber \\
&  & \mbox{} - \frac{3N_c^2N_f(N_c+2N_f)^2+N_c(N_c+2N_f)(18N_f^2-7N_f+12)+16N_f^2-8N_f}{N_f^2} \delta^{c8} J^2 \nonumber \\
&  & \mbox{} - \frac{(N_c+N_f)[3N_cN_f(N_c+2N_f)+18N_f^2-7N_f+12]}{2N_f} d^{c8e} \{J^2,T^e\} + 3 (N_c+N_f) \{T^8,\{J^r,G^{rc}\}\} \nonumber \\
&  & \mbox{} + \frac{6N_cN_f(N_c+2N_f)+39N_f^2+50N_f+24}{4N_f} \{J^2,\{T^c,T^8\}\} \nonumber \\
&  & \mbox{} - \frac{2N_cN_f(N_c+2N_f)+19N_f^2+24N_f+16}{N_f} \{J^2,\{G^{rc},G^{r8}\}\} \nonumber \\
&  & \mbox{} + \frac{2N_cN_f(N_c+2N_f)+7N_f^2-2N_f-8}{2N_f} d^{c8e} \{J^2,\{J^r,G^{re}\}\} \nonumber \\
&  & \mbox{} - \frac{4N_cN_f(N_c+2N_f)+25N_f^2+12N_f+16}{N_f} f^{cab} f^{8be} \{J^2,\{G^{ra},G^{re}\}\} - (N_f+4) d^{cab} d^{8be} \{J^2,\{G^{ra},G^{re}\}\} \nonumber \\
&  & \mbox{} - (N_f+4) \{\{J^r,G^{rc}\},\{J^m,G^{m8}\}\} - (N_c+N_f) d^{c8e} \{J^2,\{J^2,T^e\}\} + (N_c+N_f) \{J^2,\{T^8,\{J^r,G^{rc}\}\}\} \nonumber \\
&  & \mbox{} + 2 \{J^2,\{J^2,\{T^c,T^8\}\}\} - 4 \{J^2,\{J^2,\{G^{rc},G^{r8}\}\}\} - 4 f^{cab} f^{8be} \{J^2,\{J^2,\{G^{ra},G^{re}\}\}\} \nonumber \\
&  & \mbox{} - 2 \{J^2,\{\{J^r,G^{rc}\},\{J^m,G^{m8}\}\}\},
\end{eqnarray}

\begin{eqnarray}
&  & d^{ab8} \{\mathcal{O}_3^{ja},[\{J^r,G^{rc}\},[J^2,\mathcal{O}_3^{jb}]]\} = \frac{6N_c(N_c+2N_f)}{N_f} \delta^{c8} + 6 (N_c+N_f) d^{c8e} T^e + 6 \{T^c,T^8\} - 8 (N_f+1) \{G^{rc},G^{r8}\} \nonumber \\
&  & \mbox{} - 2 (N_f-2) d^{c8e} \{J^r,G^{re}\} \nonumber \\
&  & \mbox{} - \frac{N_c(N_c+2N_f)[9N_c^2N_f(N_c+2N_f)^2+12N_cN_f(N_c+2N_f)(6N_f-1)+168N_f^2-124N_f+48]+32N_f(2N_f-1)}{4N_f^2} \delta^{c8} J^2 \nonumber \\
&  & \mbox{} - \frac{(N_c+N_f)[9N_c^2N_f(N_c+2N_f)^2+12N_cN_f(N_c+2N_f)(6N_f-1)+168N_f^2-124N_f+48]}{8N_f} d^{c8e} \{J^2,T^e\} \nonumber \\
&  & \mbox{} + 3 (N_c+N_f) \{T^8,\{J^r,G^{rc}\}\} \nonumber \\
&  & \mbox{} + \frac{9N_c^2N_f(N_c+2N_f)^2+12N_cN_f(N_c+2N_f)(6N_f-1)+174N_f^2+100N_f+48}{8N_f} \{J^2,\{T^c,T^8\}\} \nonumber \\
&  & \mbox{} - \frac{3N_c^2N_f(N_c+2N_f)^2+4N_cN_f(N_c+2N_f)(6N_f-1)+86N_f^2+48N_f+32}{2N_f} \{J^2,\{G^{rc},G^{r8}\}\} \nonumber \\
&  & \mbox{} + \frac{3N_c^2N_f(N_c+2N_f)^2+4N_cN_f(Nc+2N_f)(6N_f-1)+38N_f^2-4N_f-16}{4N_f} d^{c8e} \{J^2,\{J^r,G^{re}\}\} \nonumber \\
&  & \mbox{} - \frac{3N_c^2N_f(N_c+2N_f)^2+4N_cN_f(N_c+2N_f)(6N_f-1)+57N_f^2-4N_f+16}{N_f} f^{cab} f^{8be} \{J^2,\{G^{ra},G^{re}\}\} \nonumber \\
&  & \mbox{} - (N_f+4) d^{cab} d^{8be} \{J^2,\{G^{ra},G^{re}\}\} - (N_f+4) \{\{J^r,G^{rc}\},\{J^m,G^{m8}\}\} \nonumber \\
&  & \mbox{} - \frac14 (N_c+N_f)[3N_c(N_c+2N_f)+4(6N_f-1)] d^{c8e} \{J^2,\{J^2,T^e\}\} + 4 (N_c+N_f) \{J^2,\{T^8,\{J^r,G^{rc}\}\}\} \nonumber \\
&  & \mbox{} + \frac14 [3N_c(N_c+2N_f)+24N_f+40] \{J^2,\{J^2,\{T^c,T^8\}\}\} - [N_c(N_c+2N_f)+12N_f+20] \{J^2,\{J^2,\{G^{rc},G^{r8}\}\}\} \nonumber \\
&  & \mbox{} + \frac14 [2N_c(N_c+2N_f)+9N_f] d^{c8e} \{J^2,\{J^2,\{J^r,G^{re}\}\}\} - 2 [N_c(N_c+2N_f)+8N_f+6] f^{cab} f^{8be} \{J^2,\{J^2,\{G^{ra},G^{re}\}\}\} \nonumber \\
&  & \mbox{} -(N_f+6) \{J^2,\{\{J^r,G^{rc}\},\{J^m,G^{m8}\}\}\} - \frac12 (N_c+N_f) d^{c8e} \{J^2,\{J^2,\{J^2,T^e\}\}\} \nonumber \\
&  & \mbox{} + \frac12(N_c+N_f) \{J^2,\{J^2,\{T^8,\{J^r,G^{rc}\}\}\}\} + \{J^2,\{J^2,\{J^2,\{T^c,T^8\}\}\}\} - 2 \{J^2,\{J^2,\{J^2,\{G^{rc},G^{r8}\}\}\}\} \nonumber \\
&  & \mbox{} - 2 f^{cab} f^{8be} \{J^2,\{J^2,\{J^2,\{G^{ra},G^{re}\}\}\}\} - \{J^2,\{J^2,\{\{J^r,G^{rc}\},\{J^m,G^{m8}\}\}\}\},
\end{eqnarray}

\subsection{$\{A^{j8},[\{J^r,G^{rc}\},[J^2,A^{j8}]]\}$}

\begin{eqnarray}
&  & \{G^{j8},[\{J^r,G^{rc}\},[J^2,G^{j8}]]\} = \frac14 f^{c8e} f^{8eg} \{J^r,G^{rg}\} - d^{c8e} d^{8eg} \{J^r,G^{rg}\} + \frac{2}{N_f} \delta^{c8} \{J^r,G^{r8}\} - \frac{2}{N_f} \delta^{88} \{J^r,G^{rc}\} \nonumber \\
&  & \mbox{} + 4 d^{c8e} \{G^{re},G^{r8}\} - 2 d^{88e} \{G^{rc},G^{re}\} - \frac{2}{N_f} d^{c88} J^2 + d^{c8e} \{J^2,\{G^{re},G^{r8}\}\} + \{\{J^r,G^{rc}\},\{G^{m8},G^{m8}\}\} \nonumber \\
&  & \mbox{} - \{\{G^{rc},G^{r8}\},\{J^m,G^{m8}\}\} - \frac12 d^{c8e} \{\{J^r,G^{re}\},\{J^m,G^{m8}\}\},
\end{eqnarray}

\begin{eqnarray}
&  & \{G^{j8},[\{J^r,G^{rc}\},[J^2,\mathcal{O}_3^{j8}]]\} + \{\mathcal{O}_3^{j8},[\{J^r,G^{rc}\},[J^2,G^{j8}]]\} = \frac52 f^{c8e} f^{8eg} \{J^r,G^{rg}\} - 4 d^{c8e} d^{8eg} \{J^r,G^{rg}\} \nonumber \\
&  & \mbox{} + \frac{8}{N_f} \delta^{c8} \{J^r,G^{r8}\} - \frac{8}{N_f} \delta^{88} \{J^r,G^{rc}\} + 16 d^{c8e} \{G^{re},G^{r8}\} - 8 d^{88e} \{G^{rc},G^{re}\} - \frac{8}{N_f} d^{c88} J^2 \nonumber \\
&  & \mbox{} + \frac12 f^{c8e} f^{8eg} \{J^2,\{J^r,G^{rg}\}\} - 4 d^{c8e} d^{8eg} \{J^2,\{J^r,G^{rg}\}\} + \frac{8}{N_f} \delta^{c8} \{J^2,\{J^r,G^{r8}\}\} - \frac{8}{N_f} \delta^{88} \{J^2,\{J^r,G^{rc}\}\} \nonumber \\
&  & \mbox{} + 18 d^{c8e} \{J^2,\{G^{re},G^{r8}\}\} - 6 d^{88e} \{J^2,\{G^{rc},G^{re}\}\} - \frac{8}{N_f} d^{c88} \{J^2,J^2\} + 6 \{\{J^r,G^{rc}\},\{G^{m8},G^{m8}\}\} \nonumber \\
&  & \mbox{} - 6 \{\{G^{rc},G^{r8}\},\{J^m,G^{m8}\}\} - d^{c8e} \{\{J^r,G^{re}\},\{J^m,G^{m8}\}\} - d^{88e} \{\{J^r,G^{rc}\},\{J^m,G^{me}\}\} \nonumber \\
&  & \mbox{} + 2 d^{c8e}\{J^2,\{J^2,\{G^{re},G^{r8}\}\}\} + 2 \{J^2,\{\{J^r,G^{rc}\},\{G^{m8},G^{m8}\}\}\} - 2 \{J^2,\{\{G^{rc},G^{r8}\},\{J^m,G^{m8}\}\}\} \nonumber \\
&  & \mbox{} - d^{c8e} \{J^2,\{\{J^r,G^{re}\},\{J^m,G^{m8}\}\}\},
\end{eqnarray}

\begin{eqnarray}
&  & \{\mathcal{O}_3^{j8},[\{J^r,G^{rc}\},[J^2,\mathcal{O}_3^{j8}]]\} = \frac52 f^{c8e} f^{8eg} \{J^r,G^{rg}\} - 4 d^{c8e} d^{8eg} \{J^r,G^{rg}\} + \frac{8}{N_f} \delta^{c8} \{J^r,G^{r8}\} - \frac{8}{N_f} \delta^{88} \{J^r,G^{rc}\} \nonumber \\
&  & \mbox{} + 16 d^{c8e} \{G^{re},G^{r8}\} - 8 d^{88e} \{G^{rc},G^{re}\} - \frac{8}{N_f} d^{c88} J^2 + 3 f^{c8e} f^{8eg} \{J^2,\{J^r,G^{rg}\}\} - 8 d^{c8e} d^{8eg} \{J^2,\{J^r,G^{rg}\}\} \nonumber \\
&  & \mbox{} + \frac{16}{N_f} \delta^{c8} \{J^2,\{J^r,G^{r8}\}\} - \frac{16}{N_f} \delta^{88} \{J^2,\{J^r,G^{rc}\}\} + 34 d^{c8e} \{J^2,\{G^{re},G^{r8}\}\} - 14 d^{88e} \{J^2,\{G^{rc},G^{re}\}\} \nonumber \\
&  & \mbox{} - \frac{16}{N_f} d^{c88} \{J^2,J^2\} + 6 \{\{J^r,G^{rc}\},\{G^{m8},G^{m8}\}\} - 6 \{\{G^{rc},G^{r8}\},\{J^m,G^{m8}\}\} - d^{c8e} \{\{J^r,G^{re}\},\{J^m,G^{m8}\}\} \nonumber \\
&  & \mbox{} - d^{88e} \{\{J^r,G^{rc}\},\{J^m,G^{me}\}\} + \frac14 f^{c8e} f^{8eg} \{J^2,\{J^2,\{J^r,G^{rg}\}\}\} - 3 d^{c8e} d^{8eg} \{J^2,\{J^2,\{J^r,G^{rg}\}\}\} \nonumber \\
&  & \mbox{} + \frac{6}{N_f} \delta^{c8} \{J^2,\{J^2,\{J^r,G^{r8}\}\}\} - \frac{6}{N_f} \delta^{88} \{J^2,\{J^2,\{J^r,G^{rc}\}\}\} + 16 d^{c8e}\{J^2,\{J^2,\{G^{re},G^{r8}\}\}\} \nonumber \\
&  & \mbox{} - 4 d^{88e} \{J^2,\{J^2,\{G^{rc},G^{re}\}\}\} - \frac{6}{N_f} d^{c88} \{J^2,\{J^2,J^2\}\} + 8 \{J^2,\{\{J^r,G^{rc}\},\{G^{m8},G^{m8}\}\}\} \nonumber \\
&  & \mbox{} - 8 \{J^2,\{\{G^{rc},G^{r8}\},\{J^m,G^{m8}\}\}\} - 2 d^{c8e} \{J^2,\{\{J^r,G^{re}\},\{J^m,G^{m8}\}\}\} - d^{88e} \{J^2,\{\{J^r,G^{rc}\},\{J^m,G^{me}\}\}\} \nonumber \\
&  & \mbox{} + d^{c8e} \{J^2,\{J^2,\{J^2,\{G^{re},G^{r8}\}\}\}\} + \{J^2,\{J^2,\{\{J^r,G^{rc}\},\{G^{m8},G^{m8}\}\}\}\}
- \{J^2,\{J^2,\{\{G^{rc},G^{r8}\},\{J^m,G^{m8}\}\}\}\} \nonumber \\
&  & \mbox{} - \frac12 d^{c8e} \{J^2,\{J^2,\{\{J^r,G^{re}\},\{J^m,G^{m8}\}\}\}\},
\end{eqnarray}

\subsection{$[A^{ja},[[J^2,[J^2,A^{ja}]],\{J^r,G^{rc}\}]] - \frac12 [[J^2,A^{ja}], [[J^2,A^{ja}],\{J^r,G^{rc}\}]]$}

\begin{eqnarray}
&  & [G^{ja},[[J^2,[J^2,G^{ja}]],\{J^r,G^{rc}\}]] - \frac12 [[J^2,G^{ja}], [[J^2,G^{ja}],\{J^r,G^{rc}\}]] = - 9 (N_c+N_f) T^c \nonumber \\
&  & - \frac94 [N_c(N_c+2N_f)-4N_f] \{J^r,G^{rc}\} - \frac{21}{4} (N_c+N_f) \{J^2,T^c\} + \frac92 (N_f+2) \{J^2,\{J^r,G^{rc}\}\},
\end{eqnarray}

\begin{eqnarray}
&  & [G^{ja},[[J^2,[J^2,\mathcal{O}_3^{ja}]],\{J^r,G^{rc}\}]] + [\mathcal{O}_3^{ja},[[J^2,[J^2,G^{ja}]],\{J^r,G^{rc}\}]] - \frac12 [[J^2,G^{ja}],[[J^2,\mathcal{O}_3^{ja}],\{J^r,G^{rc}\}]] \nonumber \\
&  & \mbox{} - \frac12 [[J^2,\mathcal{O}_3^{ja}], [[J^2,G^{ja}],\{J^r,G^{rc}\}]] = - 36(N_c+N_f) T^c - 9 [N_c(N_c+2N_f)-4N_f] \{J^r,G^{rc}\} \nonumber \\
&  & \mbox{} - 57 (N_c+N_f) \{J^2,T^c\} - \frac32 [5N_c(N_c+2N_f)-38N_f-24] \{J^2,\{J^r,G^{rc}\}\} - \frac{27}{2} (N_c+N_f) \{J^2,\{J^2,T^c\}\} \nonumber \\
&  & \mbox{} + 3 (3N_f+10) \{J^2,\{J^2,\{J^r,G^{rc}\}\}\},
\end{eqnarray}

\begin{eqnarray}
&  & [\mathcal{O}_3^{ja},[[J^2,[J^2,\mathcal{O}_3^{ja}]],\{J^r,G^{rc}\}]] - \frac12 [[J^2,\mathcal{O}_3^{ja}], [[J^2,\mathcal{O}_3^{ja}],\{J^r,G^{rc}\}]] = - 36 (N_c+N_f) T^c \nonumber \\
&  & \mbox{} - 9 [N_c(N_c+2N_f)-4N_f] \{J^r,G^{rc}\} - 93 (N_c+N_f) \{J^2,T^c\} - \frac12 [33N_c(N_c+2N_f)-186N_f-72] \{J^2,\{J^r,G^{rc}\}\} \nonumber \\
&  & \mbox{} - \frac{123}{2} (N_c+N_f) \{J^2,\{J^2,T^c\}\} - \frac34 [7N_c(N_c+2N_f)-76N_f-88] \{J^2,\{J^2,\{J^r,G^{rc}\}\}\} \nonumber \\
&  & \mbox{} - \frac{33}{4} (N_c+N_f) \{J^2,\{J^2,\{J^2,T^c\}\}\} + \frac12 (9N_f+42) \{J^2,\{J^2,\{J^2,\{J^r,G^{rc}\}\}\}\},
\end{eqnarray}

\subsection{$d^{ab8} \left([A^{ja},[[J^2,[J^2,A^{jb}]],\{J^r,G^{rc}\}]] - \frac12 [[J^2,A^{ja}], [[J^2,A^{jb}],\{J^r,G^{rc}\}]]\right)$}

\begin{eqnarray}
&  & d^{ab8} \left([G^{ja},[[J^2,[J^2,G^{jb}]],\{J^r,G^{rc}\}]] - \frac12 [[J^2,G^{ja}], [[J^2,G^{jb}],\{J^r,G^{rc}\}]] \right) = - \frac{9N_c(N_c+2N_f)}{2N_f} \delta^{c8} \nonumber \\
&  & \mbox{} - \frac92 (N_c+N_f) d^{c8e} T^e - \frac92 \{T^c,T^8\} + 6 (N_f+1) \{G^{rc},G^{r8}\} + \frac32 (N_f-2) d^{c8e} \{J^r,G^{re}\} \nonumber \\
&  & \mbox{} + \frac{3 [N_c(N_c+2N_f)(6N_f^2+5N_f+12)+16N_f^2-8N_f]}{4N_f^2} \delta^{c8} J^2 + \frac{3(N_c+N_f)(6N_f^2+5N_f+12)}{8N_f} d^{c8e} \{J^2,T^e\} \nonumber \\
&  & \mbox{} - \frac94 (N_c+N_f) \{T^8,\{J^r,G^{rc}\}\} - \frac{3(15N_f^2+50N_f+24)}{16N_f} \{J^2,\{T^c,T^8\}\} + \frac{3(7N_f^2+24N_f+16)}{4N_f} \{J^2,\{G^{rc},G^{r8}\}\} \nonumber \\
&  & \mbox{} - \frac{3(N_f^2-N_f-4)}{4N_f} d^{c8e} \{J^2,\{J^r,G^{re}\}\} + \frac{3(9N_f^2+20N_f+16)}{4N_f} f^{cab} f^{8be} \{J^2,\{G^{ra},G^{re}\}\} \nonumber \\
&  & \mbox{} + \frac34 (N_f+4) d^{cab} d^{8be} \{J^2,\{G^{ra},G^{re}\}\} + \frac34 (N_f+4) \{\{J^r,G^{rc}\},\{J^m,G^{m8}\}\},
\end{eqnarray}

\begin{eqnarray}
&  & d^{ab8} \Big( [G^{ja},[[J^2,[J^2,\mathcal{O}_3^{jb}]],\{J^r,G^{rc}\}]] + [\mathcal{O}_3^{ja},[[J^2,[J^2,G^{jb}]],\{J^r,G^{rc}\}]] - \frac12 [[J^2,G^{ja}],[[J^2,\mathcal{O}_3^{jb}],\{J^r,G^{rc}\}]] \nonumber \\
&  & \mbox{} - \frac12 [[J^2,\mathcal{O}_3^{ja}], [[J^2,G^{jb}],\{J^r,G^{rc}\}]] \Big) = - \frac{18N_c(N_c+2N_f)}{N_f} \delta^{c8} - 18 (N_c+N_f) d^{c8e} T^e - 18 \{T^c,T^8\} \nonumber \\
&  & \mbox{} + 24 (N_f+1) \{G^{rc},G^{r8}\} + 6 (N_f-2) d^{c8e} \{J^r,G^{re}\} \nonumber \\
&  & \mbox{} + \frac{3[9N_c^2N_f(N_c+2N_f)^2(N_f+1)+2N_c(N_c+2N_f)(30N_f^2-19N_f+12)+16N_f(2N_f-1)]}{2N_f^2} \delta^{c8} J^2 \nonumber \\
&  & \mbox{} + \frac{3[9N_cN_f(N_c+N_f)(N_c+2N_f)(N_f+1)+2(N_c+N_f)(30N_f^2-19N_f+12)]}{4N_f} d^{c8e} \{J^2,T^e\} \nonumber \\
&  & \mbox{} - 9 (N_c+N_f) \{T^8,\{J^r,G^{rc}\}\} - \frac{3[9N_cN_f(N_c+2N_f)(N_f+1)+63N_f^2+50N_f+24]}{4N_f} \{J^2,\{T^c,T^8\}\} \nonumber \\
&  & \mbox{} + \frac{3[3N_cN_f(N_c+2N_f)(N_f+1)+31N_f^2+24N_f+16]}{N_f} \{J^2,\{G^{rc},G^{r8}\}\} \nonumber \\
&  & \mbox{} - \frac{3[3N_cN_f(N_c+2N_f)(N_f+1)+13N_f^2-2N_f-8]}{2N_f} d^{c8e} \{J^2,\{J^r,G^{re}\}\} \nonumber \\
&  & \mbox{} + \frac{3 [6N_cN_f(N_c+2N_f)(N_f+1)+41N_f^2+4N_f+16]}{N_f} f^{cab} f^{8be} \{J^2,\{G^{ra},G^{re}\}\} \nonumber \\
&  & \mbox{} + 3 (N_f+4) d^{cab} d^{8be} \{J^2,\{G^{ra},G^{re}\}\} + 3 (N_f+4) \{\{J^r,G^{rc}\},\{J^m,G^{m8}\}\} \nonumber \\
&  & \mbox{} + \frac92 (N_c+N_f)(N_f+1) d^{c8e} \{J^2,\{J^2,T^e\}\} - \frac{15}{2} (N_c+N_f) \{J^2,\{T^8,\{J^r,G^{rc}\}\}\} - \frac92 (N_f+4) \{J^2,\{J^2,\{T^c,T^8\}\}\} \nonumber \\
&  & \mbox{} + 9 (N_f+4) \{J^2,\{J^2,\{G^{rc},G^{r8}\}\}\} - \frac32N_f d^{c8e} \{J^2,\{J^2,\{J^r,G^{re}\}\}\} + 6 (2N_f+5) f^{cab} f^{8be} \{J^2,\{J^2,\{G^{ra},G^{re}\}\}\} \nonumber \\
&  & + \frac32 (N_f+8) \{J^2,\{\{J^r,G^{rc}\},\{J^m,G^{m8}\}\}\},
\end{eqnarray}

\begin{eqnarray}
&  & d^{ab8} \left( [\mathcal{O}_3^{ja},[[J^2,[J^2,\mathcal{O}_3^{jb}]],\{J^r,G^{rc}\}]] - \frac12 [[J^2,\mathcal{O}_3^{ja}], [[J^2,\mathcal{O}_3^{jb}],\{J^r,G^{rc}\}]] \right) = - \frac{18N_c (N_c+2N_f)}{N_f} \delta^{c8} \nonumber \\
&  & \mbox{} - 18 (N_c+N_f) d^{c8e} T^e - 18 \{T^c,T^8\} + 24 (N_f+1) \{G^{rc},G^{r8}\} + 6 (N_f-2) d^{c8e} \{J^r,G^{re}\} \nonumber \\
&  & \mbox{} + \frac{3[9N_c^3 N_f(N_c+2N_f)^3(3N_f+5)+12N_c^2N_f(N_c+2 N_f)^2(27N_f-13)+8N_c(N_c+2N_f)(54N_f^2-43N_f+12)+64N_f(2N_f-1)]}{8N_f^2} \delta^{c8} J^2 \nonumber \\
&  & \mbox{} + \frac{3(N_c+N_f)[9N_c^2N_f(N_c+2N_f)^2(3N_f+5)+12N_cN_f(N_c+2N_f)(27N_f-13)+8(54N_f^2-43N_f+12)]}{16 N_f} d^{c8e} \{J^2,T^e\} \nonumber \\ 
&  & \mbox{} - 9 (N_c+N_f) \{T^8,\{J^r,G^{rc}\}\} \nonumber \\
&  & \mbox{} - \frac{3[9N_c^2N_f(N_c+2N_f)^2(3N_f+5)+12N_cN_f(N_c+2N_f)(27N_f-13)+4(111N_f^2+50N_f+24)]}{16N_f} \{J^2,\{T^c,T^8\}\} \nonumber \\
&  & \mbox{} + \frac{9N_c^2 N_f(N_c+2N_f)^2(3N_f+5)+12N_cN_f(N_c+2N_f)(27N_f-13)+12(55N_f^2+24N_f+16)}{4N_f} \{J^2,\{G^{rc},G^{r8}\}\} \nonumber \\
&  & \mbox{} - \frac{3[3N_c^2N_f(N_c+2N_f)^2(3N_f+5)+4N_cN_f(N_c+2N_f)(27N_f-13)+4(25N_f^2-2N_f-8)]}{8N_f} d^{c8e} \{J^2,\{J^r,G^{re}\}\} \nonumber \\
&  & \mbox{} + \frac{3[3N_c^2N_f(N_c+2N_f)^2(3N_f+5)+4N_cN_f(N_c+2N_f)(27N_f-13)+2(73N_f^2-12N_f+16)]}{2N_f} f^{cab} f^{8be} \{J^2,\{G^{ra},G^{re}\}\} \nonumber \\
&  & \mbox{} + 3 (N_f+4) d^{cab} d^{8be} \{J^2,\{G^{ra},G^{re}\}\} + 3 (N_f+4) \{\{J^r,G^{rc}\},\{J^m,G^{m8}\}\} \nonumber \\
&  & \mbox{} + \frac38 [3N_c(N_c+N_f)(N_c+2N_f)(3N_f+5)+4 (N_c+N_f) (27N_f-13)] d^{c8e} \{J^2,\{J^2,T^e\}\} \nonumber \\
&  & \mbox{} - \frac{33}{2} (N_c+N_f) \{J^2,\{T^8,\{J^r,G^{rc}\}\}\} - \frac38 [3N_c(N_c+2N_f)(3N_f+5)+108N_f+112] \{J^2,\{J^2,\{T^c,T^8\}\}\} \nonumber \\
&  & \mbox{} + \frac12 [3N_c(N_c+2N_f)(3N_f+5)+162N_f+168] \{J^2,\{J^2,\{G^{rc},G^{r8}\}\}\} \nonumber \\
&  & \mbox{} - \frac34 [N_c(N_c+2N_f)(3N_f+5)+22N_f] d^{c8e} \{J^2,\{J^2,\{J^r,G^{re}\}\}\} \nonumber \\
&  & \mbox{} + 3 [N_c(N_c+2N_f)(3N_f+5)+36N_f+10] f^{cab} f^{8be} \{J^2,\{J^2,\{G^{ra},G^{re}\}\}\} \nonumber \\
&  & \mbox{} + \frac12 (9N_f+48) \{J^2,\{\{J^r,G^{rc}\},\{J^m,G^{m8}\}\}\} + \frac34 (N_c+N_f)(3N_f+5) d^{c8e} \{J^2,\{J^2,\{J^2,T^e\}\}\} \nonumber \\
&  & \mbox{} - \frac{21}{4} (N_c+N_f) \{J^2,\{J^2,\{T^8,\{J^r,G^{rc}\}\}\}\} - \frac14 (9N_f+48) \{J^2,\{J^2,\{J^2,\{T^c,T^8\}\}\}\} \nonumber \\
&  & \mbox{} + \frac12 (9N_f+48) \{J^2,\{J^2,\{J^2,\{G^{rc},G^{r8}\}\}\}\} - \frac34 N_f d^{c8e} \{J^2,\{J^2,\{J^2,\{J^r,G^{re}\}\}\}\} \nonumber \\
&  & \mbox{} + (6N_f+21) f^{cab} f^{8be} \{J^2,\{J^2,\{J^2,\{G^{ra},G^{re}\}\}\}\} + \frac14 (3N_f+36) \{J^2,\{J^2,\{\{J^r,G^{rc}\},\{J^m,G^{m8}\}\}\}\},
\end{eqnarray}

\subsection{$[A^{j8},[[J^2,[J^2,A^{j8}]],\{J^r,G^{rc}\}]] - \frac12 [[J^2,A^{j8}], [[J^2,A^{j8}],\{J^r,G^{rc}\}]]$}

\begin{eqnarray}
&  & [G^{j8},[[J^2,[J^2,G^{j8}]],\{J^r,G^{rc}\}]] - \frac12 [[J^2,G^{j8}], [[J^2,G^{j8}],\{J^r,G^{rc}\}]] = - \frac{15}{8} f^{c8e} f^{8eg} \{J^r,G^{rg}\} + 3 d^{c8e} d^{8eg} \{J^r,G^{rg}\} \nonumber \\
&  & \mbox{} - \frac{6}{N_f} \delta^{c8} \{J^r,G^{r8}\} + \frac{6}{N_f} \delta^{88} \{J^r,G^{rc}\} - 12 d^{c8e} \{G^{re},G^{r8}\} + 6 d^{88e} \{G^{rc},G^{re}\} + \frac{6}{N_f} d^{c88} J^2 \nonumber \\
&  & \mbox{} + \frac32 d^{c8e} d^{8eg} \{J^2,\{J^r,G^{rg}\}\} - \frac{3}{N_f} \delta^{c8} \{J^2,\{J^r,G^{r8}\}\} + \frac{3}{N_f} \delta^{88} \{J^2,\{J^r,G^{rc}\}\} - \frac{15}{2} d^{c8e} \{J^2,\{G^{re},G^{r8}\}\} \nonumber \\
&  & \mbox{} + \frac32 d^{88e} \{J^2,\{G^{rc},G^{re}\}\} + \frac{3}{N_f} d^{c88} \{J^2,J^2\} - \frac92 \{\{J^r,G^{rc}\},\{G^{m8},G^{m8}\}\} + \frac92 \{\{G^{rc},G^{r8}\},\{J^m,G^{m8}\}\} \nonumber \\
&  & \mbox{} + \frac34 d^{c8e} \{\{J^r,G^{re}\},\{J^m,G^{m8}\}\} + \frac34 d^{88e} \{\{J^r,G^{rc}\},\{J^m,G^{me}\}\},
\end{eqnarray}

\begin{eqnarray}
&  & [G^{j8},[[J^2,[J^2,\mathcal{O}_3^{j8}]],\{J^r,G^{rc}\}]] + [\mathcal{O}_3^{j8},[[J^2,[J^2,G^{j8}]],\{J^r,G^{rc}\}]] - \frac12 [[J^2,G^{j8}],[[J^2,\mathcal{O}_3^{j8}],\{J^r,G^{rc}\}]] \nonumber \\
&  & \mbox{} - \frac12 [[J^2,\mathcal{O}_3^{j8}], [[J^2,G^{j8}],\{J^r,G^{rc}\}]] = - \frac{15}{2} f^{c8e} f^{8eg} \{J^r,G^{rg}\} + 12 d^{c8e} d^{8eg} \{J^r,G^{rg}\} - \frac{24}{N_f} \delta^{c8} \{J^r,G^{r8}\} \nonumber \\
&  & \mbox{} + \frac{24}{N_f} \delta^{88} \{J^r,G^{rc}\} - 48 d^{c8e} \{G^{re},G^{r8}\} + 24 d^{88e} \{G^{rc},G^{re}\} + \frac{24}{N_f} d^{c88} J^2 - \frac{21}{4} f^{c8e} f^{8eg} \{J^2,\{J^r,G^{rg}\}\} \nonumber \\
&  & \mbox{} + 18 d^{c8e} d^{8eg} \{J^2,\{J^r,G^{rg}\}\} - \frac{36}{N_f} \delta^{c8} \{J^2,\{J^r,G^{r8}\}\} + \frac{36}{N_f} \delta^{88} \{J^2,\{J^r,G^{rc}\}\} - 78 d^{c8e} \{J^2,\{G^{re},G^{r8}\}\} \nonumber \\
&  & \mbox{} + 30 d^{88e} \{J^2,\{G^{rc},G^{re}\}\} + \frac{36}{N_f} d^{c88} \{J^2,J^2\} - 18 \{\{J^r,G^{rc}\},\{G^{m8},G^{m8}\}\} + 18 \{\{G^{rc},G^{r8}\},\{J^m,G^{m8}\}\} \nonumber \\
&  & \mbox{} + 3 d^{c8e} \{\{J^r,G^{re}\},\{J^m,G^{m8}\}\} + 3 d^{88e} \{\{J^r,G^{rc}\},\{J^m,G^{me}\}\} + 3 d^{c8e} d^{8eg} \{J^2,\{J^2,\{J^r,G^{rg}\}\}\} \nonumber \\
&  & \mbox{} - \frac{6}{N_f} \delta^{c8} \{J^2,\{J^2,\{J^r,G^{r8}\}\}\} + \frac{6}{N_f} \delta^{88} \{J^2,\{J^2,\{J^r,G^{rc}\}\}\} - 21 d^{c8e}\{J^2,\{J^2,\{G^{re},G^{r8}\}\}\} \nonumber \\
&  & \mbox{} + 3 d^{88e} \{J^2,\{J^2,\{G^{rc},G^{re}\}\}\} + \frac{6}{N_f} d^{c88} \{J^2,\{J^2,J^2\}\} - 15 \{J^2,\{\{J^r,G^{rc}\},\{G^{m8},G^{m8}\}\}\} \nonumber \\
&  & \mbox{} + 15 \{J^2,\{\{G^{rc},G^{r8}\},\{J^m,G^{m8}\}\}\} + \frac92 d^{c8e} \{J^2,\{\{J^r,G^{re}\},\{J^m,G^{m8}\}\}\} \nonumber \\
&  & \mbox{} + \frac32 d^{88e} \{J^2,\{\{J^r,G^{rc}\},\{J^m,G^{me}\}\}\},
\end{eqnarray}

\begin{eqnarray}
&  & [\mathcal{O}_3^{j8},[[J^2,[J^2,\mathcal{O}_3^{j8}]],\{J^r,G^{rc}\}]] - \frac12 [[J^2,\mathcal{O}_3^{j8}], [[J^2,\mathcal{O}_3^{j8}],\{J^r,G^{rc}\}]] = - \frac{15}{2} f^{c8e} f^{8eg} \{J^r,G^{rg}\} + 12 d^{c8e} d^{8eg} \{J^r,G^{rg}\} \nonumber \\
&  & \mbox{} - \frac{24}{N_f} \delta^{c8} \{J^r,G^{r8}\} + \frac{24}{N_f} \delta^{88} \{J^r,G^{rc}\} - 48 d^{c8e} \{G^{re},G^{r8}\} + 24 d^{88e} \{G^{rc},G^{re}\} + \frac{24}{N_f} d^{c88} J^2 \nonumber \\
&  & \mbox{} - \frac{51}{4} f^{c8e} f^{8eg} \{J^2,\{J^r,G^{rg}\}\} + 30 d^{c8e} d^{8eg} \{J^2,\{J^r,G^{rg}\}\} - \frac{60}{N_f} \delta^{c8} \{J^2,\{J^r,G^{r8}\}\} + \frac{60}{N_f} \delta^{88} \{J^2,\{J^r,G^{rc}\}\} \nonumber \\
&  & \mbox{} - 126 d^{c8e} \{J^2,\{G^{re},G^{r8}\}\} + 54 d^{88e} \{J^2,\{G^{rc},G^{re}\}\} + \frac{60}{N_f} d^{c88} \{J^2,J^2\} - 18 \{\{J^r,G^{rc}\},\{G^{m8},G^{m8}\}\} \nonumber \\
&  & \mbox{} + 18 \{\{G^{rc},G^{r8}\},\{J^m,G^{m8}\}\} + 3 d^{c8e} \{\{J^r,G^{re}\},\{J^m,G^{m8}\}\} + 3 d^{88e} \{\{J^r,G^{rc}\},\{J^m,G^{me}\}\} \nonumber \\
&  & \mbox{} - \frac{27}{8} f^{c8e} f^{8eg} \{J^2,\{J^2,\{J^r,G^{rg}\}\}\} + 18 d^{c8e} d^{8eg} \{J^2,\{J^2,\{J^r,G^{rg}\}\}\} - \frac{36}{N_f} \delta^{c8} \{J^2,\{J^2,\{J^r,G^{r8}\}\}\} \nonumber \\
&  & \mbox{} + \frac{36}{N_f} \delta^{88} \{J^2,\{J^2,\{J^r,G^{rc}\}\}\} - 87 d^{c8e} \{J^2,\{J^2,\{G^{re},G^{r8}\}\}\} + 27 d^{88e} \{J^2,\{J^2,\{G^{rc},G^{re}\}\}\} \nonumber \\
&  & \mbox{} + \frac{36}{N_f} d^{c88} \{J^2,\{J^2,J^2\}\} - 33 \{J^2,\{\{J^r,G^{rc}\},\{G^{m8},G^{m8}\}\}\} + 33 \{J^2,\{\{G^{rc},G^{r8}\},\{J^m,G^{m8}\}\}\} \nonumber \\
&  & \mbox{} + \frac{15}{2} d^{c8e} \{J^2,\{\{J^r,G^{re}\},\{J^m,G^{m8}\}\}\} + \frac92 d^{88e} \{J^2,\{\{J^r,G^{rc}\},\{J^m,G^{me}\}\}\} \nonumber \\
&  & \mbox{} + \frac32 d^{c8e} d^{8eg} \{J^2,\{J^2,\{J^2,\{J^r,G^{rg}\}\}\}\} - \frac{3}{N_f} \delta^{c8} \{J^2,\{J^2,\{J^2,\{J^r,G^{r8}\}\}\}\} + \frac{3}{N_f} \delta^{88} \{J^2,\{J^2,\{J^2,\{J^r,G^{rc}\}\}\}\} \nonumber \\
&  & \mbox{} - \frac{27}{2} d^{c8e} \{J^2,\{J^2,\{J^2,\{G^{re},G^{r8}\}\}\}\} + \frac32 d^{88e} \{J^2,\{J^2,\{J^2,\{G^{rc},G^{re}\}\}\}\} + \frac{3}{N_f} d^{c88} \{J^2,\{J^2,\{J^2,J^2\}\}\} \nonumber \\
&  & \mbox{} - \frac{21}{2} \{J^2,\{J^2,\{\{J^r,G^{rc}\},\{G^{m8},G^{m8}\}\}\}\} + \frac{21}{2} \{J^2,\{J^2,\{\{G^{rc},G^{r8}\},\{J^m,G^{m8}\}\}\}\} \nonumber \\
&  & \mbox{} + \frac{15}{4} d^{c8e} \{J^2,\{J^2,\{\{J^r,G^{re}\},\{J^m,G^{m8}\}\}\}\} + \frac34 d^{88e} \{J^2,\{J^2,\{\{J^r,G^{rc}\},\{J^m,G^{me}\}\}\}\},
\end{eqnarray}

\section{\label{sec:opcoeff}Operator coefficients}

The final expression for $\delta \mathcal{A}_{\mathrm{2a}}^{(s)}$, Eq.~(\ref{eq:delta2a}), is given by the sum of three terms, namely, (\ref{eq:delta1}), (\ref{eq:delta8}), and (\ref{eq:delta27}), each of which is given by the sum of products of a coefficient times an operator from the operator bases (\ref{eq:bx}), (\ref{eq:by}), and (\ref{eq:bz}), respectively. All the pertinent coefficients, organized for the different flavor representations, read
\begin{equation}
a_{1}^{\mathbf{1}} = h_1 \left[ \frac98 a_1^2 + \frac12 a_1c_3 + \frac{1}{18} c_3^2 \right] + h_2 \left[ - \frac32 a_1^2 - \frac23 a_1c_3 - \frac{2}{27} c_3^2 \right], \nonumber
\end{equation}

\begin{equation}
a_{2}^{\mathbf{1}} = h_1 \left[ \frac12 a_1b_2 + \frac23 a_1b_3 - a_1c_3 - \frac19 c_3^2 \right] + h_2 \left[ \frac{25}{24} a_1^2 + \frac13 a_1b_2 + \frac{11}{18} a_1b_3 - \frac{5}{12} a_1c_3 - \frac{5}{108} c_3^2 \right], \nonumber
\end{equation}

\begin{equation}
a_{3}^{\mathbf{1}} = h_1 \left[ \frac{1}{18} a_1b_3 + \frac13 a_1c_3 + \frac{1}{12} b_2^2 + \frac{5}{54} c_3^2 \right] + h_2 \left[ \frac{1}{36} a_1b_2 - \frac19 a_1b_3 - \frac{7}{18} a_1c_3 - \frac{19}{162} c_3^2 \right], \nonumber
\end{equation}

\begin{equation}
a_{4}^{\mathbf{1}} = h_1 \left[ \frac19 b_2b_3 + \frac{2}{27} b_3^2 - \frac{5}{54} c_3^2 \right] + h_2 \left[ \frac{1}{18} a_1b_3 + \frac{5}{18} a_1c_3 + \frac{1}{36} b_2^2 + \frac{2}{27} b_2b_3 + \frac{11}{162} b_3^2 + \frac{1}{648} c_3^2 \right], \nonumber
\end{equation}

\begin{equation}
a_{5}^{\mathbf{1}} = h_1 \left[ \frac{1}{162} b_3^2 + \frac{2}{81} c_3^2 \right] + h_2 \left[ \frac{1}{162} b_2b_3 - \frac{1}{81} b_3^2 - \frac{1}{36} c_3^2 \right], \nonumber
\end{equation}

\begin{equation}
a_{6}^{\mathbf{1}} = h_2 \left[ \frac{1}{162} b_3^2 + \frac{19}{972} c_3^2 \right], \nonumber
\end{equation}

\begin{equation}
a_{7}^{\mathbf{1}} = 0, \nonumber
\end{equation}

\begin{equation}
a_{8}^{\mathbf{1}} = 0, \nonumber
\end{equation}

\begin{equation}
b_{1}^{\mathbf{1}} = h_1 \left[ \frac34 a_1^2 + \frac13 a_1c_3 + \frac{1}{27} c_3^2 \right] + h_2 \left[ - a_1^2 - \frac{4}{9} a_1c_3 - \frac{4}{81} c_3^2 \right], \nonumber
\end{equation}

\begin{equation}
b_{2}^{\mathbf{1}} = h_1 \left[ - a_1^2 - \frac23 a_1c_3 - \frac{2}{27} c_3^2 \right] + h_2 \left[ - \frac16 a_1^2 - \frac{5}{18} a_1c_3 - \frac{5}{162} c_3^2 \right], \nonumber
\end{equation}

\begin{equation}
b_{3}^{\mathbf{1}} = h_1 \left[ \frac16 a_1^2 + \frac{7}{18} a_1c_3 + \frac{13}{162} c_3^2 \right] + h_2 \left[ - \frac16 a_1^2 - \frac{13}{27} a_1c_3 - \frac{25}{243} c_3^2 \right], \nonumber
\end{equation}

\begin{equation}
b_{4}^{\mathbf{1}} = h_1 \left[ - \frac29 a_1c_3 - \frac{8}{81} c_3^2 \right] + h_2 \left[ \frac19 a_1^2 + \frac{4}{27} a_1c_3 - \frac{7}{486} c_3^2 \right], \nonumber
\end{equation}

\begin{equation}
b_{5}^{\mathbf{1}} = h_1 \left[ \frac{1}{27} a_1c_3 + \frac{37}{972} c_3^2 \right] + h_2 \left[ - \frac{1}{27} a_1c_3 - \frac{11}{243} c_3^2 \right], \nonumber
\end{equation}

\begin{equation}
b_{6}^{\mathbf{1}} = h_1 \left[ - \frac{1}{81} c_3^2 \right] + h_2 \left[ \frac{2}{81} a_1c_3 + \frac{31}{1458} c_3^2 \right], \nonumber
\end{equation}

\begin{equation}
b_{7}^{\mathbf{1}} = h_1 \left[ \frac{1}{486} c_3^2 \right] + h_2 \left[ - \frac{1}{486} c_3^2 \right], \nonumber
\end{equation}

\begin{equation}
b_{8}^{\mathbf{1}} = h_2 \left[ \frac{1}{729} c_3^2 \right], \nonumber
\end{equation}

\begin{equation}
c_{1}^{\mathbf{1}} = h_1 \left[ \frac14 a_1^2 + \frac19 a_1c_3 + \frac{1}{81} c_3^2 \right] + h_2 \left[ - \frac13 a_1^2 - \frac{4}{27} a_1c_3 - \frac{4}{243} c_3^2 \right], \nonumber
\end{equation}

\begin{equation}
c_{2}^{\mathbf{1}} = h_1 \left[ - \frac12 a_1^2 - \frac29 a_1c_3 - \frac{2}{81} c_3^2 \right] + h_2 \left[ - \frac{5}{24} a_1^2 - \frac{5}{54} a_1c_3 - \frac{5}{486} c_3^2 \right], \nonumber
\end{equation}

\begin{equation}
c_{3}^{\mathbf{1}} = h_1 \left[ \frac16 a_1^2 + \frac{5}{27} a_1c_3 + \frac{8}{243} c_3^2 \right] + h_2 \left[ - \frac{7}{36} a_1^2 - \frac{19}{81} a_1c_3 - \frac{31}{729} c_3^2 \right], \nonumber
\end{equation}

\begin{equation}
c_{4}^{\mathbf{1}} = h_1 \left[ - \frac{5}{27} a_1c_3 - \frac{11}{243} c_3^2 \right] + h_2 \left[ \frac{5}{36} a_1^2 + \frac{1}{324} a_1c_3 - \frac{29}{2916} c_3^2 \right], \nonumber
\end{equation}

\begin{equation}
c_{5}^{\mathbf{1}} = h_1 \left[ \frac{4}{81} a_1c_3 + \frac{67}{2916} c_3^2 \right] + h_2 \left[ - \frac{1}{18} a_1c_3 - \frac{41}{1458} c_3^2 \right], \nonumber
\end{equation}

\begin{equation}
c_{6}^{\mathbf{1}} = h_1 \left[ - \frac{7}{486} c_3^2 \right] + h_2 \left[ \frac{19}{486} a_1c_3 + \frac{127}{17496} c_3^2 \right], \nonumber
\end{equation}

\begin{equation}
c_{7}^{\mathbf{1}} = h_1 \left[ \frac{5}{1458} c_3^2 \right] + h_2 \left[ - \frac{11}{2916} c_3^2 \right], \nonumber
\end{equation}

\begin{equation}
c_{8}^{\mathbf{1}} = h_2 \left[ \frac{23}{8748} c_3^2 \right], \nonumber
\end{equation}

\begin{equation}
a_{1}^{\mathbf{8}} = h_2 \left[ - \frac98 a_1^2 - \frac12 a_1c_3 - \frac{1}{18} c_3^2 \right], \nonumber
\end{equation}

\begin{equation}
a_{2}^{\mathbf{8}} = h_1 \left[ \frac{9}{16} a_1^2 + \frac14 a_1c_3 + \frac{1}{36} c_3^2 \right] + h_2 \left[ - \frac34 a_1^2 - \frac13 a_1c_3 - \frac{1}{27} c_3^2 \right], \nonumber
\end{equation}

\begin{equation}
a_{3}^{\mathbf{8}} = h_2 \left[ - \frac18 a_1^2 - \frac{1}{18} a_1c_3 - \frac{1}{162} c_3^2 \right], \nonumber
\end{equation}

\begin{equation}
a_{4}^{\mathbf{8}} = h_2 \left[ \frac23 a_1^2 + \frac{8}{27} a_1c_3 + \frac{8}{243} c_3^2 \right], \nonumber
\end{equation}

\begin{equation}
a_{5}^{\mathbf{8}} = h_1 \left[ \frac14 a_1b_2 + \frac13 a_1b_3 - \frac12 a_1c_3 - \frac{1}{18} c_3^2 \right] + h_2 \left[ \frac{13}{48} a_1^2 + \frac16 a_1b_2 + \frac{1}{18} a_1b_3 + \frac{1}{54} a_1c_3 + \frac{1}{486} c_3^2 \right], \nonumber
\end{equation}

\begin{equation}
a_{6}^{\mathbf{8}} = h_2 \left[ \frac{2}{9} a_1^2 - \frac16 a_1b_3 + \frac{769}{324} a_1c_3 + \frac{3}{2} b_3^2 + \frac{15187}{2916} c_3^2 \right], \nonumber
\end{equation}

\begin{equation}
a_{7}^{\mathbf{8}} = h_1 \left[ \frac{1}{36} a_1b_3 + \frac16 a_1c_3 + \frac{1}{24} b_2^2 + \frac{5}{108} c_3^2 \right] + h_2 \left[ \frac{1}{72} a_1b_2 - \frac{1}{18} a_1b_3 + \frac34 a_1c_3 + \frac12 b_3^2 + \frac{187}{108} c_3^2 \right], \nonumber
\end{equation}

\begin{equation}
a_{8}^{\mathbf{8}} = h_1 \left[ - \frac{1}{18} a_1b_3 + \frac{1}{12} a_1c_3 + \frac{1}{108} c_3^2 \right] + h_2 \left[ - \frac{1}{36} a_1b_2 \right], \nonumber
\end{equation}

\begin{equation}
a_{9}^{\mathbf{8}} = h_1 \left[ \frac{1}{18} a_1b_3 - \frac{1}{12} a_1c_3 - \frac{1}{108} c_3^2 \right] + h_2 \left[ \frac{1}{36} a_1b_2 + \frac19 a_1b_3 - \frac16 a_1c_3 - \frac{1}{54} c_3^2 \right], \nonumber
\end{equation}

\begin{equation}
a_{10}^{\mathbf{8}} = h_2 \left[ \frac{7}{216} a_1b_3 - \frac{103}{432} a_1c_3 - \frac{1}{12} b_3^2 - \frac{1219}{3888} c_3^2 \right], \nonumber
\end{equation}

\begin{equation}
a_{11}^{\mathbf{8}} = h_2 \left[ - \frac{1}{54} a_1b_3 + \frac{151}{324} a_1c_3 + \frac19 b_3^2 + \frac{1339}{2916} c_3^2 \right], \nonumber
\end{equation}

\begin{equation}
a_{12}^{\mathbf{8}} = h_1 \left[ \frac{1}{18} b_2b_3 + \frac{1}{27} b_3^2 - \frac{5}{108} c_3^2 \right] + h_2 \left[ \frac{1}{108} a_1b_3 - \frac{1}{162} a_1c_3 + \frac{1}{72} b_2^2 + \frac{1}{27} b_2b_3 - \frac{4}{81} b_3^2 - \frac{1075}{5832} c_3^2 \right], \nonumber
\end{equation}

\begin{equation}
a_{13}^{\mathbf{8}} = h_2 \left[ - \frac{1}{18} a_1b_3 + \frac{157}{324} a_1c_3 + \frac{2}{9} b_3^2 + \frac{2341}{2916} c_3^2 \right], \nonumber
\end{equation}

\begin{equation}
a_{14}^{\mathbf{8}} = h_2 \left[ - \frac{1}{18} a_1b_3 + \frac{7}{108} a_1c_3 + \frac{7}{972} c_3^2 \right], \nonumber
\end{equation}

\begin{equation}
a_{15}^{\mathbf{8}} = h_2 \left[ - \frac{1}{18} a_1b_3 + \frac{7}{108} a_1c_3 + \frac{7}{972} c_3^2 \right], \nonumber
\end{equation}

\begin{equation}
a_{16}^{\mathbf{8}} = h_1 \left[ \frac{1}{324} b_3^2 + \frac{1}{81} c_3^2 \right] + h_2 \left[ \frac{1}{324} b_2b_3 + \frac{1}{81} b_3^2 + \frac{1}{27} c_3^2 \right], \nonumber
\end{equation}

\begin{equation}
a_{17}^{\mathbf{8}} = h_1 \left[ - \frac{1}{162} b_3^2 + \frac{5}{648} c_3^2 \right] + h_2 \left[ - \frac{1}{162} b_2b_3 \right], \nonumber
\end{equation}

\begin{equation}
a_{18}^{\mathbf{8}} = h_1 \left[ \frac{1}{162} b_3^2 - \frac{5}{648} c_3^2 \right] + h_2 \left[ \frac{1}{162} b_2b_3 + \frac{1}{81} b_3^2 - \frac{5}{324} c_3^2 \right], \nonumber
\end{equation}

\begin{equation}
a_{19}^{\mathbf{8}} = h_2 \left[ - \frac{1}{243} b_3^2 - \frac{7}{648} c_3^2 \right], \nonumber
\end{equation}

\begin{equation}
a_{20}^{\mathbf{8}} = h_2 \left[ \frac{1}{81} b_3^2 + \frac{7}{324} c_3^2 \right], \nonumber
\end{equation}

\begin{equation}
a_{21}^{\mathbf{8}} = h_2 \left[ \frac{1}{324} b_3^2 - \frac{1}{648} c_3^2 \right], \nonumber
\end{equation}

\begin{equation}
a_{22}^{\mathbf{8}} = h_2 \left[ \frac{2}{243} b_3^2 + \frac{11}{486} c_3^2 \right], \nonumber
\end{equation}

\begin{equation}
a_{23}^{\mathbf{8}} = h_2 \left[ - \frac{1}{162} b_3^2 + \frac{11}{1944} c_3^2 \right], \nonumber
\end{equation}

\begin{equation}
a_{24}^{\mathbf{8}} = 0, \nonumber
\end{equation}

\begin{equation}
a_{25}^{\mathbf{8}} = 0, \nonumber
\end{equation}

\begin{equation}
a_{26}^{\mathbf{8}} = 0, \nonumber
\end{equation}

\begin{equation}
a_{27}^{\mathbf{8}} = 0, \nonumber
\end{equation}

\begin{equation}
a_{28}^{\mathbf{8}} = 0, \nonumber
\end{equation}

\begin{equation}
a_{29}^{\mathbf{8}} = 0, \nonumber
\end{equation}

\begin{equation}
a_{30}^{\mathbf{8}} = 0, \nonumber
\end{equation}

\begin{equation}
a_{31}^{\mathbf{8}} = 0, \nonumber
\end{equation}

\begin{equation}
b_{1}^{\mathbf{8}} = h_2 \left[ - \frac34 a_1^2 - \frac13 a_1c_3 - \frac{1}{27} c_3^2 \right], \nonumber
\end{equation}

\begin{equation}
b_{2}^{\mathbf{8}} = h_1 \left[ \frac38 a_1^2 + \frac16 a_1c_3 + \frac{1}{54} c_3^2 \right] + h_2 \left[ - \frac12 a_1^2 - \frac29 a_1c_3 - \frac{2}{81} c_3^2 \right], \nonumber
\end{equation}

\begin{equation}
b_{3}^{\mathbf{8}} = h_2 \left[ - \frac{1}{12} a_1^2 - \frac{1}{27} a_1c_3 - \frac{1}{243} c_3^2 \right], \nonumber
\end{equation}

\begin{equation}
b_{4}^{\mathbf{8}} = h_2 \left[ \frac{4}{9} a_1^2 + \frac{16}{81} a_1c_3 + \frac{16}{729} c_3^2 \right], \nonumber
\end{equation}

\begin{equation}
b_{5}^{\mathbf{8}} = h_1 \left[ - \frac12 a_1^2 - \frac13 a_1c_3 - \frac{1}{27} c_3^2 \right] + h_2 \left[ \frac{5}{72} a_1^2 + \frac{1}{81} a_1c_3 + \frac{1}{729} c_3^2 \right], \nonumber
\end{equation}

\begin{equation}
b_{6}^{\mathbf{8}} = h_2 \left[ \frac{32}{27} a_1^2 + \frac{1802}{243} a_1c_3 + \frac{336715}{17496} c_3^2 \right], \nonumber
\end{equation}

\begin{equation}
b_{7}^{\mathbf{8}} = h_1 \left[ \frac{1}{12} a_1^2 + \frac{7}{36} a_1c_3 + \frac{13}{324} c_3^2 \right] + h_2 \left[ \frac13 a_1^2 + \frac{22}{9} a_1c_3 + \frac{1385}{216} c_3^2 \right], \nonumber
\end{equation}

\begin{equation}
b_{8}^{\mathbf{8}} = h_1 \left[ \frac{1}{12} a_1^2 + \frac{1}{18} a_1c_3 + \frac{1}{162} c_3^2 \right], \nonumber
\end{equation}

\begin{equation}
b_{9}^{\mathbf{8}} = h_1 \left[ - \frac{1}{12} a_1^2 - \frac{1}{18} a_1c_3 - \frac{1}{162} c_3^2 \right] + h_2 \left[ - \frac16 a_1^2 - \frac19 a_1c_3 - \frac{1}{81} c_3^2 \right], \nonumber
\end{equation}

\begin{equation}
b_{10}^{\mathbf{8}} = h_2 \left[ - \frac18 a_1^2 - \frac{337}{648} a_1c_3 - \frac{12707}{11664} c_3^2 \right], \nonumber
\end{equation}

\begin{equation}
b_{11}^{\mathbf{8}} = h_2 \left[ \frac{13}{54} a_1^2 + \frac{421}{486} a_1c_3 + \frac{13019}{8748} c_3^2 \right], \nonumber
\end{equation}

\begin{equation}
b_{12}^{\mathbf{8}} = h_1 \left[ - \frac{1}{9} a_1c_3 - \frac{4}{81} c_3^2 \right] + h_2 \left[ \frac{1}{54} a_1^2 - \frac{211}{972} a_1c_3 - \frac{12383}{17496} c_3^2 \right], \nonumber
\end{equation}

\begin{equation}
b_{13}^{\mathbf{8}} = h_2 \left[ \frac{13}{54} a_1^2 + \frac{601}{486} a_1c_3 + \frac{6293}{2187} c_3^2 \right], \nonumber
\end{equation}

\begin{equation}
b_{14}^{\mathbf{8}} = h_2 \left[ \frac{1}{18} a_1^2 + \frac{7 }{162} a_1c_3 + \frac{7}{1458} c_3^2 \right], \nonumber
\end{equation}

\begin{equation}
b_{15}^{\mathbf{8}} = h_2 \left[ \frac{1}{18} a_1^2 + \frac{7 }{162} a_1c_3 + \frac{7}{1458} c_3^2 \right], \nonumber
\end{equation}

\begin{equation}
b_{16}^{\mathbf{8}} = h_1 \left[ \frac{1}{54} a_1c_3 + \frac{37}{1944} c_3^2 \right] + h_2 \left[ \frac{1}{27} a_1c_3 + \frac{149}{972} c_3^2 \right], \nonumber
\end{equation}

\begin{equation}
b_{17}^{\mathbf{8}} = h_1 \left[ \frac{1}{54} a_1c_3 + \frac{2}{243} c_3^2 \right], \nonumber
\end{equation}

\begin{equation}
b_{18}^{\mathbf{8}} = h_1 \left[ - \frac{1}{54} a_1c_3 - \frac{2}{243} c_3^2 \right] + h_2 \left[ - \frac{1}{27} a_1c_3 - \frac{4}{243} c_3^2 \right], \nonumber
\end{equation}

\begin{equation}
b_{19}^{\mathbf{8}} = h_2 \left[ - \frac{1}{81} a_1c_3 - \frac{193}{5832} c_3^2 \right], \nonumber
\end{equation}

\begin{equation}
b_{20}^{\mathbf{8}} = h_2 \left[ \frac{2}{81} a_1c_3 + \frac{83}{1458} c_3^2 \right], \nonumber
\end{equation}

\begin{equation}
b_{21}^{\mathbf{8}} = h_1 \left[ - \frac{1}{162} c_3^2 \right] + h_2 \left[ - \frac{1}{72} c_3^2 \right], \nonumber
\end{equation}

\begin{equation}
b_{22}^{\mathbf{8}} = h_2 \left[ \frac{2}{81} a_1c_3 + \frac{19}{243} c_3^2 \right], \nonumber
\end{equation}

\begin{equation}
b_{23}^{\mathbf{8}} = h_2 \left[ \frac{1}{81} a_1c_3 + \frac{1}{162} c_3^2 \right], \nonumber
\end{equation}

\begin{equation}
b_{24}^{\mathbf{8}} = h_1 \left[ \frac{1}{972} c_3^2 \right] + h_2 \left[ \frac{1}{486} c_3^2 \right], \nonumber
\end{equation}

\begin{equation}
b_{25}^{\mathbf{8}} = h_1 \left[ \frac{1}{972} c_3^2 \right], \nonumber
\end{equation}

\begin{equation}
b_{26}^{\mathbf{8}} = h_1 \left[ - \frac{1}{972} c_3^2 \right] + h_2 \left[ - \frac{1}{486} c_3^2 \right], \nonumber
\end{equation}

\begin{equation}
b_{27}^{\mathbf{8}} = h_2 \left[ - \frac{1}{1458} c_3^2 \right], \nonumber
\end{equation}

\begin{equation}
b_{28}^{\mathbf{8}} = h_2 \left[ \frac{1}{729} c_3^2 \right], \nonumber
\end{equation}

\begin{equation}
b_{29}^{\mathbf{8}} = 0, \nonumber
\end{equation}

\begin{equation}
b_{30}^{\mathbf{8}} = h_2 \left[ \frac{1}{729} c_3^2 \right], \nonumber
\end{equation}

\begin{equation}
b_{31}^{\mathbf{8}} = h_2 \left[ \frac{1}{1458} c_3^2 \right], \nonumber
\end{equation}

\begin{equation}
c_{1}^{\mathbf{8}} = h_2 \left[ - \frac14 a_1^2 - \frac19 a_1c_3 - \frac{1}{81} c_3^2 \right], \nonumber
\end{equation}

\begin{equation}
c_{2}^{\mathbf{8}} = h_1 \left[ \frac18 a_1^2 + \frac{1}{18} a_1c_3 + \frac{1}{162} c_3^2 \right] + h_2 \left[ - \frac16 a_1^2 - \frac{2}{27} a_1c_3 - \frac{2}{243} c_3^2 \right], \nonumber
\end{equation}

\begin{equation}
c_{3}^{\mathbf{8}} = h_2 \left[ - \frac{1}{36} a_1^2 - \frac{1}{81} a_1c_3 - \frac{1}{729} c_3^2 \right], \nonumber
\end{equation}

\begin{equation}
c_{4}^{\mathbf{8}} = h_2 \left[ \frac{4}{27} a_1^2 + \frac{16}{243} a_1c_3 + \frac{16}{2187} c_3^2 \right], \nonumber
\end{equation}

\begin{equation}
c_{5}^{\mathbf{8}} = h_1 \left[ - \frac14 a_1^2 - \frac19 a_1c_3 - \frac{1}{81} c_3^2 \right] + h_2 \left[ \frac{1}{108} a_1^2 + \frac{1}{243} a_1c_3 + \frac{1}{2187} c_3^2 \right], \nonumber
\end{equation}

\begin{equation}
c_{6}^{\mathbf{8}} = h_2 \left[ \frac{769}{648} a_1^2 + \frac{15187}{1458} a_1c_3 + \frac{1550905}{52488} c_3^2 \right], \nonumber
\end{equation}

\begin{equation}
c_{7}^{\mathbf{8}} = h_1 \left[ \frac{1}{12} a_1^2 + \frac{5}{54} a_1c_3 + \frac{4}{243} c_3^2 \right] + h_2 \left[ \frac38 a_1^2 + \frac{187}{54} a_1c_3 + \frac{19145}{1944} c_3^2 \right], \nonumber
\end{equation}

\begin{equation}
c_{8}^{\mathbf{8}} = h_1 \left[ \frac{1}{24} a_1^2 + \frac{1}{54} a_1c_3 + \frac{1}{486} c_3^2 \right], \nonumber
\end{equation}

\begin{equation}
c_{9}^{\mathbf{8}} = h_1 \left[ - \frac{1}{24} a_1^2 - \frac{1}{54} a_1c_3 - \frac{1}{486} c_3^2 \right] + h_2 \left[ - \frac{1}{12} a_1^2 - \frac{1}{27} a_1c_3 - \frac{1}{243} c_3^2 \right], \nonumber
\end{equation}

\begin{equation}
c_{10}^{\mathbf{8}} = h_2 \left[ - \frac{103}{864} a_1^2 - \frac{1219}{1944} a_1c_3 - \frac{57725}{34992} c_3^2 \right], \nonumber
\end{equation}

\begin{equation}
c_{11}^{\mathbf{8}} = h_2 \left[ \frac{151}{648} a_1^2 + \frac{1339}{1458} a_1c_3 + \frac{58109}{26244} c_3^2 \right], \nonumber
\end{equation}

\begin{equation}
c_{12}^{\mathbf{8}} = h_1 \left[ - \frac{5}{54} a_1c_3 - \frac{11}{486} c_3^2 \right] + h_2 \left[ - \frac{1}{324} a_1^2 - \frac{1075}{2916} a_1c_3 - \frac{57365}{52488} c_3^2 \right], \nonumber
\end{equation}

\begin{equation}
c_{13}^{\mathbf{8}} = h_2 \left[ \frac{157}{648} a_1^2 + \frac{2341}{1458} a_1c_3 + \frac{28790}{6561} c_3^2 \right], \nonumber
\end{equation}

\begin{equation}
c_{14}^{\mathbf{8}} = h_2 \left[ \frac{7}{216} a_1^2 + \frac{7}{486} a_1c_3 + \frac{7}{4374} c_3^2 \right], \nonumber
\end{equation}

\begin{equation}
c_{15}^{\mathbf{8}} = h_2 \left[ \frac{7}{216} a_1^2 + \frac{7}{486} a_1c_3 + \frac{7}{4374} c_3^2 \right], \nonumber
\end{equation}

\begin{equation}
c_{16}^{\mathbf{8}} = h_1 \left[ \frac{2}{81} a_1c_3 + \frac{67}{5832} c_3^2 \right] + h_2 \left[ \frac{2}{27} a_1c_3 + \frac{703}{2916} c_3^2 \right], \nonumber
\end{equation}

\begin{equation}
c_{17}^{\mathbf{8}} = h_1 \left[ \frac{5}{324} a_1c_3 + \frac{11}{2916} c_3^2 \right], \nonumber
\end{equation}

\begin{equation}
c_{18}^{\mathbf{8}} = h_1 \left[ - \frac{5}{324} a_1c_3 - \frac{11}{2916} c_3^2 \right] + h_2 \left[ - \frac{5}{162} a_1c_3 - \frac{11}{1458} c_3^2 \right], \nonumber
\end{equation}

\begin{equation}
c_{19}^{\mathbf{8}} = h_2 \left[ - \frac{7}{324} a_1c_3 - \frac{785}{17496} c_3^2 \right], \nonumber
\end{equation}

\begin{equation}
c_{20}^{\mathbf{8}} = h_2 \left[ \frac{7 }{162} a_1c_3 + \frac{149}{2187} c_3^2 \right], \nonumber
\end{equation}

\begin{equation}
c_{21}^{\mathbf{8}} = h_1 \left[ - \frac{7}{972} c_3^2 \right] + h_2 \left[ - \frac{1}{324} a_1c_3 - \frac{37}{1458} c_3^2 \right], \nonumber
\end{equation}

\begin{equation}
c_{22}^{\mathbf{8}} = h_2 \left[ \frac{11}{243} a_1c_3 + \frac{248}{2187} c_3^2 \right], \nonumber
\end{equation}

\begin{equation}
c_{23}^{\mathbf{8}} = h_2 \left[ \frac{11}{972} a_1c_3 + \frac{25}{8748} c_3^2 \right], \nonumber
\end{equation}

\begin{equation}
c_{24}^{\mathbf{8}} = h_1 \left[ \frac{5}{2916} c_3^2 \right] + h_2 \left[ \frac{7}{1458} c_3^2 \right], \nonumber
\end{equation}

\begin{equation}
c_{25}^{\mathbf{8}} = h_1 \left[ \frac{7}{5832} c_3^2 \right], \nonumber
\end{equation}

\begin{equation}
c_{26}^{\mathbf{8}} = h_1 \left[ - \frac{7}{5832} c_3^2 \right] + h_2 \left[ - \frac{7}{2916} c_3^2 \right], \nonumber
\end{equation}

\begin{equation}
c_{27}^{\mathbf{8}} = h_2 \left[ - \frac{25}{17496} c_3^2 \right], \nonumber
\end{equation}

\begin{equation}
c_{28}^{\mathbf{8}} = h_2 \left[ \frac{25}{8748} c_3^2 \right], \nonumber
\end{equation}

\begin{equation}
c_{29}^{\mathbf{8}} = h_2 \left[ - \frac{1}{5832} c_3^2 \right], \nonumber
\end{equation}

\begin{equation}
c_{30}^{\mathbf{8}} = h_2 \left[ \frac{13}{4374} c_3^2 \right], \nonumber
\end{equation}

\begin{equation}
c_{31}^{\mathbf{8}} = h_2 \left[ \frac{5}{5832} c_3^2 \right], \nonumber
\end{equation}

\begin{equation}
a_{1}^{\mathbf{27}} = h_1 \left[ \frac38 a_1^2 + \frac{1}{24} a_1b_3 + \frac{5}{48} a_1c_3 + \frac{5}{432} c_3^2 \right] + h_2 \left[ \frac{1}{48} a_1b_2 \right], \nonumber
\end{equation}

\begin{equation}
a_{2}^{\mathbf{27}} = h_1 \left[ \frac16 a_1b_2 \right] + h_2 \left[ \frac{1}{24} a_1^2 + \frac{1}{36} a_1b_3 - \frac{5}{216} a_1c_3 - \frac{5}{1944} c_3^2 \right], \nonumber
\end{equation}

\begin{equation}
a_{3}^{\mathbf{27}} = h_2 \left[ \frac{1}{12} a_1^2 + \frac{1}{27} a_1c_3 + \frac{1}{243} c_3^2 \right], \nonumber
\end{equation}

\begin{equation}
a_{4}^{\mathbf{27}} = h_2 \left[ - \frac{1}{18} a_1^2 - \frac{2}{81} a_1c_3 - \frac{2}{729} c_3^2 \right], \nonumber
\end{equation}

\begin{equation}
a_{5}^{\mathbf{27}} = h_2 \left[ \frac{1}{18} a_1^2 + \frac{2}{81} a_1c_3 + \frac{2}{729} c_3^2 \right], \nonumber
\end{equation}

\begin{equation}
a_{6}^{\mathbf{27}} = h_2 \left[ - \frac13 a_1^2 - \frac{4}{27} a_1c_3 - \frac{4}{243} c_3^2 \right], \nonumber
\end{equation}

\begin{equation}
a_{7}^{\mathbf{27}} = h_2 \left[ \frac16 a_1^2 + \frac{2}{27} a_1c_3 + \frac{2}{243} c_3^2 \right], \nonumber
\end{equation}

\begin{equation}
a_{8}^{\mathbf{27}} = h_2 \left[ \frac{1}{18} a_1^2 + \frac{2}{81} a_1c_3 + \frac{2}{729} c_3^2 \right], \nonumber
\end{equation}

\begin{equation}
a_{9}^{\mathbf{27}} = h_1 \left[ \frac19 a_1b_3 - \frac16 a_1c_3 - \frac{1}{54} c_3^2 \right] + h_2 \left[ \frac{1}{18} a_1b_2 \right], \nonumber
\end{equation}

\begin{equation}
a_{10}^{\mathbf{27}} = h_1 \left[ \frac{1}{18} a_1b_3 + \frac{1}{18} a_1c_3 + \frac{1}{36} b_2^2 + \frac{1}{216} b_3^2 + \frac{49}{2592} c_3^2 \right] + h_2 \left[ \frac{1}{36} a_1b_2 + \frac{1}{216} b_2b_3 \right], \nonumber
\end{equation}

\begin{equation}
a_{11}^{\mathbf{27}} = h_1 \left[ \frac{1}{27} b_2b_3 \right] + h_2 \left[ \frac{1}{54} a_1b_3 + \frac{1}{108} b_2^2 + \frac{1}{324} b_3^2 - \frac{7}{3888} c_3^2 \right], \nonumber
\end{equation}

\begin{equation}
a_{12}^{\mathbf{27}} = h_2 \left[ \frac{1}{54} a_1c_3 + \frac{1}{162} c_3^2 \right], \nonumber
\end{equation}

\begin{equation}
a_{13}^{\mathbf{27}} = h_2 \left[ - \frac{1}{81} a_1c_3 - \frac{1}{243} c_3^2 \right], \nonumber
\end{equation}

\begin{equation}
a_{14}^{\mathbf{27}} = h_2 \left[ \frac{1}{81} a_1c_3 + \frac{1}{243} c_3^2 \right], \nonumber
\end{equation}

\begin{equation}
a_{15}^{\mathbf{27}} = h_2 \left[ - \frac{1}{27} a_1b_3 - \frac{5}{54} a_1c_3 - \frac{13}{486} c_3^2 \right], \nonumber
\end{equation}

\begin{equation}
a_{16}^{\mathbf{27}} = h_2 \left[ \frac{1}{27} a_1b_3 + \frac{1}{54} a_1c_3 + \frac{5}{486} c_3^2 \right], \nonumber
\end{equation}

\begin{equation}
a_{17}^{\mathbf{27}} = h_2 \left[ \frac{1}{81} a_1c_3 + \frac{1}{243} c_3^2 \right], \nonumber
\end{equation}

\begin{equation}
a_{18}^{\mathbf{27}} = h_2 \left[ \frac{1}{27} a_1b_3 - \frac{1}{18} a_1c_3 - \frac{1}{162} c_3^2 \right], \nonumber
\end{equation}

\begin{equation}
a_{19}^{\mathbf{27}} = h_2 \left[ - \frac{1}{27} a_1b_3 + \frac{1}{18} a_1c_3 + \frac{1}{162} c_3^2 \right], \nonumber
\end{equation}

\begin{equation}
a_{20}^{\mathbf{27}} = h_2 \left[ \frac{1}{54} a_1b_3 + \frac{1}{108} a_1c_3 + \frac{1}{972} c_3^2 \right], \nonumber
\end{equation}

\begin{equation}
a_{21}^{\mathbf{27}} = h_2 \left[ - \frac{1}{54} a_1b_3 + \frac{1}{108} a_1c_3 + \frac{1}{972} c_3^2 \right], \nonumber
\end{equation}

\begin{equation}
a_{22}^{\mathbf{27}} = h_1 \left[ \frac{1}{81} b_3^2 - \frac{5}{324} c_3^2 \right] + h_2 \left[ \frac{1}{81} b_2b_3 \right], \nonumber
\end{equation}

\begin{equation}
a_{23}^{\mathbf{27}} = h_1 \left[ \frac{1}{162} b_3^2 + \frac{1}{324} c_3^2 \right] + h_2 \left[ \frac{1}{162} b_2b_3 \right], \nonumber
\end{equation}

\begin{equation}
a_{24}^{\mathbf{27}} = h_2 \left[ \frac{1}{486} b_3^2 \right], \nonumber
\end{equation}

\begin{equation}
a_{25}^{\mathbf{27}} = h_2 \left[ \frac{1}{972} c_3^2 \right], \nonumber
\end{equation}

\begin{equation}
a_{26}^{\mathbf{27}} = h_2 \left[ - \frac{1}{1458} c_3^2 \right], \nonumber
\end{equation}

\begin{equation}
a_{27}^{\mathbf{27}} = h_2 \left[ \frac{1}{1458} c_3^2 \right], \nonumber
\end{equation}

\begin{equation}
a_{28}^{\mathbf{27}} = h_2 \left[ - \frac{1}{243} b_3^2 - \frac{7}{972} c_3^2 \right], \nonumber
\end{equation}

\begin{equation}
a_{29}^{\mathbf{27}} = h_2 \left[ \frac{1}{243} b_3^2 + \frac{1}{972} c_3^2 \right], \nonumber
\end{equation}

\begin{equation}
a_{30}^{\mathbf{27}} = h_2 \left[ \frac{1}{1458} c_3^2 \right], \nonumber
\end{equation}

\begin{equation}
a_{31}^{\mathbf{27}} = h_2 \left[ \frac{1}{243} b_3^2 - \frac{5}{972} c_3^2 \right], \nonumber
\end{equation}

\begin{equation}
a_{32}^{\mathbf{27}} = h_2 \left[ - \frac{1}{243} b_3^2 + \frac{5}{972} c_3^2 \right], \nonumber
\end{equation}

\begin{equation}
a_{33}^{\mathbf{27}} = h_2 \left[ \frac{1}{486} b_3^2 + \frac{1}{648} c_3^2 \right], \nonumber
\end{equation}

\begin{equation}
a_{34}^{\mathbf{27}} = h_2 \left[ - \frac{1}{486} b_3^2 + \frac{1}{1944} c_3^2 \right], \nonumber
\end{equation}

\begin{equation}
a_{35}^{\mathbf{27}} = 0, \nonumber
\end{equation}

\begin{equation}
a_{36}^{\mathbf{27}} = 0, \nonumber
\end{equation}

\begin{equation}
a_{37}^{\mathbf{27}} = 0, \nonumber
\end{equation}

\begin{equation}
a_{38}^{\mathbf{27}} = 0, \nonumber
\end{equation}

\begin{equation}
a_{39}^{\mathbf{27}} = 0, \nonumber
\end{equation}

\begin{equation}
a_{40}^{\mathbf{27}} = 0, \nonumber
\end{equation}

\begin{equation}
a_{41}^{\mathbf{27}} = 0, \nonumber
\end{equation}

\begin{equation}
a_{42}^{\mathbf{27}} = 0, \nonumber
\end{equation}

\begin{equation}
a_{43}^{\mathbf{27}} = 0, \nonumber
\end{equation}

\begin{equation}
a_{44}^{\mathbf{27}} = 0, \nonumber
\end{equation}

\begin{equation}
a_{45}^{\mathbf{27}} = 0, \nonumber
\end{equation}

\begin{equation}
a_{46}^{\mathbf{27}} = 0, \nonumber
\end{equation}

\begin{equation}
b_{1}^{\mathbf{27}} = h_1 \left[ \frac{3}{16} a_1^2 + \frac{5}{72} a_1c_3 + \frac{5}{648} c_3^2 \right], \nonumber
\end{equation}

\begin{equation}
b_{2}^{\mathbf{27}} = h_2 \left[ - \frac{1}{72} a_1^2 - \frac{5}{324} a_1c_3 - \frac{5}{2916} c_3^2 \right], \nonumber
\end{equation}

\begin{equation}
b_{3}^{\mathbf{27}} = h_2 \left[ \frac{1}{18} a_1^2 + \frac{2}{81} a_1c_3 + \frac{2}{729} c_3^2 \right], \nonumber
\end{equation}

\begin{equation}
b_{4}^{\mathbf{27}} = h_2 \left[ - \frac{1}{27} a_1^2 - \frac{4}{243} a_1c_3 - \frac{4}{2187} c_3^2 \right], \nonumber
\end{equation}

\begin{equation}
b_{5}^{\mathbf{27}} = h_2 \left[ \frac{1}{27} a_1^2 + \frac{4}{243} a_1c_3 + \frac{4}{2187} c_3^2 \right], \nonumber
\end{equation}

\begin{equation}
b_{6}^{\mathbf{27}} = h_2 \left[ - \frac{2}{9} a_1^2 - \frac{8}{81} a_1c_3 - \frac{8}{729} c_3^2 \right], \nonumber
\end{equation}

\begin{equation}
b_{7}^{\mathbf{27}} = h_2 \left[ \frac19 a_1^2 + \frac{4}{81} a_1c_3 + \frac{4}{729} c_3^2 \right], \nonumber
\end{equation}

\begin{equation}
b_{8}^{\mathbf{27}} = h_2 \left[ \frac{1}{27} a_1^2 + \frac{4}{243} a_1c_3 + \frac{4}{2187} c_3^2 \right], \nonumber
\end{equation}

\begin{equation}
b_{9}^{\mathbf{27}} = h_1 \left[ - \frac16 a_1^2 - \frac19 a_1c_3 - \frac{1}{81} c_3^2 \right], \nonumber
\end{equation}

\begin{equation}
b_{10}^{\mathbf{27}} = h_1 \left[ \frac{17}{216} a_1c_3 + \frac{4}{243} c_3^2 \right], \nonumber
\end{equation}

\begin{equation}
b_{11}^{\mathbf{27}} = h_2 \left[ - \frac{1}{324} a_1c_3 - \frac{1}{486} c_3^2 \right], \nonumber
\end{equation}

\begin{equation}
b_{12}^{\mathbf{27}} = h_2 \left[ \frac{2}{81} a_1c_3 + \frac{4}{729} c_3^2 \right], \nonumber
\end{equation}

\begin{equation}
b_{13}^{\mathbf{27}} = h_2 \left[ - \frac{4}{243} a_1c_3 - \frac{8}{2187} c_3^2 \right], \nonumber
\end{equation}

\begin{equation}
b_{14}^{\mathbf{27}} = h_2 \left[ \frac{4}{243} a_1c_3 + \frac{8}{2187} c_3^2 \right], \nonumber
\end{equation}

\begin{equation}
b_{15}^{\mathbf{27}} = h_2 \left[ - \frac{1}{18} a_1^2 - \frac19 a_1c_3 - \frac{17}{729} c_3^2 \right], \nonumber
\end{equation}

\begin{equation}
b_{16}^{\mathbf{27}} = h_2 \left[ \frac{1}{27} a_1c_3 + \frac{7}{729} c_3^2 \right], \nonumber
\end{equation}

\begin{equation}
b_{17}^{\mathbf{27}} = h_2 \left[ \frac{4}{243} a_1c_3 + \frac{8}{2187} c_3^2 \right], \nonumber
\end{equation}

\begin{equation}
b_{18}^{\mathbf{27}} = h_2 \left[ - \frac{1}{18} a_1^2 - \frac{1}{27} a_1c_3 - \frac{1}{243} c_3^2 \right], \nonumber
\end{equation}

\begin{equation}
b_{19}^{\mathbf{27}} = h_2 \left[ \frac{1}{18} a_1^2 + \frac{1}{27} a_1c_3 + \frac{1}{243} c_3^2 \right], \nonumber
\end{equation}

\begin{equation}
b_{20}^{\mathbf{27}} = h_2 \left[ \frac{1}{36} a_1^2 + \frac{1}{162} a_1c_3 + \frac{1}{1458} c_3^2 \right], \nonumber
\end{equation}

\begin{equation}
b_{21}^{\mathbf{27}} = h_2 \left[ \frac{1}{162} a_1c_3 + \frac{1}{1458} c_3^2 \right], \nonumber
\end{equation}

\begin{equation}
b_{22}^{\mathbf{27}} = h_1 \left[ - \frac{1}{27} a_1c_3 - \frac{4}{243} c_3^2 \right], \nonumber
\end{equation}

\begin{equation}
b_{23}^{\mathbf{27}} = h_1 \left[ \frac{25}{3888} c_3^2 \right], \nonumber
\end{equation}

\begin{equation}
b_{24}^{\mathbf{27}} = h_2 \left[ - \frac{1}{5832} c_3^2 \right], \nonumber
\end{equation}

\begin{equation}
b_{25}^{\mathbf{27}} = h_2 \left[ \frac{1}{486} c_3^2 \right], \nonumber
\end{equation}

\begin{equation}
b_{26}^{\mathbf{27}} = h_2 \left[ - \frac{1}{729} c_3^2 \right], \nonumber
\end{equation}

\begin{equation}
b_{27}^{\mathbf{27}} = h_2 \left[ \frac{1}{729} c_3^2 \right], \nonumber
\end{equation}

\begin{equation}
b_{28}^{\mathbf{27}} = h_2 \left[ - \frac{1}{81} a_1c_3 - \frac{8}{729} c_3^2 \right], \nonumber
\end{equation}

\begin{equation}
b_{29}^{\mathbf{27}} = h_2 \left[ \frac{2}{729} c_3^2 \right], \nonumber
\end{equation}

\begin{equation}
b_{30}^{\mathbf{27}} = h_2 \left[ \frac{1}{729} c_3^2 \right], \nonumber
\end{equation}

\begin{equation}
b_{31}^{\mathbf{27}} = h_2 \left[ - \frac{1}{81} a_1c_3 - \frac{4}{729} c_3^2 \right], \nonumber
\end{equation}

\begin{equation}
b_{32}^{\mathbf{27}} = h_2 \left[ \frac{1}{81} a_1c_3 + \frac{4}{729} c_3^2 \right], \nonumber
\end{equation}

\begin{equation}
b_{33}^{\mathbf{27}} = h_2 \left[ \frac{1}{162} a_1c_3 + \frac{1}{729} c_3^2 \right], \nonumber
\end{equation}

\begin{equation}
b_{34}^{\mathbf{27}} = h_2 \left[ \frac{1}{1458} c_3^2 \right], \nonumber
\end{equation}

\begin{equation}
b_{35}^{\mathbf{27}} = h_1 \left[ - \frac{1}{486} c_3^2 \right], \nonumber
\end{equation}

\begin{equation}
b_{36}^{\mathbf{27}} = 0, \nonumber
\end{equation}

\begin{equation}
b_{37}^{\mathbf{27}} = 0, \nonumber
\end{equation}

\begin{equation}
b_{38}^{\mathbf{27}} = 0, \nonumber
\end{equation}

\begin{equation}
b_{39}^{\mathbf{27}} = 0, \nonumber
\end{equation}

\begin{equation}
b_{40}^{\mathbf{27}} = h_2 \left[ - \frac{1}{1458} c_3^2 \right], \nonumber
\end{equation}

\begin{equation}
b_{41}^{\mathbf{27}} = 0, \nonumber
\end{equation}

\begin{equation}
b_{42}^{\mathbf{27}} = 0, \nonumber
\end{equation}

\begin{equation}
b_{43}^{\mathbf{27}} = h_2 \left[ - \frac{1}{1458} c_3^2 \right], \nonumber
\end{equation}

\begin{equation}
b_{44}^{\mathbf{27}} = h_2 \left[ \frac{1}{1458} c_3^2 \right], \nonumber
\end{equation}

\begin{equation}
b_{45}^{\mathbf{27}} = h_2 \left[ \frac{1}{2916} c_3^2 \right], \nonumber
\end{equation}

\begin{equation}
b_{46}^{\mathbf{27}} = 0, \nonumber
\end{equation}

\begin{equation}
c_{1}^{\mathbf{27}} = h_1 \left[ \frac{5}{96} a_1^2 + \frac{5}{216} a_1c_3 + \frac{5}{1944} c_3^2 \right], \nonumber
\end{equation}

\begin{equation}
c_{2}^{\mathbf{27}} = h_2 \left[ - \frac{5}{432} a_1^2 - \frac{5}{972} a_1c_3 - \frac{5}{8748} c_3^2 \right], \nonumber
\end{equation}

\begin{equation}
c_{3}^{\mathbf{27}} = h_2 \left[ \frac{1}{54} a_1^2 + \frac{2}{243} a_1c_3 + \frac{2}{2187} c_3^2 \right], \nonumber
\end{equation}

\begin{equation}
c_{4}^{\mathbf{27}} = h_2 \left[ - \frac{1}{81} a_1^2 - \frac{4}{729} a_1c_3 - \frac{4}{6561} c_3^2 \right], \nonumber
\end{equation}

\begin{equation}
c_{5}^{\mathbf{27}} = h_2 \left[ \frac{1}{81} a_1^2 + \frac{4}{729} a_1c_3 + \frac{4}{6561} c_3^2 \right], \nonumber
\end{equation}

\begin{equation}
c_{6}^{\mathbf{27}} = h_2 \left[ - \frac{2}{27} a_1^2 - \frac{8}{243} a_1c_3 - \frac{8}{2187} c_3^2 \right], \nonumber
\end{equation}

\begin{equation}
c_{7}^{\mathbf{27}} = h_2 \left[ \frac{1}{27} a_1^2 + \frac{4}{243} a_1c_3 + \frac{4}{2187} c_3^2 \right], \nonumber
\end{equation}

\begin{equation}
c_{8}^{\mathbf{27}} = h_2 \left[ \frac{1}{81} a_1^2 + \frac{4}{729} a_1c_3 + \frac{4}{6561} c_3^2 \right], \nonumber
\end{equation}

\begin{equation}
c_{9}^{\mathbf{27}} = h_1 \left[ - \frac{1}{12} a_1^2 - \frac{1}{27} a_1c_3 - \frac{1}{243} c_3^2 \right], \nonumber
\end{equation}

\begin{equation}
c_{10}^{\mathbf{27}} = h_1 \left[ \frac{1}{36} a_1^2 + \frac{49}{1296} a_1c_3 + \frac{79}{11664} c_3^2 \right], \nonumber
\end{equation}

\begin{equation}
c_{11}^{\mathbf{27}} = h_2 \left[ - \frac{7}{1944} a_1c_3 - \frac{17}{17496} c_3^2 \right], \nonumber
\end{equation}

\begin{equation}
c_{12}^{\mathbf{27}} = h_2 \left[ \frac{1}{108} a_1^2 + \frac{1}{81} a_1c_3 + \frac{5}{2187} c_3^2 \right], \nonumber
\end{equation}

\begin{equation}
c_{13}^{\mathbf{27}} = h_2 \left[ - \frac{1}{162} a_1^2 - \frac{2}{243} a_1c_3 - \frac{10}{6561} c_3^2 \right], \nonumber
\end{equation}

\begin{equation}
c_{14}^{\mathbf{27}} = h_2 \left[ \frac{1}{162} a_1^2 + \frac{2}{243} a_1c_3 + \frac{10}{6561} c_3^2 \right], \nonumber
\end{equation}

\begin{equation}
c_{15}^{\mathbf{27}} = h_2 \left[ - \frac{5}{108} a_1^2 - \frac{13}{243} a_1c_3 - \frac{7}{729} c_3^2 \right], \nonumber
\end{equation}

\begin{equation}
c_{16}^{\mathbf{27}} = h_2 \left[ \frac{1}{108} a_1^2 + \frac{5}{243} a_1c_3 + \frac{1}{243} c_3^2 \right], \nonumber
\end{equation}

\begin{equation}
c_{17}^{\mathbf{27}} = h_2 \left[ \frac{1}{162} a_1^2 + \frac{2}{243} a_1c_3 + \frac{10}{6561} c_3^2 \right], \nonumber
\end{equation}

\begin{equation}
c_{18}^{\mathbf{27}} = h_2 \left[ - \frac{1}{36} a_1^2 - \frac{1}{81} a_1c_3 - \frac{1}{729} c_3^2 \right], \nonumber
\end{equation}

\begin{equation}
c_{19}^{\mathbf{27}} = h_2 \left[ \frac{1}{36} a_1^2 + \frac{1}{81} a_1c_3 + \frac{1}{729} c_3^2 \right], \nonumber
\end{equation}

\begin{equation}
c_{20}^{\mathbf{27}} = h_2 \left[ \frac{1}{216} a_1^2 + \frac{1}{486} a_1c_3 + \frac{1}{4374} c_3^2 \right], \nonumber
\end{equation}

\begin{equation}
c_{21}^{\mathbf{27}} = h_2 \left[ \frac{1}{216} a_1^2 + \frac{1}{486} a_1c_3 + \frac{1}{4374} c_3^2 \right], \nonumber
\end{equation}

\begin{equation}
c_{22}^{\mathbf{27}} = h_1 \left[ - \frac{5}{162} a_1c_3 - \frac{11}{1458} c_3^2 \right], \nonumber
\end{equation}

\begin{equation}
c_{23}^{\mathbf{27}} = h_1 \left[ \frac{1}{162} a_1c_3 + \frac{11}{2592} c_3^2 \right], \nonumber
\end{equation}

\begin{equation}
c_{24}^{\mathbf{27}} = h_2 \left[ - \frac{1}{3888} c_3^2 \right], \nonumber
\end{equation}

\begin{equation}
c_{25}^{\mathbf{27}} = h_2 \left[ \frac{1}{486} a_1c_3 + \frac{1}{729} c_3^2 \right], \nonumber
\end{equation}

\begin{equation}
c_{26}^{\mathbf{27}} = h_2 \left[ - \frac{1}{729} a_1c_3 - \frac{2}{2187} c_3^2 \right], \nonumber
\end{equation}

\begin{equation}
c_{27}^{\mathbf{27}} = h_2 \left[ \frac{1}{729} a_1c_3 + \frac{2}{2187} c_3^2 \right], \nonumber
\end{equation}

\begin{equation}
c_{28}^{\mathbf{27}} = h_2 \left[ - \frac{7}{486} a_1c_3 - \frac{29}{4374} c_3^2 \right], \nonumber
\end{equation}

\begin{equation}
c_{29}^{\mathbf{27}} = h_2 \left[ \frac{1}{486} a_1c_3 + \frac{1}{486} c_3^2 \right], \nonumber
\end{equation}

\begin{equation}
c_{30}^{\mathbf{27}} = h_2 \left[ \frac{1}{729} a_1c_3 + \frac{2}{2187} c_3^2 \right], \nonumber
\end{equation}

\begin{equation}
c_{31}^{\mathbf{27}} = h_2 \left[ - \frac{5}{486} a_1c_3 - \frac{11}{4374} c_3^2 \right], \nonumber
\end{equation}

\begin{equation}
c_{32}^{\mathbf{27}} = h_2 \left[ \frac{5}{486} a_1c_3 + \frac{11}{4374} c_3^2 \right], \nonumber
\end{equation}

\begin{equation}
c_{33}^{\mathbf{27}} = h_2 \left[ \frac{1}{324} a_1c_3 + \frac{5}{8748} c_3^2 \right], \nonumber
\end{equation}

\begin{equation}
c_{34}^{\mathbf{27}} = h_2 \left[ \frac{1}{972} a_1c_3 + \frac{1}{2916} c_3^2 \right], \nonumber
\end{equation}

\begin{equation}
c_{35}^{\mathbf{27}} = h_1 \left[ - \frac{7}{2916} c_3^2 \right], \nonumber
\end{equation}

\begin{equation}
c_{36}^{\mathbf{27}} = h_1 \left[ \frac{1}{2916} c_3^2 \right], \nonumber
\end{equation}

\begin{equation}
c_{37}^{\mathbf{27}} = h_2 \left[ \frac{1}{8748} c_3^2 \right], \nonumber
\end{equation}

\begin{equation}
c_{38}^{\mathbf{27}} = h_2 \left[ - \frac{1}{13122} c_3^2 \right], \nonumber
\end{equation}

\begin{equation}
c_{39}^{\mathbf{27}} = h_2 \left[ \frac{1}{13122} c_3^2 \right], \nonumber
\end{equation}

\begin{equation}
c_{40}^{\mathbf{27}} = h_2 \left[ - \frac{1}{972} c_3^2 \right], \nonumber
\end{equation}

\begin{equation}
c_{41}^{\mathbf{27}} = h_2 \left[ \frac{1}{8748} c_3^2 \right], \nonumber
\end{equation}

\begin{equation}
c_{42}^{\mathbf{27}} = h_2 \left[ \frac{1}{13122} c_3^2 \right], \nonumber
\end{equation}

\begin{equation}
c_{43}^{\mathbf{27}} = h_2 \left[ - \frac{7}{8748} c_3^2 \right], \nonumber
\end{equation}

\begin{equation}
c_{44}^{\mathbf{27}} = h_2 \left[ \frac{7}{8748} c_3^2 \right], \nonumber
\end{equation}

\begin{equation}
c_{45}^{\mathbf{27}} = h_2 \left[ \frac{5}{17496} c_3^2 \right], \nonumber
\end{equation}

\begin{equation}
c_{46}^{\mathbf{27}} = h_2 \left[ \frac{1}{17496} c_3^2 \right], \nonumber
\end{equation}


\begin{thebibliography}{99}

%\cite{'tHooft:1974hx}
\bibitem{thooft}
G.~'t Hooft,
%"A Two-Dimensional Model For Mesons,"
Nucl.\ Phys.\ B {\bf 75}, 461 (1974).
%%CITATION = NUPHA,B75,461;%%

%\cite{Witten:1979kh}
\bibitem{witten}
E.~Witten,
%"Baryons In The 1/N Expansion,"
Nucl.\ Phys.\ B {\bf 160}, 57 (1979).
%%CITATION = NUPHA,B160,57;%%

%\cite{Jenkins:1995gc}
\bibitem{jen96}
E.~Jenkins,
%``Chiral Lagrangian for Baryons in the $1/N_c$ Expansion,''
Phys.\ Rev.\ D {\bf 53}, 2625 (1996).
%[arXiv:hep-ph/9509433].
%%CITATION = HEP-PH 9509433;%%

\bibitem{rfm06}
 R.~Flores-Mendieta and C.~P.~Hofmann,
 %``Renormalization of the baryon axial vector current in large-N(c) chiral perturbation theory,''
 Phys.\ Rev.\ D {\bf 74}, 094001 (2006).
 %[hep-ph/0609120].
 %%CITATION = HEP-PH/0609120;%%
 %8 citations counted in INSPIRE as of 10 Jul 2013

\bibitem{rfm12}
 R.~Flores-Mendieta, M.~A.~Hernandez-Ruiz and C.~P.~Hofmann,
 %``Renormalization of the baryon axial vector current in large-N_c chiral perturbation theory: Effects of the decuplet-octet mass difference and flavor symmetry breaking,''
 Phys.\ Rev.\ D {\bf 86}, 094041 (2012).
 %[arXiv:1210.8445 [hep-ph]].
 %%CITATION = ARXIV:1210.8445;%%
 %2 citations counted in INSPIRE as of 30 May 2013

%\cite{FloresMendieta:2009rq}
\bibitem{rfm09}
 R.~Flores-Mendieta,
 %``Baryon magnetic moments in large-N_c chiral perturbation theory,''
 Phys.\ Rev.\ D {\bf 80}, 094014 (2009).
 %[arXiv:0910.1103 [hep-ph]].
 %%CITATION = PHRVA,D80,094014;%%

%\cite{Ahuatzin:2010ef}
\bibitem{rfm14} 
 G.~Ahuatzin, R.~Flores-Mendieta, M.~A.~Hernandez-Ruiz, and C.P.~Hofmann,
 %``Baryon magnetic moments in large-$N_c$ chiral perturbation theory: Effects of the decuplet-octet mass difference and flavor symmetry breaking,''
 Phys.\ Rev.\ D {\bf 89}, 034012 (2014).
 %doi:10.1103/PhysRevD.89.034012
 %[arXiv:1011.5268 [hep-ph]].
 %%CITATION = doi:10.1103/PhysRevD.89.034012;%%
 %15 citations counted in INSPIRE as of 04 Nov 2018

%\cite{Flores-Mendieta:2014vaa}
\bibitem{rfm14a}
 R.~Flores-Mendieta and J.~L.~Goity,
 %``Baryon vector current in the chiral and 1/$N_c$ expansions,''
 Phys.\ Rev.\ D {\bf 90}, 114008 (2014).
% [arXiv:1407.0926 [hep-ph]].
 %%CITATION = ARXIV:1407.0926;%%

%\cite{Flores-Mendieta:2015wir}
\bibitem{rfm15} 
  R.~Flores-Mendieta and M.~A.~Rivera-Ruiz,
  %``Dirac form factors and electric charge radii of baryons in the combined chiral and 1/N$_c$ expansions,''
  Phys.\ Rev.\ D {\bf 92}, 094026 (2015).
%  doi:10.1103/PhysRevD.92.094026
%  [arXiv:1511.02932 [hep-ph]].
  %%CITATION = doi:10.1103/PhysRevD.92.094026;%%
  %4 citations counted in INSPIRE as of 28 Nov 2018

%\cite{Bijnens:1985kj}
\bibitem{bij}
  J.~Bijnens, H.~Sonoda and M.~B.~Wise,
  %``On the Validity of Chiral Perturbation Theory for Weak Hyperon Decays,''
  Nucl.\ Phys.\ B {\bf 261}, 185 (1985).
%  doi:10.1016/0550-3213(85)90569-3
  %%CITATION = doi:10.1016/0550-3213(85)90569-3;%%
  %109 citations counted in INSPIRE as of 14 Dec 2018

%\cite{Jenkins:1991bt}
\bibitem{jen92} 
 E.~E.~Jenkins,
 %``Hyperon nonleptonic decays in chiral perturbation theory,''
 Nucl.\ Phys.\ B {\bf 375}, 561 (1992).
% doi:10.1016/0550-3213(92)90111-N
 %%CITATION = doi:10.1016/0550-3213(92)90111-N;%%
 %71 citations counted in INSPIRE as of 30 Aug 2018

%\cite{Borasoy:1998ku}
\bibitem{b99} 
  B.~Borasoy and B.~R.~Holstein,
  %``Nonleptonic hyperon decays in chiral perturbation theory,''
  Eur.\ Phys.\ J.\ C {\bf 6}, 85 (1999).
%  doi:10.1007/s100529800896, 10.1007/s100520050323
%  [hep-ph/9805430].
  %%CITATION = doi:10.1007/s100529800896, 10.1007/s100520050323;%%
  %25 citations counted in INSPIRE as of 01 Dec 2018

%\cite{AbdElHady:1999mj}
\bibitem{abd} 
  A.~Abd El-Hady and J.~Tandean,
  %``Hyperon nonleptonic decays in chiral perturbation theory reexamined,''
  Phys.\ Rev.\ D {\bf 61}, 114014 (2000).
%  doi:10.1103/PhysRevD.61.114014
%  [hep-ph/9908498].
  %%CITATION = doi:10.1103/PhysRevD.61.114014;%%
  %11 citations counted in INSPIRE as of 01 Dec 2018

%\cite{Tanabashi:2018oca}
\bibitem{part} 
  M.~Tanabashi {\it et al.} [Particle Data Group],
  %``Review of Particle Physics,''
  Phys.\ Rev.\ D {\bf 98}, 030001 (2018).
%  doi:10.1103/PhysRevD.98.030001
  %%CITATION = doi:10.1103/PhysRevD.98.030001;%%
  %553 citations counted in INSPIRE as of 29 Nov 2018

%\cite{Dashen:1993as}
\bibitem{dm315}
R.~F.~Dashen and A.~V.~Manohar,
%``Baryon - pion couplings from large N(c) QCD,''
Phys.\ Lett.\ B {\bf 315}, 425 (1993); 438 (1993).
%[arXiv:hep-ph/9307241].
%%CITATION = HEP-PH 9307241;%%
%%CITATION = HEP-PH 9307242;%%

%\cite{Dashen:1994qi}
\bibitem{djm95}
R.~F.~Dashen, E.~Jenkins and A.~V.~Manohar,
%``Spin flavor structure of large N(c) baryons,''
Phys.\ Rev.\ D {\bf 51}, 3697 (1995).
%[arXiv:hep-ph/9411234].
%%CITATION = HEP-PH 9411234;%%

%\cite{Gervais:1984rc}
\bibitem{gs}
J.~L.~Gervais and B.~Sakita,
%``Large-N Baryonic Soliton And Quarks,''
Phys.\ Rev.\ Lett.\ {\bf 52}, 87 (1984);
Phys.\ Rev.\ D {\bf 30}, 1795 (1984).
%%CITATION = PRLTA,52,87;%%
%%CITATION = PHRVA,D30,1795;%%

%\cite{Jenkins:1995td}
\bibitem{jl}
 E.~E.~Jenkins and R.~F.~Lebed,
 %``Baryon mass splittings in the 1/N(c) expansion,''
 Phys.\ Rev.\ D {\bf 52}, 282 (1995).
 %doi:10.1103/PhysRevD.52.282
 %[hep-ph/9502227].
 %%CITATION = doi:10.1103/PhysRevD.52.282;%%
 %155 citations counted in INSPIRE as of 30 Aug 2018
E.\ Jenkins and R.\ F.\ Lebed, Phys. Rev. D 52, 282 ͑1996͒.

%\cite{Flores-Mendieta:1998ii}
\bibitem{rfm98}
R.~Flores-Mendieta, E.~Jenkins and A.~V.~Manohar,
%``SU(3) symmetry breaking in hyperon semileptonic decays,''
Phys.\ Rev.\ D {\bf 58}, 094028 (1998).
%[arXiv:hep-ph/9805416].
%%CITATION = HEP-PH 9805416;%%

%\cite{Lee:1964zzc}
\bibitem{lee} 
 B.~W.~Lee,
% %``Transformation Properties of Nonleptonic Weak Interactions,''
 Phys.\ Rev.\ Lett.\ {\bf 12}, 83 (1964).
% doi:10.1103/PhysRevLett.12.83
 %%CITATION = doi:10.1103/PhysRevLett.12.83;%%
 %123 citations counted in INSPIRE as of 22 Aug 2018

%\cite{Sugawara:1964zz}
\bibitem{sugawara} 
 H.~Sugawara,
% %``A New Triangle Relation for Nonleptonic Hyperon Decay Amplitudes as a Consequence of the Octet Spurion and the R Symmetry,''
 Prog.\ Theor.\ Phys.\ {\bf 31}, 213 (1964).
% doi:10.1143/PTP.31.213
 %%CITATION = doi:10.1143/PTP.31.213;%%
 %94 citations counted in INSPIRE as of 22 Aug 2018

%\cite{Flores-Mendieta:2000mz}
\bibitem{fmhjm}
R.~Flores-Mendieta, C.~P.~Hofmann, E.~Jenkins and A.\ V.\ Manohar,
%``On the structure of large N(c) cancellations in baryon chiral perturbation theory,''
Phys.\ Rev.\ D {\bf 62}, 034001 (2000).
%[arXiv:hep-ph/0001218].
%%CITATION = HEP-PH 0001218;%%

%\cite{Flores-Mendieta:2017gnx}
\bibitem{fm17} 
  R.~Flores-Mendieta and R.~Padron-Stevens,
  %``Sum rules for leading vector form factors in hyperon semileptonic decays,''
  Phys.\ Rev.\ D {\bf 95}, 076018 (2017).
  %doi:10.1103/PhysRevD.95.076018
  %[arXiv:1704.07429 [hep-ph]].
  %%CITATION = doi:10.1103/PhysRevD.95.076018;%%
  %1 citations counted in INSPIRE as of 09 Mar 2019

\end{thebibliography}
\end{document}